\RequirePackage{rotating}
\documentclass[fleqn,usenatbib]{mnras}

\usepackage{newtxtext,newtxmath}
\usepackage[T1]{fontenc}

\DeclareRobustCommand{\VAN}[3]{#2}
\let\VANthebibliography\thebibliography
\def\thebibliography{\DeclareRobustCommand{\VAN}[3]{##3}\VANthebibliography}

\usepackage{graphicx}	% Including figure files
\usepackage{amsmath}	% Advanced maths commands

\usepackage{float}
\usepackage{lscape}
\usepackage{rotating}
\usepackage{caption}
\usepackage{subcaption} %not needed if you are not using subfigure
\usepackage{morefloats}
\usepackage[flushleft]{threeparttable} %Set up three part table with notes
\usepackage{multicol} %Set up figures with multiple columns 
\usepackage{hyperref}
\hypersetup{colorlinks=true, linkcolor=blue, urlcolor=blue, citecolor=blue}
\usepackage{pdflscape} %Make Landscape tables
\usepackage{longtable} %Tables which go on more than one page

\usepackage{changepage}
\usepackage{array}

\usepackage{fancyvrb}

\usepackage{wrapfig}
\usepackage{lscape}
\usepackage{rotating}
\usepackage{epstopdf}

\newcommand{\DustT}{\texttt{Dustribution}}

\title[Milky Way All sky 3D dust density and extinction]{All-sky three-dimensional dust density and extinction Maps of the Milky Way out to 2.8~kpc}% \n 
\author[T. E. Dharmawardena et al.]{T. E. Dharmawardena$^{1,2,3}$\thanks{E-mail: tdharmawardena@flatironinstitute.org}\thanks{NASA Hubble Fellow},
C.A.L. Bailer-Jones$^{1}$,
M. Fouesneau $^{1}$,
D. Foreman-Mackey$^{2}$,
P. Coronica$^{4}$, \newauthor
T. Colnaghi$^{4}$,
T. M\"{u}ller$^{1}$ and 
A. G. Wilson$^{5}$
\\
$^{1}$Max Plank Institute for Astronomy (MPIA), Königstuhl 17, 69117 Heidelberg, Germany.\\
$^{2}$Center for Computational Astrophysics, Flatiron Institute, 162 5th Ave, New York, NY 10010, USA.\\
$^{3}$Center for Cosmology and Particle Physics, Department of Physics, 726 Broadway, Room 1005, New York University, New York, NY 10003, USA.\\
$^{4}$Max Planck Computing and Data Facility, Gießenbachstraße 2, 85748, Garching, Germany.\\
$^{5}$Courant Institute of Mathematical Sciences, New York University, 251 Mercer St, New York, NY 10012, USA.
}

\date{Accepted XXX. Received YYY; in original form ZZZ}

\pubyear{2015}

\begin{document}
\label{firstpage}
\pagerange{\pageref{firstpage}--\pageref{lastpage}}
\maketitle

%##################

% Abstract of the paper
\begin{abstract}
Three-dimensional dust density maps are crucial for understanding the structure of the interstellar medium of the Milky Way and the processes that shape it. However, constructing these maps requires large datasets and the methods used to analyse them are computationally expensive and difficult to scale up. As a result it is has only recently become possible to map kiloparsec-scale regions of our Galaxy at parsec-scale grid sampling. We present all-sky three-dimensional dust density and extinction maps of the Milky Way out to 2.8~kpc in distance from the Sun using the fast and scalable Gaussian Process algorithm \DustT. The sampling of the three-dimensional map is $l,b,d = 1^{\circ} \times1^{\circ} \times 1.7$~pc. The input extinction and distance catalogue contains 120 million stars with photometry and astrometry from Gaia DR2, 2MASS and AllWISE. This combines the strengths of optical and infrared data to probe deeper into the dusty regions of the Milky Way. We compare our maps with other published 3D dust maps. All maps quantitatively agree at the $0.001$~mag~pc$^{-1}$ scale with many qualitatively similar features, although each map also has its own features. We recover Galactic features previously identified in the literature. Moreover, we also see a large under-density that may correspond to an inter-arm or -spur gap towards the Galactic Centre. %, and also see evidence for low-density clouds within the Local Bubble.  

\end{abstract}
%, finding good agreement

% Select between one and six entries from the list of approved keywords.
% Don't make up new ones.

%%%%%%%%%%%%%%%%%%%%%%%%%%%%%%%%%%%%%%%%%%%%%%%%%%

%%%%%%%%%%%%%%%%% BODY OF PAPER %%%%%%%%%%%%%%%%%%

\section{Introduction}
%Still need to add something about Gaia here!
The three-dimensional (3D) structure of the Milky Way is a key component in astronomical studies. One of the best tracers for obtaining the 3D structures of the Milky Way is dust. The absorption and scattering of dust combined with distances to the obscured stars allow us to study the distribution of dust within the Milky Way and hence obtain a full picture in 3D. Three-dimensional dust extinction density maps (from here on simply referred to as dust density maps) of the Milky Way are crucial for understanding the structure of the Galactic interstellar medium and the processes that shaped it, including the coupling between dust and gas in the Milky Way, as well as its dynamics and processes in the cycle of matter which governs star- and planet-formation. For example, while global turbulence has been successfully measured in nearby galaxies from {\sc hi} emission \citep[e.g.][]{Dutta2013}, as we lie in the Galactic Plane a three-dimensional picture is essential to properly understand the turbulent behaviour of the Galactic interstellar medium (ISM) on both large and small scales, which at present is only available through dust maps under assumptions coupling between dust and gas. Similarly, 3D dust maps provide a much cleaner estimate of the dust masses of clouds, which in turn can improve our understanding of variations in the dust-to-gas ratio and the impact on star formation \citep{Dharmawardena2022}. Three dimensional dust maps also aid in many other aspects of research in astronomy, for example to subtract the reddening effect of the Milky Way's dust on extragalactic studies or estimate extinctions or distances to Galactic sources.   

Developments in modern statistical and machine-learning techniques have led to great strides in mapping the 3D structure of the dust in our Galaxy. Works such as \citet{Leike2020, Babusiaux2020, Lallement2019, Green2019_Bayestars19, Sale2018} exploit Bayesian inference, inversion, and Gaussian processes to produce both 3D dust density and extinction maps using various datasets and for different regions of the Milky Way. 

Mapping the Milky Way in 3D would be impossible without all-sky optical and infrared imaging and astrometric surveys. The Gaia survey \citep{2016Prusti} is of particular importance, providing optical photometry and more importantly parallaxes to over one billion stars from which we can estimate distances. Combining these optical data with infrared surveys such as 2MASS and WISE allows us to derive extinction to stars probing denser dusty regions.

To map dust in 3D, we must rely on measurements of extinction (which are an integral along the line-of-sight) and the distances to the stars being used to measure extinction. Gaussian processes are particularly useful in 3D mapping, because they allow us to smoothly are able to mitigate effects such as the finger-of-god effect by correlating all points in space. In \citet{Dharmawardena2022} we exploited Gaussian processes and variational inference to develop our novel 3D dust mapping algorithm \DustT. This method mitigated finger-of-god effects and allowed us to map both the small-scale and large-scale structure of the Galactic molecular clouds in short (hours to days) computational times. \citet{Dharmawardena2022} and \citet{Dharmawardena2023} use \DustT\ to map Galactic molecular clouds at distances of up to 2.5 kpc, revealing their structure in detail. 

In this work we present 3D dust density and extinction maps of the Milky Way out to 2.8 kpc with a sampling of 1.7 pc along the line-of-sight. The maps are constructed using the \DustT\ algorithm combined with a catalogue of dust extinctions and distances derived from Gaia DR2, 2MASS and ALLWISE surveys. We highlight features observed in the maps and compare it to works from literature for validation. The \DustT\ code is publicly available in GitHub at \url{https://github.com/Thavisha/Dustribution}. The predicted 3D density and extinction data sets of the Milky Way are available interactively via the \DustT\ website at \url{www.mwdust.com}, where they can also be downloaded from. These data sets are also available to download via Zenodo at \url{https://doi.org/10.5281/zenodo.11448780}.

\section{Deriving 3D dust density and extinction maps with \DustT}
\label{sec:Method_3D}

\subsection{\DustT\ Code}

In \citet{Dharmawardena2022} we introduced and validated our algorithm \DustT\ and in \citet{Dharmawardena2023}, we made improvements to it. Here, we provide a summary of our method, highlighting in particular further improvements made to it for the present work. In contrast to inversion methods used in some studies \citep[e.g.][]{Lallement2014, Lallement2019, Vergely2022}, \DustT\ uses a forward model to infer the 3D dust density and line-of-sight extinction of any user-selected region from arbitrary catalogues of stellar extinctions, distances. and angular coordinates of stars as input. Our algorithm uses latent variable GP models \citep{Rasmussen2006_GPbook} and variational inference \citep{Bishop2006_VIandELBObook, Blei2017:VI}. This allows it to directly account for the 3D correlations between any density points throughout the mapped region in order to infer a probability distribution over the logarithm of the dust density. 

For this inference we need a prior, for which we use a Gaussian process on the logarithm of density, which ensures we only predict positive densities. The predictions from this model are integrated along all lines-of-sight to which we have observable data in order to compute model extinctions; these are compared to the observed extinctions to compute the (Gaussian) likelihood. The covariance matrix of our GP is described by a radial basis function (RBF) kernel, a Gaussian kernel which is not to be confused with the GP itself \citep{Rasmussen2006_GPbook}. The RBF kernel has four hyperparameters: three physical scale lengths in the three heliocentric Cartesian coordinates ($x$, $y$, $z$) of physical space and one exponential scale factor. In addition, our GP includes a constant mean density, giving a total of five hyperparameters.

Our model set up is such that our GP is a prior on a set of latent variables of the logarithm of the density. This prior must then be integrated to predict the joint density distribution at all points, producing an extremely large parameter space as we have one free parameter for each grid cell's density. Such a parameter space would be prohibitively expensive to compute. To overcome this, we employ an inducing point method and variational inference. Inducing points introduce latent inputs that form a rank-$m$ approximation of the GP covariance matrix for $m$ inducing points, through interpolation. Often a very small number of inducing points $m \ll n$ (where $n$ is the number of grid points $10^{5}$) provide excellent performance (e.g.\ \citealt{Quinonero-candela_sparseGP, titsias_sparsevariationalGP, Sale2018}; Figure 1 in \citealt{Uhrenholt2021} demonstrates how this represents the underlying data just as well as a non-sparse GP). By using 1000 inducing points, we accelerate the problem by orders of magnitude to make it tractable. In addition, the locations of the inducing points are optimised to best-represent the entire space with the reduced number of points rather than relying on a specific set of pre-determined locations. On the other hand, the likelihood evaluation still exploits the entire input set of stars. To further reduce the computing time, we use variational inference, which replaces the target posterior with a simpler approximate posterior by finding the parameters for this approximation that best reproduces the true posterior. This allows us to directly compute the approximate posterior and its gradient with respect to the free parameters, thus enabling the use of gradient-descent optimisers in our model \citep[similar to][]{Hensman2013_BigGPs}.  We have implemented this using the GPyTorch \citep{Gardner2018_GpyTorch} and Pyro \citep{Bingham2018_Pyro1, Phan2019_Pyro2} packages. 

Several key stages of \DustT\ are now computed using the lazy tensor interface of GPyTorch, to eliminate unnecessary operations and memory usage. Lazy computing delays the evaluation of an expression until there is a need for it, meaning expressions are evaluated only as and when results are needed. As a result, unnecessary calculations are avoided, and the memory associated with them is never allocated. We also exploit pytorch's GPU interface to parallelise the computations. Collectively, all the aforementioned improvements result in a factor of 100 reduction in typical run-times and a factor of 200 reduction in memory usage.

\citet{Dharmawardena2023} showed that \DustT\ recovers the structures of Galactic molecular clouds, and revealed previously unidentified filaments outside the main star-forming clouds within 2.5~kpc of the Sun. By integrating the dust densities, we derived cloud masses with $12\%$ statistical uncertainties. However, like all GP methods, the prior introduces some smoothing, which tends to average out the sharpest peaks in density. This issue is compounded by the poor sampling of the highest-density points, both because they are compact (and hence there is a low probability of the required background stars) and because their extinctions are so high that background stars may not be detectable. These two effects tend to reduce the inferred peak densities.

\subsection{Mapping the Milky Way in 3D}

While \DustT\ is  capable of mapping molecular clouds quickly, when applied to the whole Milky Way, the memory required to compute the covariance matrices on a sufficiently fine grid in one go is beyond that available. We therefore cannot map the entire Milky Way with a single run of \DustT. 
To overcome this, we split the Milky Way into smaller overlapping regions (hereafter ``chunks"), run \DustT\ individually on each chunk, and then combine the chunks. We detail below our procedure and choices to define these chunks.
% The central limit theorem ensures that the combination converges in the limit of an infinite number of chunks to the result we would obtain by running on all the data at once. It then becomes a trade-off between computing feasibility and accuracy.

Each chunk covers a section of the sky $18^{\circ} \times 18^{\circ}$ in angular size and 1000~pc in distance along the line-of-sight. We run the mapping process once from $0^{\circ} < l < 360^{\circ}$; $-90^{\circ} < b < 90^{\circ}$ and then again shifted by half the length and width in $l$ and $b$ and by 600~pc in distance ($d$) for each chunk, to provide overlaps for all chunks (8 overlapping chunks in total). The distance ranges are displayed in Fig.\ref{fig:MergeBoundaries}. The high degree of overlap of the chunks increases the reliability of our final map and ensures the chunks join smoothly. 

Once all the chunks are processed, we combine them into one single map using a weighted median. For each chunk, we generate 100 sample maps from the GP to be combined. At each 3D location $(l, b, d)$, we then take the weighted median of all samples from all 8 overlapping chunks. This weighting reduces edge effects that are otherwise more prominent when using a plain median. It also avoids biasing the map with spurious high-densities introduced by only one of the chunks, as a mean (weighted or otherwise) would be susceptible to. 

The weighting scheme is the product of two components - one in angular coordinates, and the other in distance. 
The weighting function in $(l, b)$ is defined as
\begin{equation}
\label{eqn:lbweight}
    w(l, b) = 1 - \left(0.9999 * \frac{\theta\left(l, b\right)}{\theta_{\rm max}}\right)
\end{equation}
where $\theta\left(l, b\right)$ is the angular distance 
%but not the haversine distance
from the centre of the chunk (in $l, b$ only), $\theta_{\rm max}$ is the maximum distance from the centre of the chunk i.e. the distance from the centre to the corner. As we are using spherical coordinates, this maximum distance is the great circle length from the centre of the ''square'' chunk to the corner. The fraction is multiplied by 0.9999 to ensure that the smallest weight is $10^{-4}$ to avoid weights of 0.
This gives the highest weight to pixels close to the (angular) centre of the chunk, and negligible weight to pixels at the very edges.

\begin{figure}
  \centering
   \includegraphics[width=0.5\textwidth] {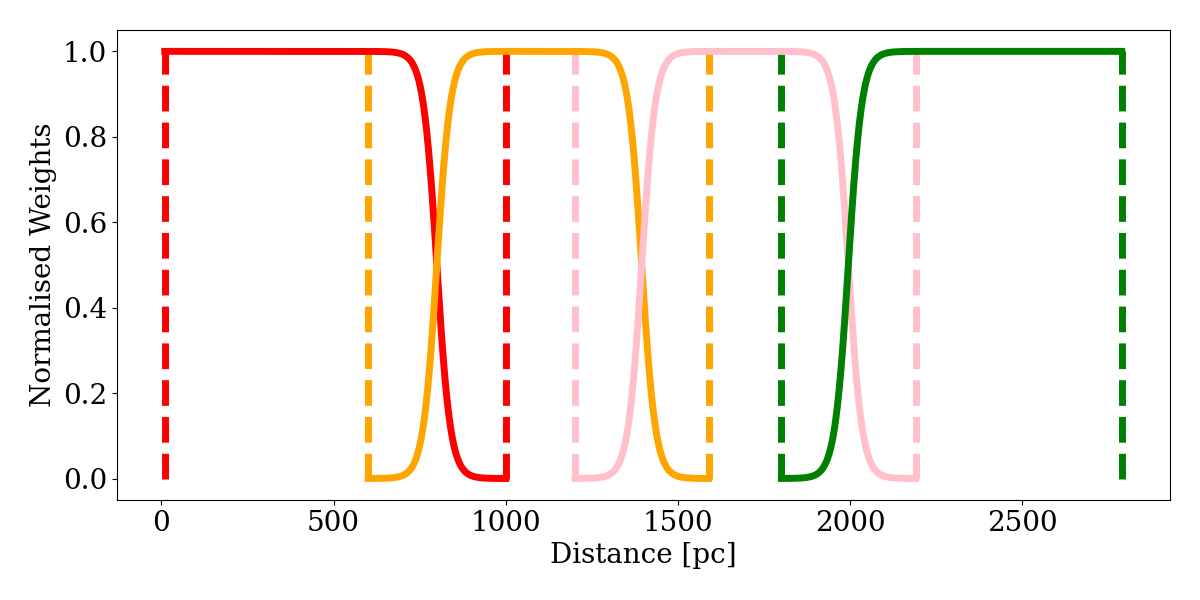}
 \caption{Representation of the normalised distance weighting scheme. Each distance range is represented by a different colour (Red: 10-1000~pc; Orange: 600-1590~pc; Pink: 1200-2190~pc; Green: 1800-2790~pc.). The solid lines represent the weights for each distance chunk; the dashed vertical lines in the corresponding colours to the solid lines show the chunk upper and lower boundaries ($d_{\rm min}$, $d_{\rm max}$).}
    \label{fig:weights}
\end{figure}

The weighting function in distance is a sigmoid function that smoothly grows from $10^{-4}$ to 1 (or conversely, shrinks from 1 to $10^{-4}$) over the region of overlap, and is precisely 0.5 at the mid-point of overlap, i.e.\
%Sigmoid Weighting Equation
\begin{equation}
\label{eqn:weightingDist}
    w(d) = \left\{ \begin{array}{ll} 
                10^{-4} + \frac{1}{1 + \exp- \frac{\left(d - \frac{\left(d_{\rm max}^{i-1} + d_{\rm min}^{i}\right)}{2}\right)}{20}
                } & \hspace{5mm} d\leq d_{\rm max}^{i-1} \\
                1 - \frac{1}{1 + \exp- \frac{\left(d - \frac{\left(d_{\rm min}^{i+1} + d_{\rm max}^{i}\right)}{2}\right)}{20}} & \hspace{5mm} d\geq d_{\rm min}^{i+1} \\
                1 & \hspace{5mm} \mathrm{otherwise} \\
                \end{array} \right.
\end{equation}
where $d_{\rm min}$ is the minimum distance value for the chunk, $d_{\rm max}$ is the maximum distance value for the chunk, $d_{\rm min}^{i+1}$ is the minimum distance value for the next chunk and $d_{\rm max}^{i-1}$ is the maximum distance value for the previous chunk. While this allows us to achieve as smooth and continuous a map as possible, we are nevertheless limited by the number of overlapping chunks we can run. Fig~\ref{fig:weights} shows the behaviour of the eqn.~\ref{eqn:weightingDist} distance weighting function. 

The final weight in 3D is given by
\begin{equation}
    w_{\rm tot} = w(d) \times w(l, b).
\end{equation}
These weights are used to determine the weighted-median map, which we take as our point estimate of the dust density in the volume we covered. 

We also determine point-wise credible intervals for the density by applying the same weighting scheme to determine the 16$^{\rm th}$- and 84$^{\rm th}$-percentile maps, analogous to an asymmetric 1-$\sigma$ interval. We use the python package \texttt{xarray} \citep{Hoyer2017_xarrayMain, Hoyer_xarrayV17} to derive the weighted quantiles using the above weighting function. Internally \texttt{xarray} normalises these weights by dividing by their sum for each pixel. 

By breaking the map into overlapping chunks and smoothly joining them, we can scale up our map to arbitrary volumes without encountering constraints on the total number of inducing points, memory usage or run time and still derive smooth continuous dust extinction and extinction density maps. Further validation of the merging technique is given in appendix \ref{sec:app:merge_validation}, along with a discussion of the limitations in the final dataset that result from this merging.

\subsection{Multiple scale length map}
\label{sec:MultiSL}

The ISM is a multi-scale medium, so a map produced with a single scale length may fail to capture all scales effectively without a prohibitively large number of inducing points. To recover structure on both small and large scales within reasonable hardware constraints,  we map each chunk twice with different initial scale lengths. We use an initial scale length of 10~pc to recover small-scale structures, and an initial scale length of 100~pc to recover larger-scale trends, both of which are optimised during the GP optimisation process. Both these maps are then merged with equal weight (otherwise following the weighting scheme described above) to recover both the dense and diffuse interstellar medium. 
%A comparison of the maps made with the two scale lengths separately is shown in appendix XX and the final map merging both scales is shown in appendix XX. 

\subsection{Input extinction and distance data set}
\label{sec:LBol}

As with \citet{Dharmawardena2022} and \citet{Dharmawardena2023}, the input data for our 3D maps comes from \citet{Fouesneau2022_LBol}. From this all-sky catalogue of ~120 million stars, we use their inferred extinction at 550~nm ($A_{0}$) and their distances as $d$ along with their reported uncertainties as well as $l$ and $b$ which are taken to be essentially error-free. For the present work, we extracted all stars within our $l,b,d$ boundaries and made no other cuts to the catalogue. We note that \DustT\ is wavelength agnostic and could use extinctions at any wavelength(s) from any catalogue. 

\citet{Fouesneau2022_LBol} simultaneously estimates distance, extinction parameters ($A_{0}$, $R_{0}$), effective temperature, luminosity, surface gravity, mass, and age for each star independently using photometry from Gaia DR2, 2MASS, and AllWISE together with with Gaia DR2 parallaxes. All sources must have all photometric bands to be included in to the catalogue. Using the Fitzpatrick extinction law \citep{Fitzpatrick1999}, PARSEC isochrone models, \citep{Chen2014_PARSEC1, Marigo2013_PARSEC2, Rosenfield2016_PARSEC3} and the ATLAS9 atmospheric library \citep{Castelli2004_ATLAS9}, their models predict the reddened spectral energy distributions (SEDs) from the input data. The results are estimated from Markov chain Monte Carlo (MCMC) samples of the posterior parameter distribution using a neural network is used to interpolate between the forwardmodels. The median typical uncertainties for extinction are 0.34 mag and the fractional parallax uncertainties are between 0.13 and 0.19. 

% MF: 
We draw attention to an important issue with using extinction estimates derived in part from optical photometry, such as here with Gaia: In very dusty regions (e.g.\ the Galactic Center or cores of molecular clouds) the observed number of stars significantly decreases, creating possible spatial gaps in highly reddened regions. Based on the selection function defined in \citet{Cantat-Gaudin2023}, we estimate that Gaia is over 90\% complete in uncrowded regions of the sky for stars with $G > 20$mag, corresponding to the faintest stars in \citet{Fouesneau2022_LBol}. 

Two reasons motivated our choice to use the \citet{Fouesneau2022_LBol} catalogue over other Gaia-DR2-based catalogues or the more recent Gaia-DR3 catalogue: 1.\ As the catalogue uses 2MASS and AllWISE photometry, the broad wavelength coverage from optical to infrared (0.3-5.0\,$\mu$m) will improve the precision of the extinction estimates, particularly at high extinction, thanks to the longer lever-arm in wavelength, which helps to break the $T-A_{\rm V}$ degeneracy, significantly improving the precision of the stellar parameters, particularly in dense dusty regions such as molecular clouds 2.\ This catalogue has been extensively validated providing reliable parameters required for our maps. Gaia DR3 improvements lie primarily in spectral information. 

At the time of writing the Gaia DR3 (spectro)photometry remains challenging to analyse with non-empirical methods (e.g.\ \citealt{Fouesneau2023_GaiaDR3}, \citealt{Babusiaux2023_GaiaDR3}) but seems promising with empirical approaches (e.g.\ \citealt{Andrae2023_GDR3}, \citealt{Rix2022} and \citealt{Zhang2023_catalogue}). However, the systematics resulting from these analyses and their impact on extinction estimates have not yet been characterised in the literature; this leads to significant uncertainty on their suitability for dust mapping, and exploring this is beyond the scope of this paper. In particular, the stellar labels used to supervise the data-driven models used so far cover only a small part of the required parameter space, and so are unable to capture information which is key to dust mapping \citep{LarocheSpeagle2023}. Once either this label problem can be overcome or new non-empirical catalogues are released which exploit the spectrophotometry and improved parallaxes of DR3, we can expect significant improvements in our types of applications. Our mapping approach effectively pools information from many stars at any given point and hence parallaxes do not contribute uniformly to the map; in regions with many stars improved parallaxes may contribute less to the resulting inference whereas they may contribute more to regions with lower stellar density. Therefore, it is not obvious that the uncertainty on the distance in the dust map is directly related to the uncertainty of the input parallaxes. Exploring these effects deserves careful scrutiny and is much beyond the scope of this paper. % lead to significant differences, in particular in extinction parameters that are not discussed in the literature and go beyond the scope of this paper.

\begin{figure}
    \centering
    \includegraphics[width=0.5\textwidth]%trim: left bottom right top
    {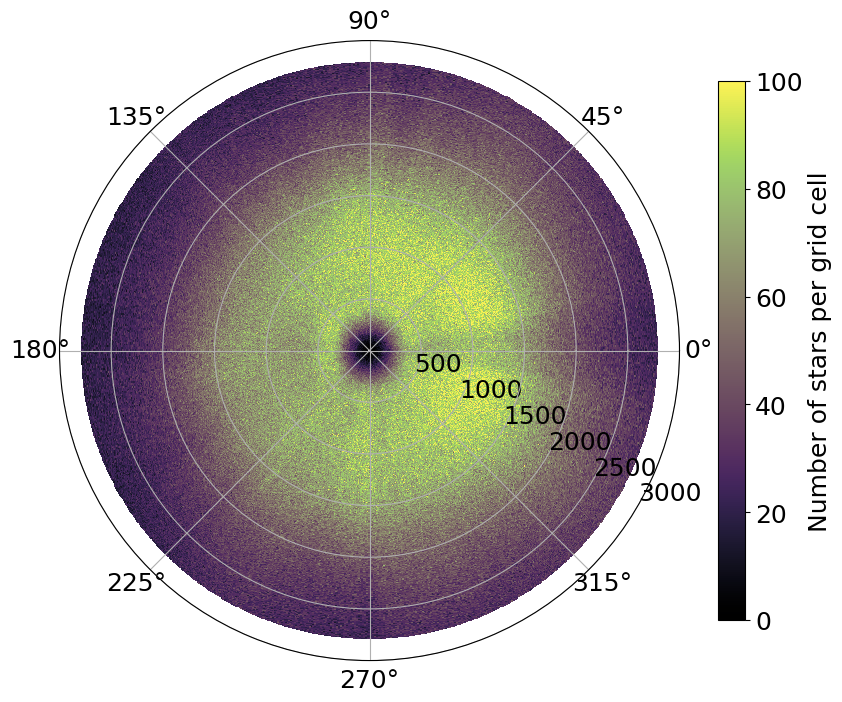}
    \caption{Number of input sources, binned in the (l, d) bins used to fit the GP. The data was first binned into the (l, b, d) bins used in our model, then summed over the b bins. The view shown is centred on the sun. The view shown is centred on the Sun with radial distance bins increasing outwards from the Sun in parsecs along the plane shown. The longitude $l$ is seen increasing counter clockwise in degrees.}
    \label{fig:SourceDensity_LBol}
\end{figure}

\begin{figure}
    \centering
    \includegraphics[width=0.5\textwidth]%trim: left bottom right top
    {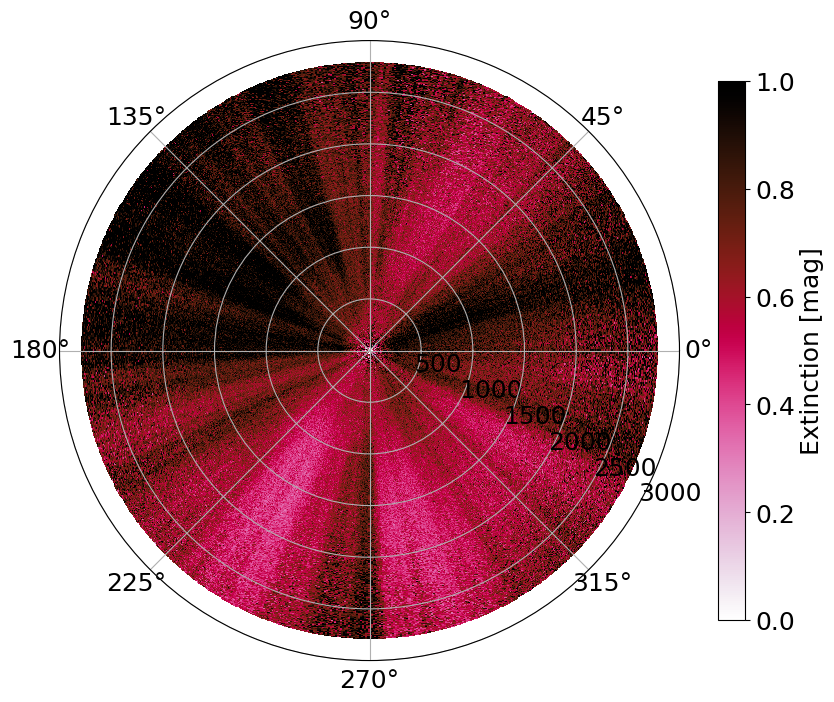}
    \caption{Input extinction data, averaged in the (l, d) bins used to fit the GP. The data was first binned into the (l, b, d) bins used in our model, then averaged over the b bins. The view shown is centred on the Sun with radial distance bins increasing outwards from the Sun in parsecs along the plane shown. The longitude $l$ is seen increasing counter clockwise in degrees.}
    \label{fig:SourceExtsBinned_LBol}
\end{figure}

\begin{figure}
    \centering
    \includegraphics[width=0.5\textwidth]%trim: left bottom right top
    {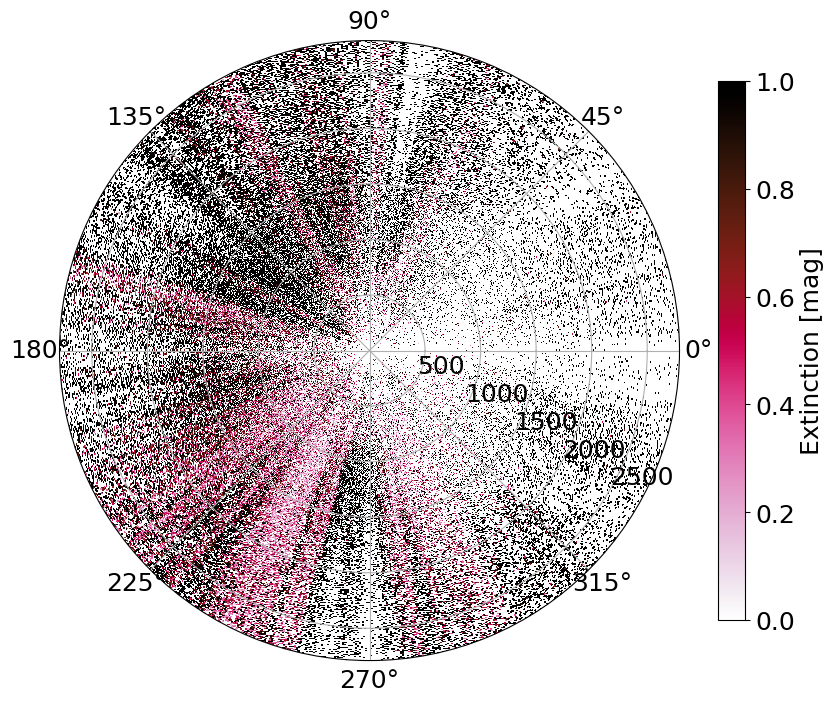}
    % \caption{Input extinction data, averaged in the (l, d) bins used to fit the GP. The data was first binned into the (l, b, d) bins used in our model, then averaged over the b bins.}
%     \label{fig:SourceExtsBinned_GalPlane_LBol}
% \end{figure}

% \begin{figure}
%     \centering
    \includegraphics[width=0.5\textwidth]%trim: left bottom right top
    {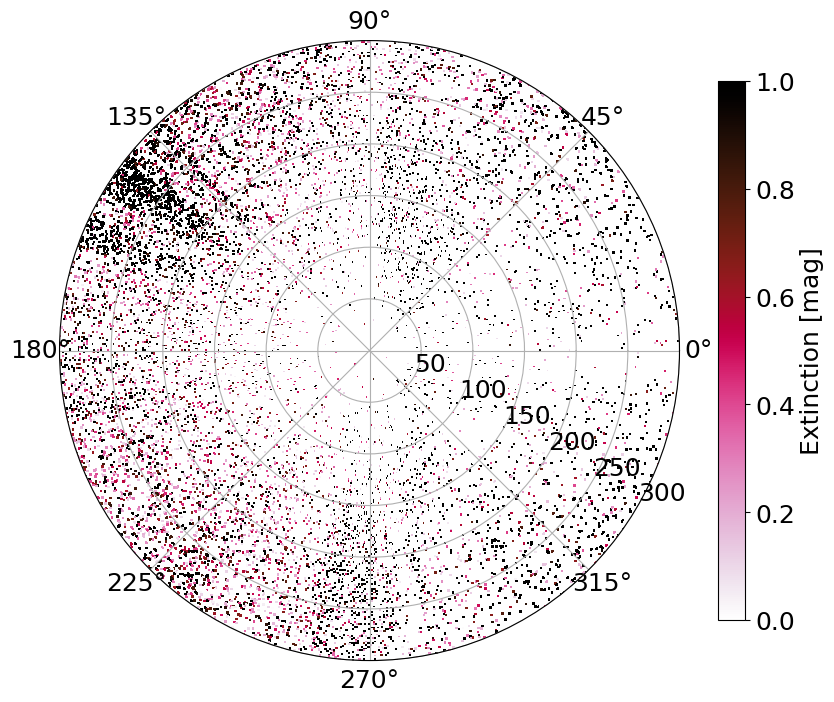}
    \caption{Input extinction data in the galactic plane centred on the sun. The data was first binned into the (l, b, d) bins used in our model, then averaged over the two b bins adjacent to b=0. \textit{Top:} the entire 2.8 kpc radius of our model. \textit{Bottom:} as top, but zoomed in to the region within 300\,pc of the Sun. The view shown is centred on the Sun with radial distance bins increasing outwards from the Sun in parsecs along the plane shown. The longitude $l$ is seen increasing counter clockwise in degrees.}
    \label{fig:SourceExtsBinned_GalPlane_LBol}
\end{figure}

Figures~\ref{fig:SourceDensity_LBol} to \ref{fig:SourceExtsBinned_GalPlane_LBol} show the number counts and mean extinctions of the input sources used in this paper, binned based on the grid used in the dust density model. In fig.~\ref{fig:SourceDensity_LBol}, we see that the source density remains high up to a distance of 2~kpc, and then drops thereafter, with slightly lower density towards the Galactic Centre (because of the selections made in the creation of the catalogue). This decreasing density translates into a decreased availability of background stars to constrain the map at large distances, suggesting that greater care should be taken in interpreting the outermost reaches of the map, particularly beyond 2~kpc. The small size of the bins close to the Sun also means that the counts per bin drops at very small distances. In total, there are 23,565 sources within 50~pc, 179,136 within 100~pc, 1,262,896 within 200~pc. In total we have 81 million sources out to 2.9~kpc.

Figure~\ref{fig:SourceExtsBinned_LBol} shows that the average input extinction rises rapidly and consistently within a few hundred parsecs of the Sun, with only a few lines of sight showing consistently low extinction. Where fig.~\ref{fig:SourceExtsBinned_LBol} averages over all $b$ bins, fig.~\ref{fig:SourceExtsBinned_GalPlane_LBol} shows only the sources close to the Galactic plane with $\left|b\right| \leq 0.4^{o}$. As expected, we see that the mean extinction rises faster in the Plane and there are fewer low-extinction directions. Zooming in to sources within 300~pc (fig.~\ref{fig:SourceExtsBinned_GalPlane_LBol}, lower panel), we see that there are already sources within 50~pc of the Sun with $A_{\rm V} \geq 1$~mag. 

\section{Results}
\label{sec:Results}

We produce two sets of maps in this work: 1. The dust density across the Milky Way in three dimensions in units of mag~pc$^{-1}$; 2. The integrated dust extinction across the Milky Way as seen from the Sun in units of mag. As stated in Sec.\ref{sec:LBol} extinction is calculated at 550 nm, i.e $A_{0}$. To convert our maps to $A_{V}$ one can assume a conversion factor of 1.003 based on the extinction curve from \citet{Gordon2023}. All maps are predicted on a grid with a sampling of 1.7~pc along the l.o.s and the $l,b$ direction is mapped at a sampling of $1^{\circ}$. 

\begin{figure*}
    \centering
    \begin{subfigure}{\textwidth}
    \includegraphics[width=\textwidth, trim=5cm 2cm 3cm 2cm, clip]{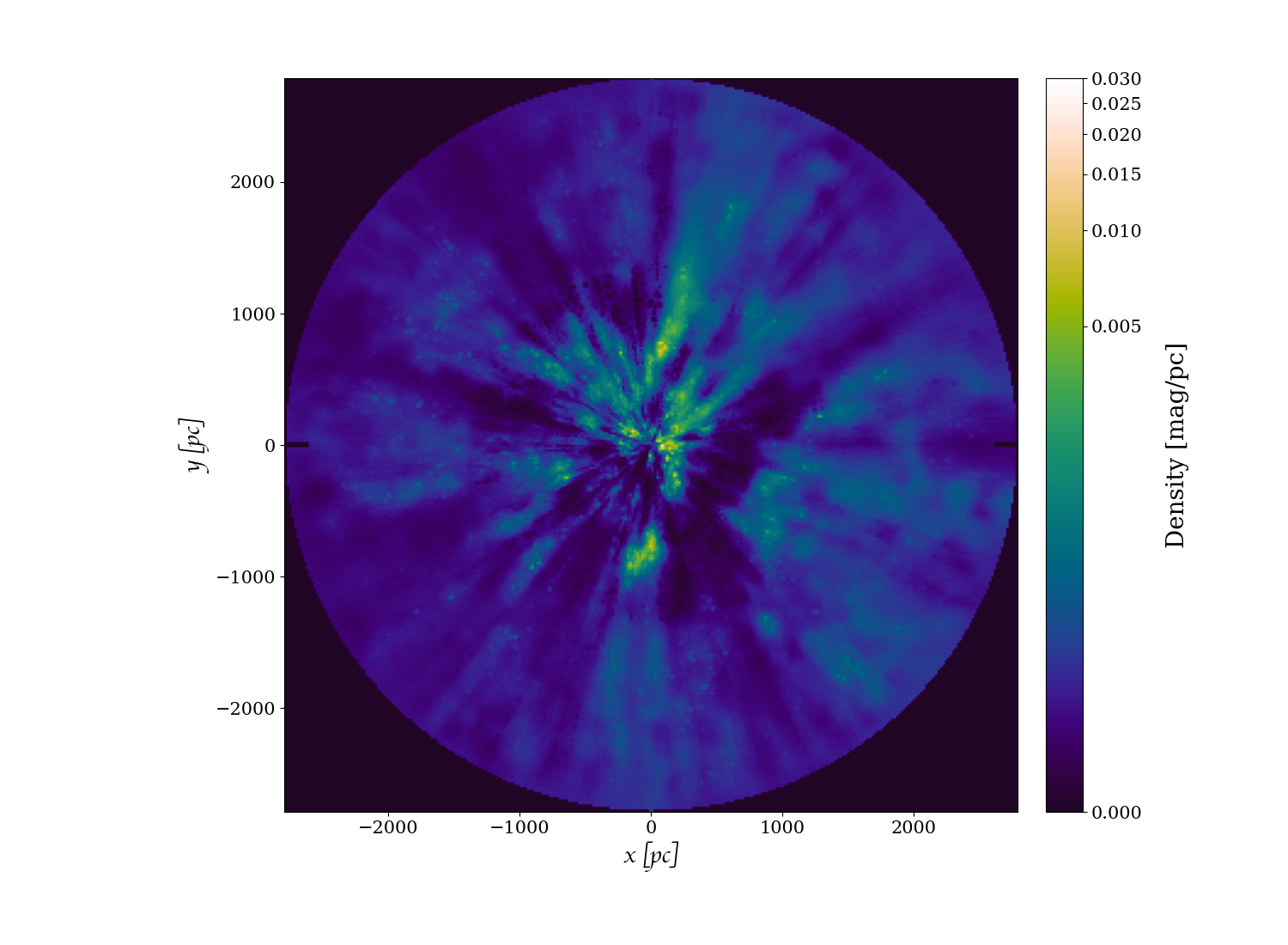} %trim: left bottom right top
    \end{subfigure}

    \begin{subfigure}{\textwidth}
    \includegraphics[width=\textwidth, trim=1cm 2cm 3cm 2cm, clip]%trim: left bottom right top
    {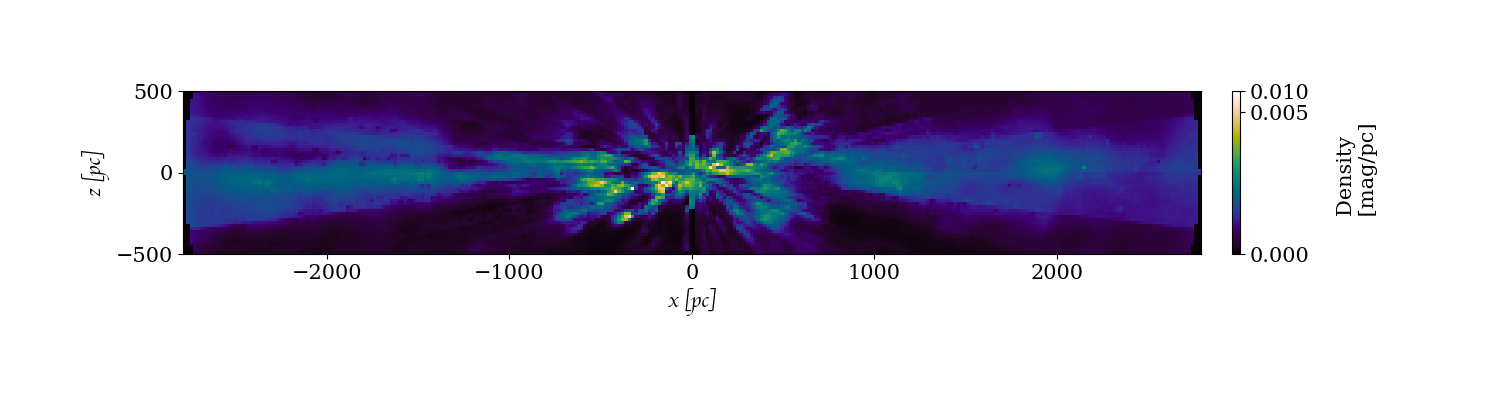}
    \end{subfigure}

    \begin{subfigure}{\textwidth}
    \includegraphics[width=\textwidth, trim=1cm 2cm 3cm 2cm, clip]%trim: left bottom right top
    {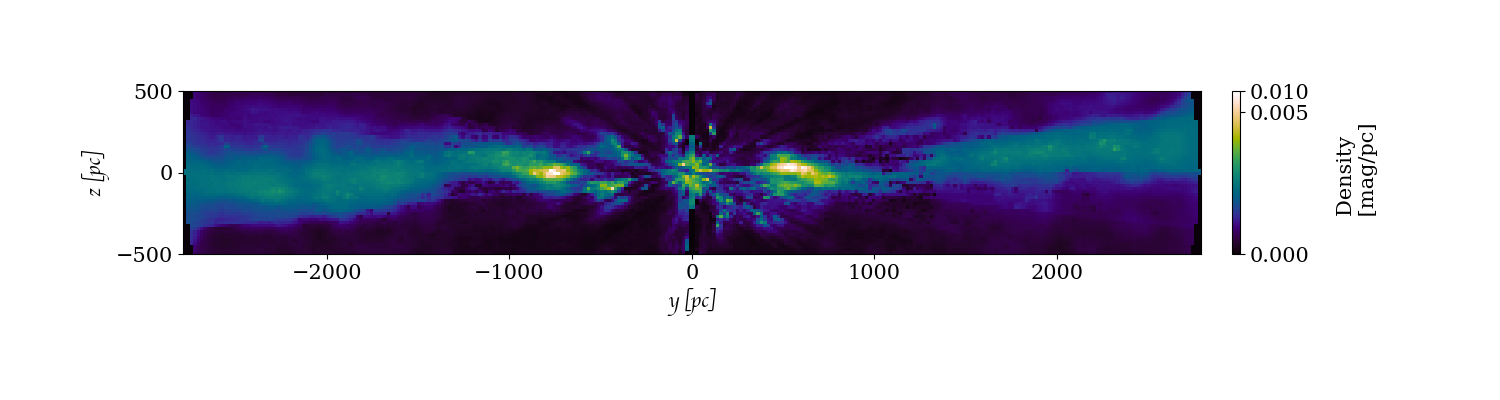}
    \end{subfigure}
    
    \caption{Heliocentric Cartesian views of the dust density, showing slices through the z=0 (top), y=0 (middle) and x=0 (bottom) planes. The Galactic centre is outside of this map at $x,y,z = +8122, 0, -20.8$.}
    
\label{fig:xy_dustdense}
\end{figure*}

In Fig.~\ref{fig:xy_dustdense}, we show the heliocentric Cartesian view of the dust density in slices through the z=0, y=0 and x=0 planes. The 16$^{\rm th}$ and 84$^{\rm th}$ percentiles of the dust density maps are shown in appendix~\ref{sec:app:Percentiles}. In Fig.~\ref{fig:xy_dustInteg} we plot the integrated dust extinction, integrated along $-500 \leq z \leq 500$~pc. We use this figure to mark Galactic features and highlight interesting over- and under-densities. Line-of-sight extinction profiles across the Galaxy are shown in Fig.~\ref{fig:ExtLos}. 

The plots shown in heliocentric Cartesian coordinates are derived by mapping our Galactic coordinates data cubes to a regular Cartesian grid using simple nearest-neighbour interpolation where the nearest neighbour is defined in Galactic coordinates. This introduces some aliasing and hence these figures should only be used for visualisation; for numerical inference the original cubes produced by \DustT\ in $l,b,d$ should be used.

To determine the size of the smallest and largest coherent dust density structures recovered in our maps, following the same prescription as \citet{Dharmawardena2023} Sec. 3, we apply \texttt{astrodendro}\footnote{\url{https://dendrograms.readthedocs.io/en/stable}}, to our dust density map. \texttt{Astrodendro} is a dendrogram-based hierarchical structure-finding tool, which performs top-down segmentation of a dataset starting from the highest data values. Following the approach taken in \citet{Dharmawardena2023}, we choose the minimum density above which cells are able to contribute to structures ($min\_value$) equal to $4 \times 10^{-4}$ mag pc$^{-1}$, the minimum difference ($min\_delta$) to be a 2 times the $min\_value$, and the minimum number of pixels in a structure ($npix$) is set to 8, the number that fit within a sphere of diameter 10~pc. The smallest density structures we recover (leaves in the dendrogram) are a few tens of parsecs (18~pc at minimum) and we recover 700 leaves, The largest structures (trunks in the dendrogram) are kilo-parsec scales (3~kpc at maximum) and we recover 500 trunks.

\begin{figure*}
    \centering
    \includegraphics[width=\textwidth, ]%trim: left bottom right top. trim=5cm 2cm 3cm 2cm, clip
{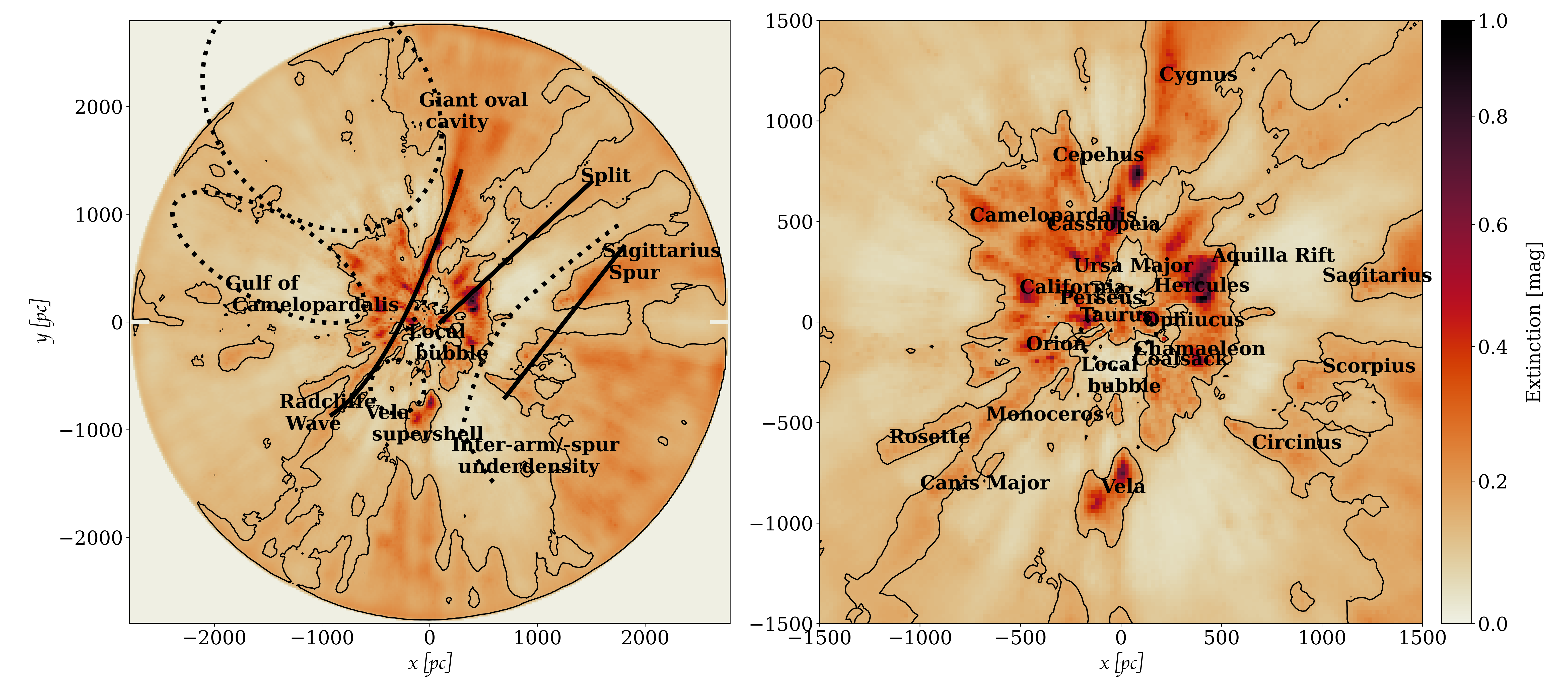}
\caption{Integrated dust extinction along the $z$ direction ($-500 \leq z \leq 500$~pc; perpendicular to the Galactic plane) as seen from above the Galactic plane. The contours in both panels represent extinction at 0.2~mag. The right panel is a zoom in of the left panel showing the region within 1500~pc of the Sun. In the left panel we highlight several large scale over-densities such as the Split in solid navy and under-densities such as the inter-arm/-spur region and cavities in dotted navy lines. Note that the cavities are marked in ovals to best match their locations, but they are unlikely to be spheroidal in reality.  In the right panel many of the known local molecular clouds are labelled in black text.}
    \label{fig:xy_dustInteg}
\end{figure*}

\begin{figure*}
    \centering
    \includegraphics[width=\textwidth, trim=3cm 2cm 3cm 0cm, clip]%trim: left bottom right top
    {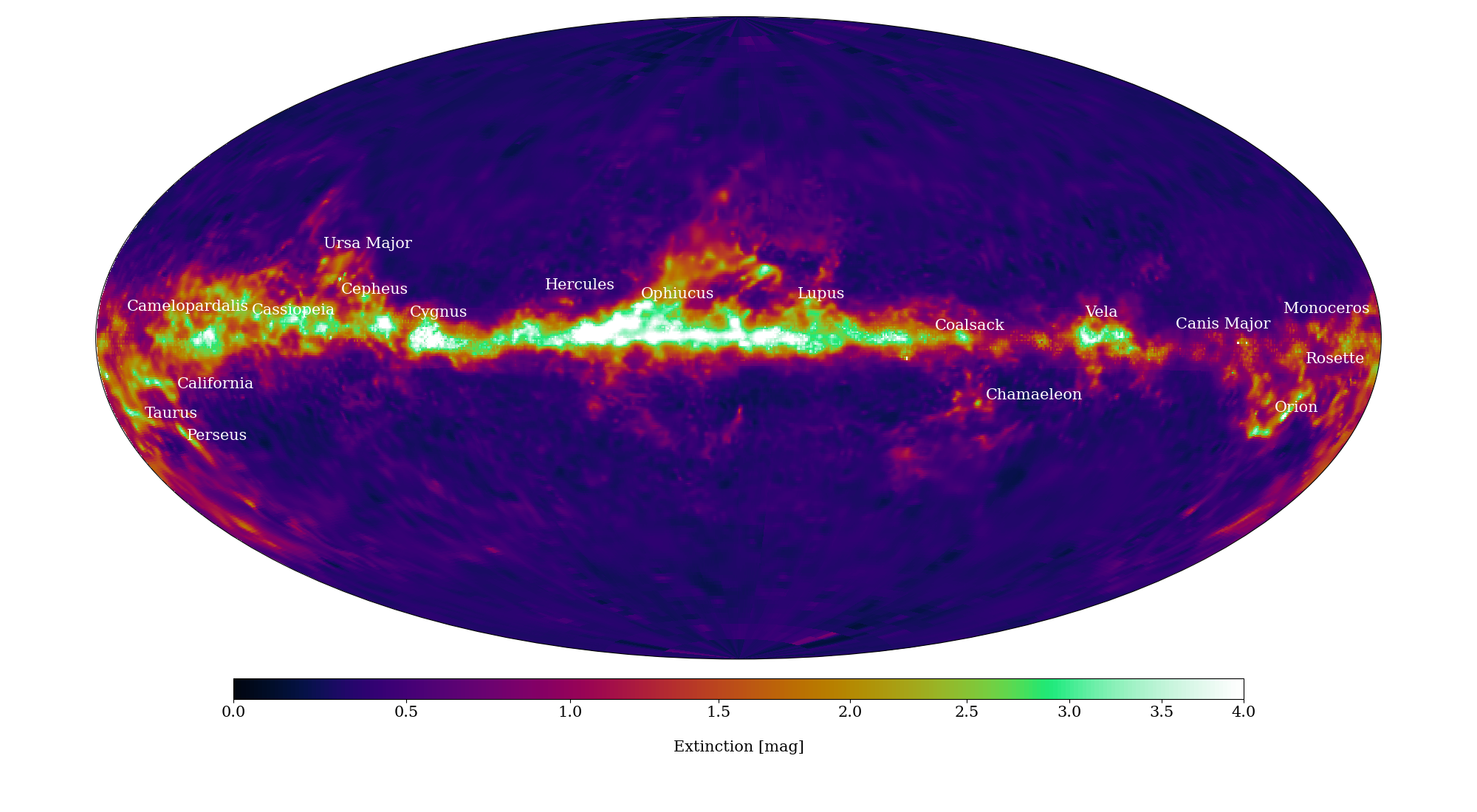}
    \caption{Integrated extinction along the line-of-sight from the Sun out to the complete mapped distance. Some of the most well known Galactic molecular clouds are highlighted, however, this is not exhaustive.}
    \label{fig:fullMWlb}
\end{figure*}

%\section{Validation of our maps}
\section{Discussion}
\label{sec:ValidationMain}

%\subsection{Validation by comparison to literature maps}
%\label{sec:LitMaps}

\subsection{Observed features} %Add some discussion of PHANGS JWST maps

In both Figs.~\ref{fig:xy_dustdense} and~\ref{fig:xy_dustInteg} we see a variety of Galactic features in both over-densities (clouds, dust lanes) and under-densities (cavities, large low density region). The nearby dense star-forming clouds are clearly visible, and their large-scale shapes match those from the detailed reconstructions of \citet{Dharmawardena2022} and \citet{Dharmawardena2023}. The large scale structures known as the Radcliffe Wave \citep{Alves2020_RadWave} and the Split \citep{Lallement2019} are both visible and marked for clarity, along with cavities between them. More distant clouds such as Circinus, Scorpius, and Sagittarius are visible towards the Galactic Centre forming parts of the larger structure, the Sagittarius spur \citep{Kuhn2021}.

We see possible evidence of the inner part of the ``giant oval cavity'' identified by \citet{Vergely2022}. However, our map does not extend to the full distance of their identified cavity and therefore we cannot confirm its presence in our results. Another large scale cavity of interest is the Gulf of Camelopardalis north of the Galactic anti-center extending $2$~kpc in length. The Vela supershell is clearly visible as a cavity in Fig.~\ref{fig:xy_dustInteg} next to the Vela molecular ridge. Interestingly, we also recover a large under-density seen as a low density curved arc in the direction of the Galactic centre, centred at $x,y \sim 0,750$~pc which could be an inter-arm/-spur region. 

\begin{figure}
    \centering
    \includegraphics[width=0.45\textwidth]%trim: left bottom right top
{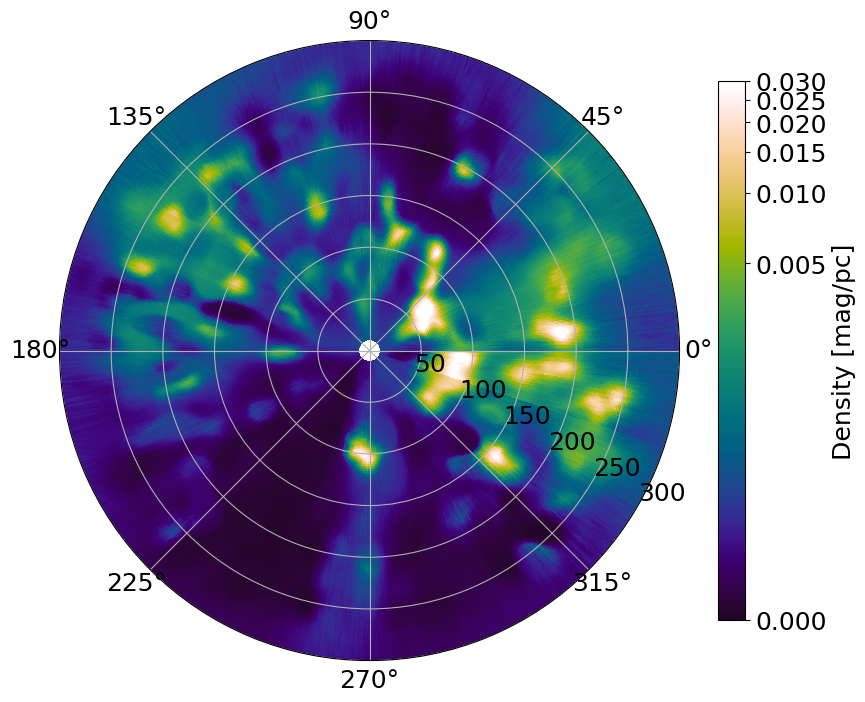}
    \caption{Zoom in on dust density of the Galactic Plane at $b$ and $z$ $=0$. The view shown is centred on the Sun with radial distance bins increasing outwards from the Sun in parsecs along the plane shown. The longitude $l$ is seen increasing counter clockwise in degrees. }
    \label{fig:LocBubDensity}
\end{figure}

It is interesting to zoom in on the immediate Solar Neighbourhood. The Sun is currently transiting a region of hot, low-density gas known as the Local Bubble, believed to have been created by a nearby supernova \citep{Cox1987, Breitschwerdt2016, MaizApellaniz2001}. Thanks to Gaia, it has recently become possible to map the edges of this bubble in detail \citep{Farhang2019}, and it has also been mapped based on 3D dust maps \citep[e.g.][]{Pelgrims2020, ONeill2024}, although different input data result in significantly different interpretations of the shape of the Bubble. To explore this region in our map, we zoom in on the region within 300~pc of the Sun in fig.~\ref{fig:LocBubDensity}, which shows the average density in the cells which have a boundary at $b=0$, effectively tracing the density in the Galactic plane. We do not see any dense dust clouds within 50~pc of the Sun, although there is a significant amount of low-density dust spread throughout the region, some of which may be a result of dust clouds being elongated along the line of sight in the GP reconstruction. However, we do not see any evidence of unphysical ``pile-up'' of dust at the inner edges of our map, unlike \citet{Edenhofer2023} who had to mask the inner 69~pc to avoid this. As a result, the \citet{ONeill2024} interpretation of the Local Bubble is forced to consider an inner boundary that is always at least 69~pc; given that the shape appears to strongly depend on the dataset used to define the Bubble, an alternative definition based on our map would result in a different shape altogether. However, creating such a Bubble map is beyond the scope of this paper. 
The most significant nearby clouds that we see are located at $x,y,z = -10, -86, 0$~pc and $x,y,z = -10, 100, 0$~pc and they have distances of 87 and 110~pc and peak densities of $0.00446$~mag~pc$^{-1}$ and $0.00225$~mag~pc$^{-1}$ respectively. If we were to define an edge for the Local Bubble based on our map, these would likely define the inner edge in their respective directions, however they lie within the edge defined in both \citet{Pelgrims2020} and \citet{ONeill2024}. Both clouds are rather low mass compared to the known nearby star-forming clouds. 
As described in Sect.~\ref{sec:LBol}, the source count within 50~pc is 23,565 allowing us to be confident in the quality of our reconstruction even at such small distances.

% Closer to the Sun we recover the Local Bubble in both Figs.\ref{fig:xy_dustdense} and \ref{fig:xy_dustInteg} (marked in the right hand panel). We shown a zoom in in Fig XXX. This figure shows the average of the density in the cells which have a boundary at $b=0$, effectively the density in the Galactic plane. XXXXXXX. We show this plot in 
% Acknowledge that some of the elopngation could come from stuff being spread out byu the GP.
% Edenhofer cuts mapp off at 69 pc, we don't because we don't see any evidence of the pile up in our map.

% We also observe small dust clouds in both figures within the Local Bubble indicating it is not completely devoid of dust. The densest point of the two most distinct clouds are located at $x,y,z = -10, -86, 0$ and $x,y,z = -10, 100, 0$ and they have a peak density of $0.00446$~mag~pc$^{-1}$ and $0.00225$~mag~pc$^{-1}$ respectively. In Fig.~\ref{fig:xy_dustInteg}, which shows the integrated extinction along $-500 \leq z \leq 500$~pc, we see evidence for dust within and surrounding the Local Bubble. Since it includes the contribution of off-plane material, it shows more material than is visible in Fig.~\ref{fig:xy_dustdense}, which may be above, below or inside the Local Bubble. 

\subsection{Validating with published Molecular clouds, YSO and Maser catalogues}

\begin{figure}
    \centering
    \includegraphics[width=0.45\textwidth, trim=6cm 3cm 3.5cm 3cm, clip]%trim: left bottom right top
{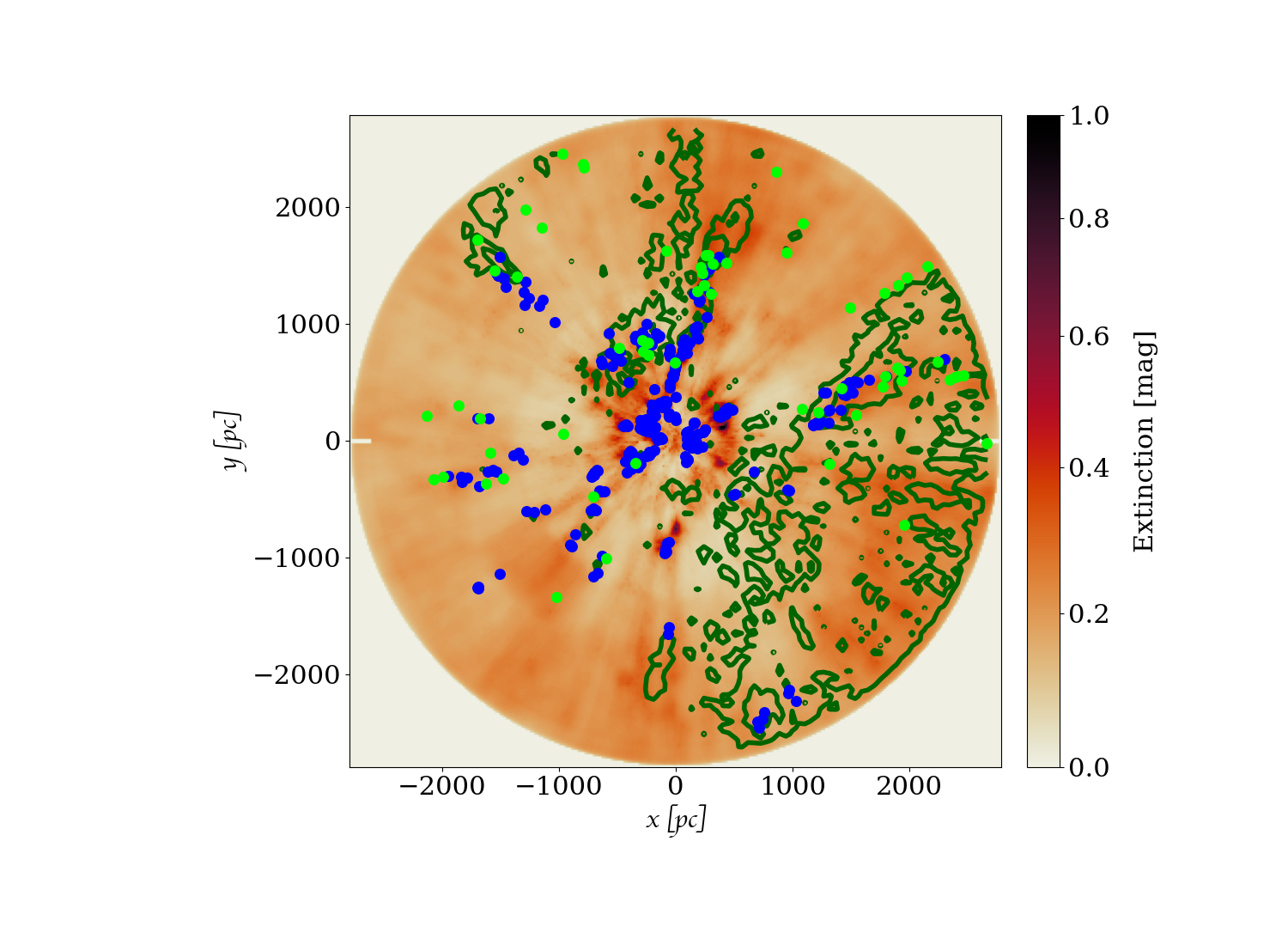}
    \caption{Comparison of this work (underlying map) to molecular clouds from \citet{Zucker2020} (blue dots), masers from \citet{Reid2019} (lime dots) and the surface density of OBA stars from \citet{Zari2021} (green contours: lowest contour is at 10$^4$ kpc$^{-2}$ or 30 stars per bin).}
    \label{fig:xy_cloudsmasersYSOs}
\end{figure}

In Fig.~\ref{fig:xy_cloudsmasersYSOs} we compare objects which trace of star formation to our integrated dust extinction map. Specifically, we overplot the molecular clouds that fall within our mapped region from the \citet{Zucker2020} molecular cloud catalogue, the catalogue of masers by \citet{Reid2019}, which are confirmed YSO masers, and the density of short-lived OBA stars from \citet{Zari2021}. Once again we see good agreement between these and the locations of our high density regions, which probably trace molecular clouds. 
Some maser sources near the edge of the map do not correspond to any dust clouds, however: these sources may be outside our map, given the uncertainties on their parallaxes. Naturally, the uncertainties between different input parallaxes or distances will result in disagreement between maps. Interestingly the density of YSOs from \citet{Zari2021} corresponds well with dense dust regions: the YSOs avoid the under-density which we have identified as a possible inter-arm/-spur region as well as the gulf of Camelopardalis. 

\subsection{Comparison to dust maps in literature}
\label{sec:LitCompMaps}
%Add loc.bubble cloud size and density

% Why is it important to compare different maps against each other
All reconstructions of dust density are subject to statistical and systematic uncertainty, which can be difficult to quantify given the absence of ground truth to compare to, or comprehensive shared benchmarks between the different approaches taken to produce the maps.
In that case, the best way to understand differences between the maps is to perform a quantitative comparison between them by transforming them to a common system so their values at the same locations can be compared, for example by interpolation.
While this necessarily alters the resolution of at least one of the maps in the comparison, potentially altering the values being compared, this remains the only way to directly compare two maps with different sampling unless a common system can be agreed on by the community.
Here, we attempt to do so for our map and two recent 3D dust density maps of the Milky Way in the literature from \citet{Vergely2022} and \citet{Edenhofer2023}.
For completeness, all three maps are shown side-by-side at native sampling in appendix~\ref{sec:app:litmaps}.

% Two of the most recent 3D dust density maps of the Milky Way in the literature are from \citet{Vergely2022} and \citet{Edenhofer2023}. 
\citet{Vergely2022} used a Bayesian inversion technique and mapped a region of $x,y,z = 10~\mathrm{kpc}\,\times 10~\mathrm{kpc} \times 800~\mathrm{pc}$ around the Sun at varying grid samplings of 10~pc to 50~pc. They used a catalogue of extinctions and densities derived from a combination of optical (Gaia DR2, SDSS, Pan-STARRS) and near-IR data (2MASS, WISE). The inner region of their map ($x,y,z = 3~\mathrm{kpc}\, \times 3~\mathrm{kpc}\, \times 800~\mathrm{pc}$) is sampled at 10~pc which is the highest sampling of their map. \citet{Edenhofer2023} used a Gaussian process and metric Gaussian variational inference to map the dust density of the Milky Way up to a radius of 1.2~kpc from the Sun. Their mapping grid used logarithmic sampling in distance, and the resulting maps where interpolated to 1~pc sampling for release to the community. They used a catalogue of extinction and distances derived from optical data (Gaia DR3 G photometry and BP/RP spectra and photometry).

To compare our maps to theirs, we compare the dust density in the $z=0$ plane for each data set. We remap their data onto our sampling grid and use the same colour scale as in Fig.~\ref{fig:xy_dustdense}. To compare the maps we need to put them in the same units. \citet{Vergely2022} give their map in units of $A_V$, while ours is in $A_0$. As noted in Sect.~\ref{sec:Results} the conversion from $A_0$ to $A_V$ is a factor of 1.003 and we use this to update our map during this comparison to $A_V$. However, \citet{Edenhofer2023} give their map in a unitless measure of extinction; they provide a conversion factor of 2.8 to obtain $A_V$, which we use to make their map directly comparable to that of \citet{Vergely2022} and this work. 

\begin{figure}
    \centering
\begin{subfigure}{0.5\textwidth}
    \includegraphics[width=\textwidth, trim=4.5cm 2cm 2.5cm 2cm, clip]%trim: left bottom right top
{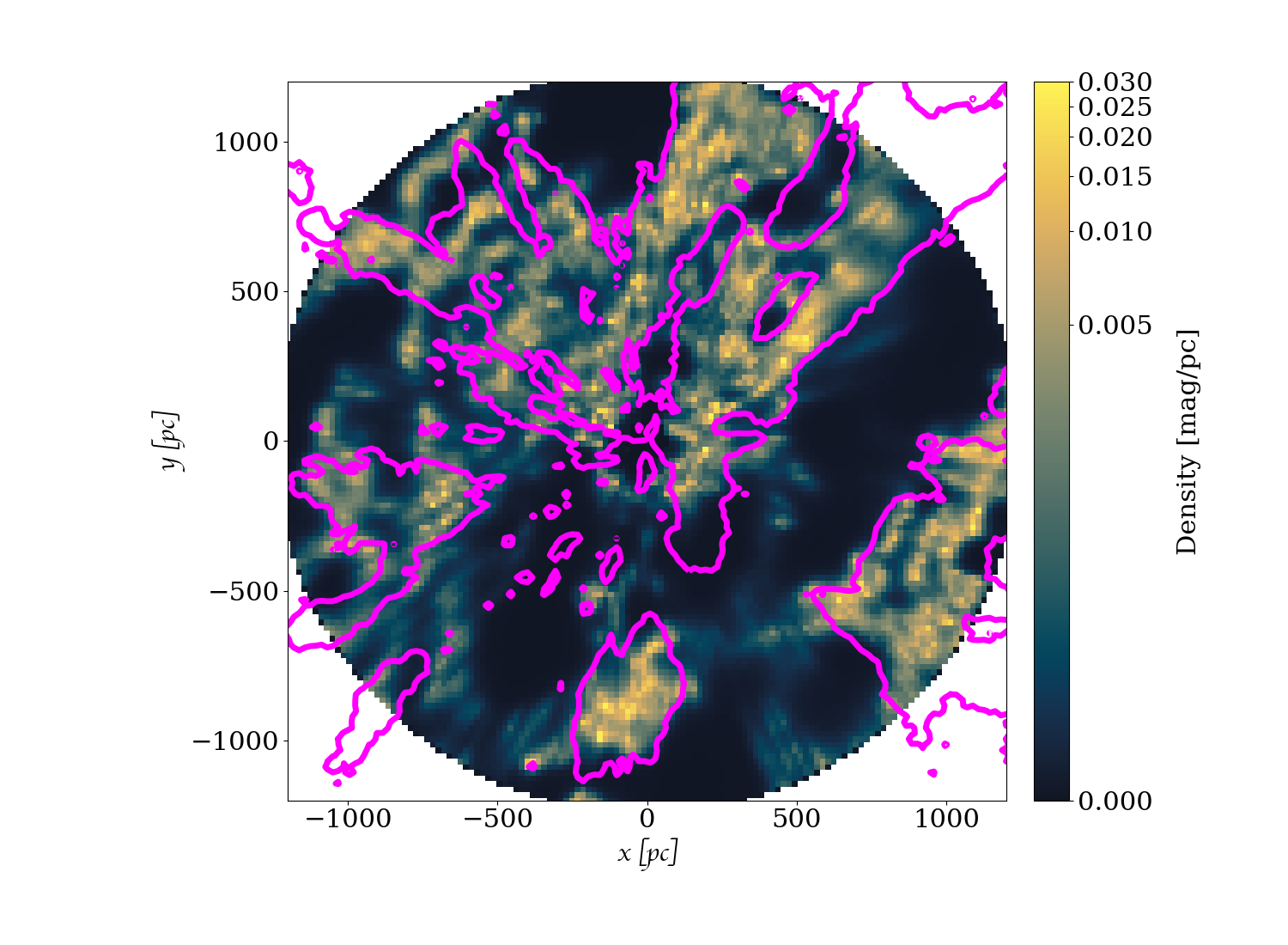}
    \end{subfigure}
\begin{subfigure}{0.5\textwidth}
    \includegraphics[width=\textwidth, trim=4.5cm 2cm 2.5cm 2cm, clip]%trim: left bottom right top
{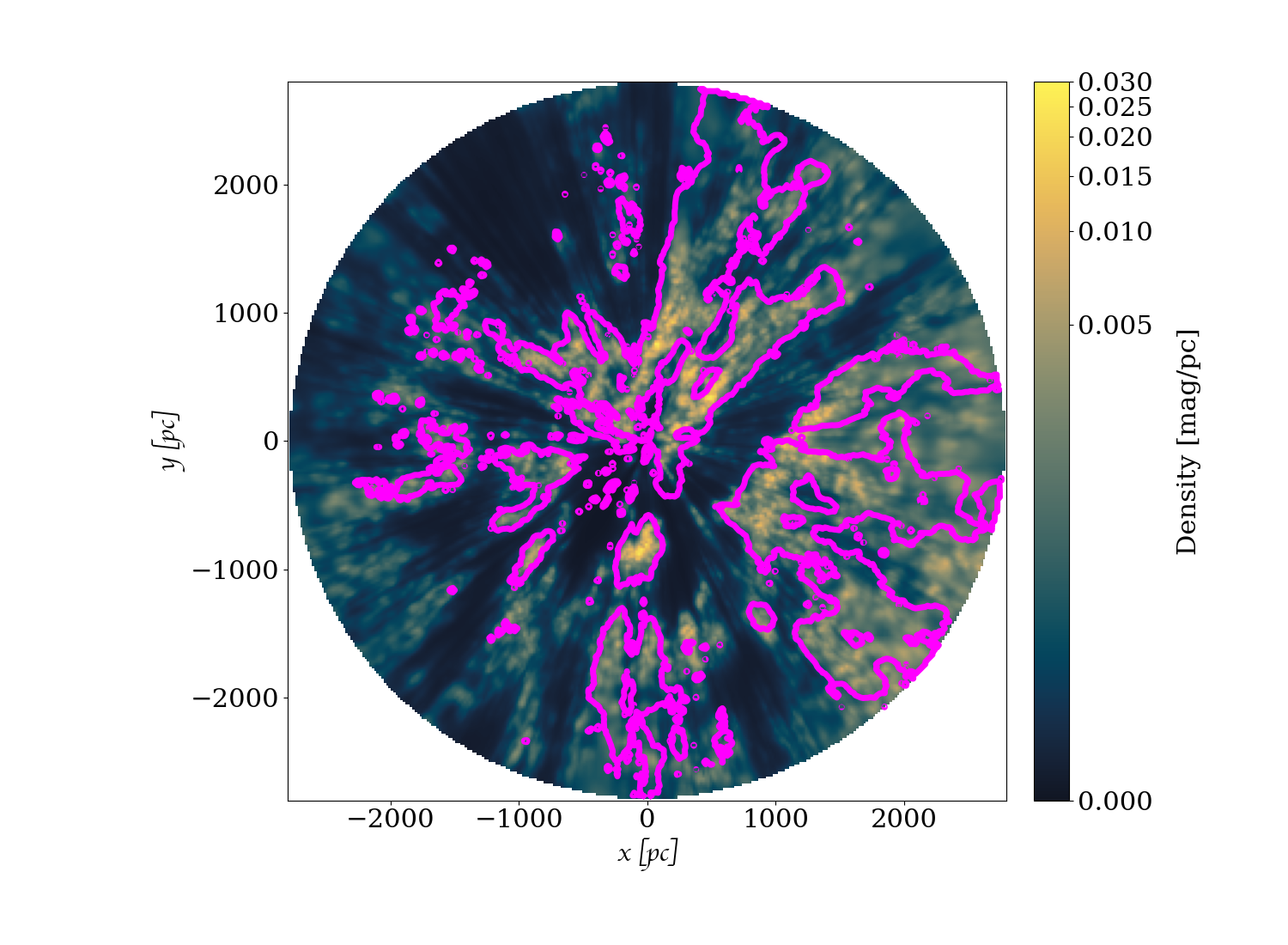}
    \end{subfigure}
    \caption{Comparison of the dust density at $z=0$ plane of this work with \citet{Edenhofer2023} (top) and \citet{Vergely2022} (bottom). The underlying maps are taken from the literature and interpolated onto our grid. The overlaid contours are the dust densities as seen in Fig.\ref{fig:xy_dustdense} derived in this work at $4\times10^{-4}$ mag pc$^{-1}$. Note that \citet{Edenhofer2023} only extends out to 1.2~kpc where as \citet{Vergely2022} is mapped to the full distance of our map.}
    \label{fig:xy_LitMapsComp}
\end{figure}

\begin{figure*}
    \centering
\begin{subfigure}{0.49\textwidth}
    \includegraphics[width=\textwidth, trim=5cm 2cm 3cm 2cm, clip]%trim: left bottom right top
{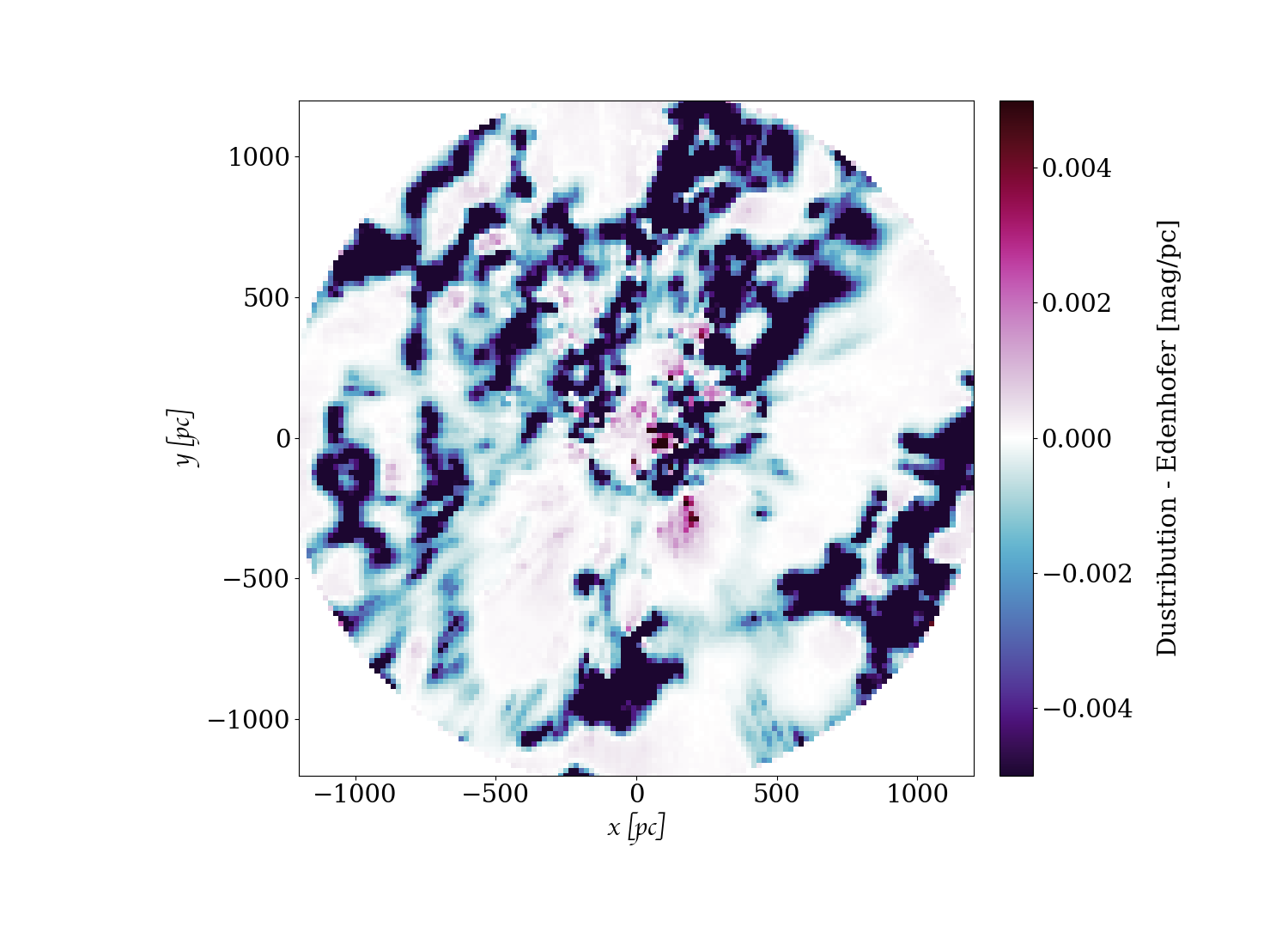}
    \end{subfigure}
\begin{subfigure}{0.49\textwidth}
    \includegraphics[width=\textwidth]%trim: left bottom right top
{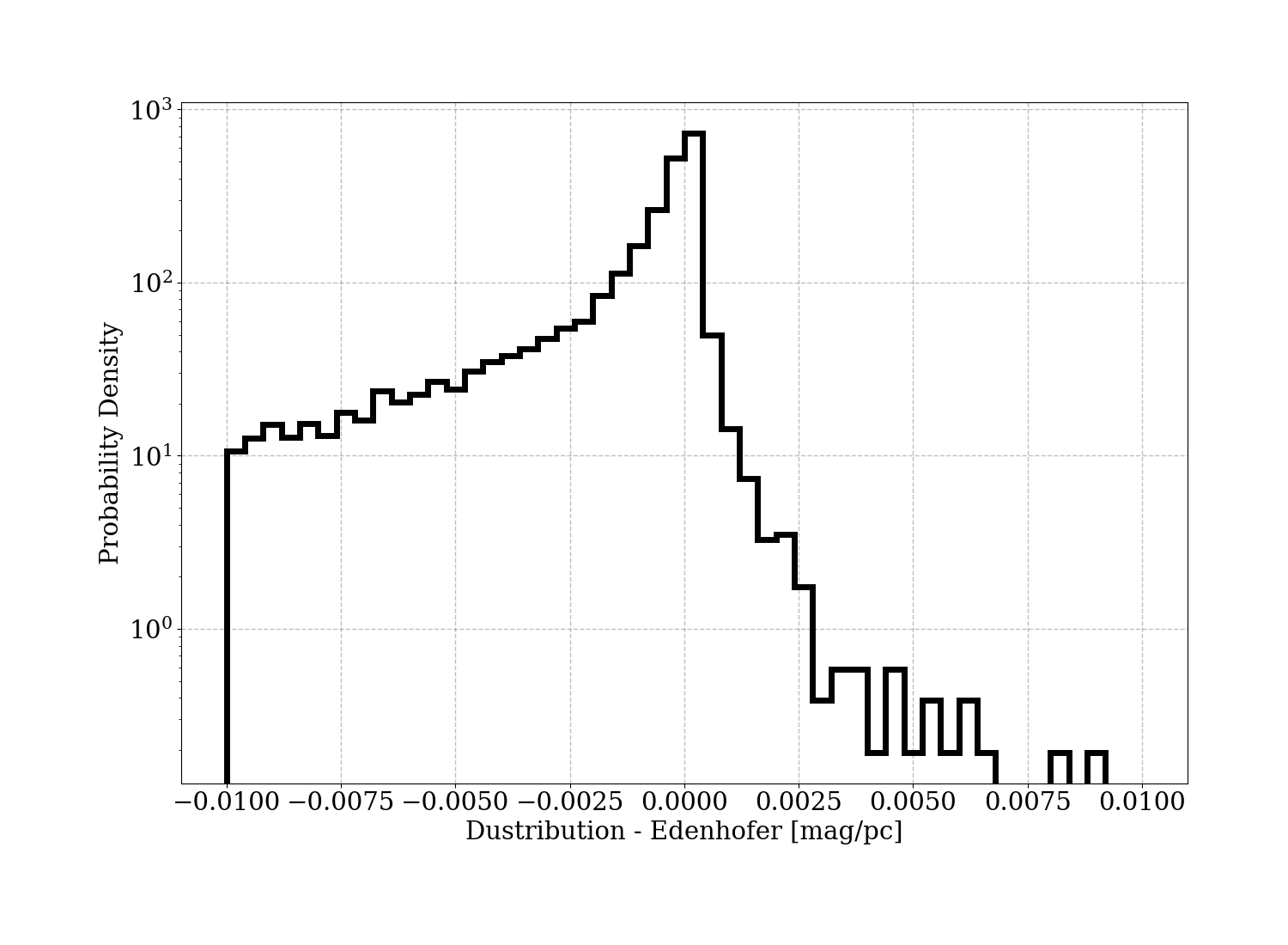}
    \end{subfigure}
\begin{subfigure}{0.49\textwidth}
    \includegraphics[width=\textwidth, trim=5cm 2cm 3cm 2cm, clip]%trim: left bottom right top
{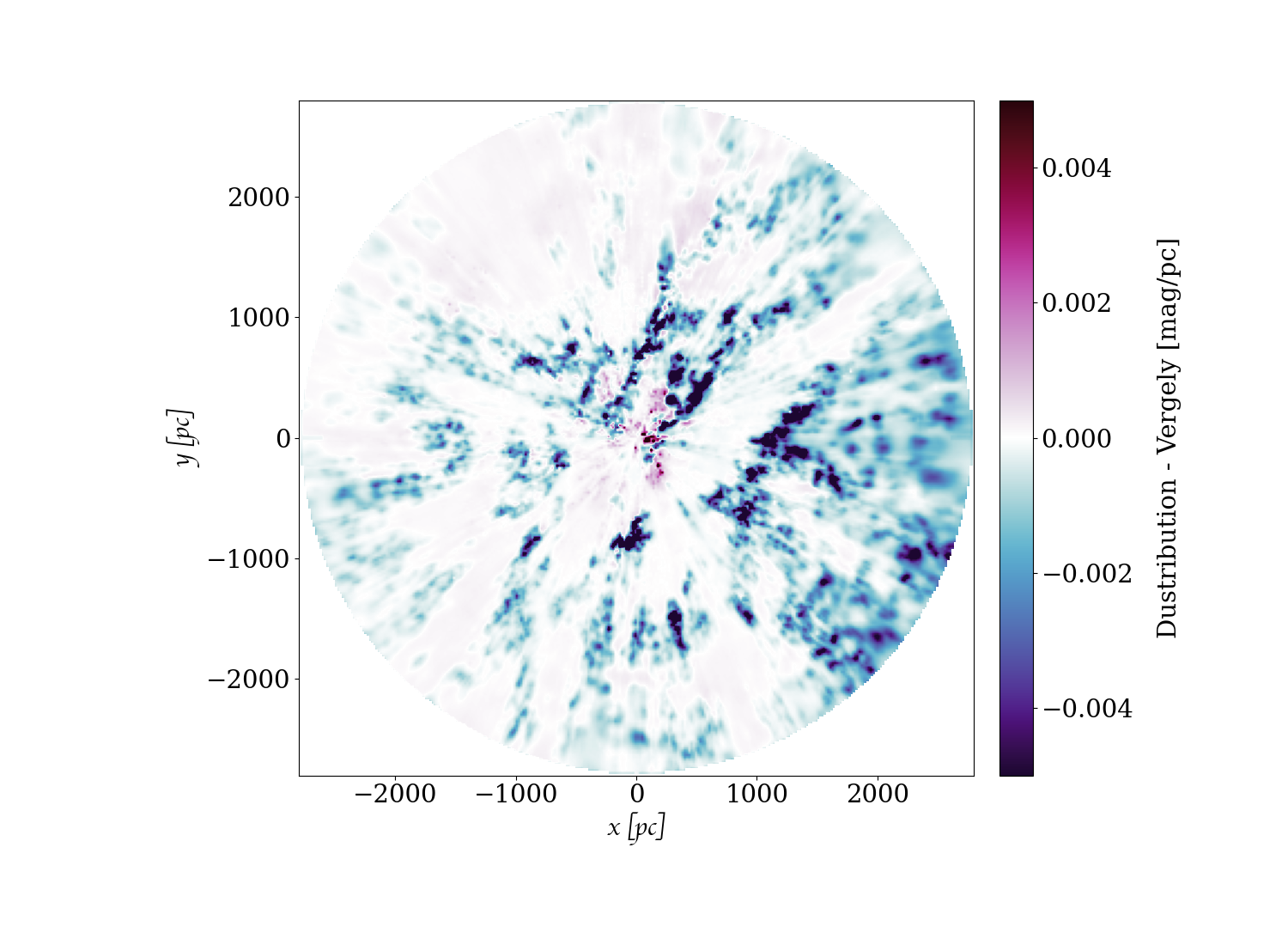}
    \end{subfigure}
\begin{subfigure}{0.49\textwidth}
    \includegraphics[width=\textwidth]%trim: left bottom right top
{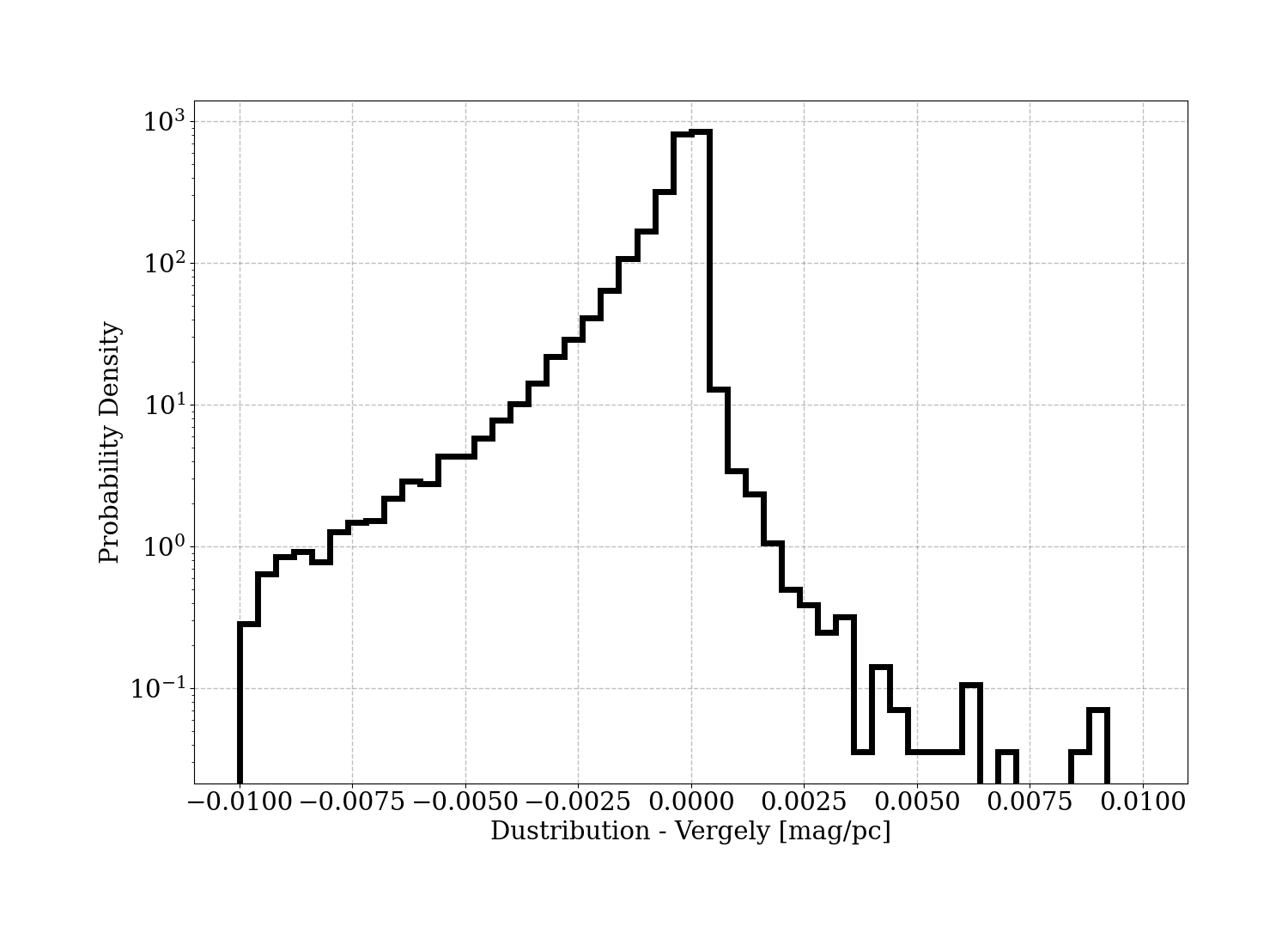}
    \end{subfigure}
\begin{subfigure}{0.49\textwidth}
    \includegraphics[width=\textwidth, trim=5cm 2cm 3cm 2cm, clip]%trim: left bottom right top
{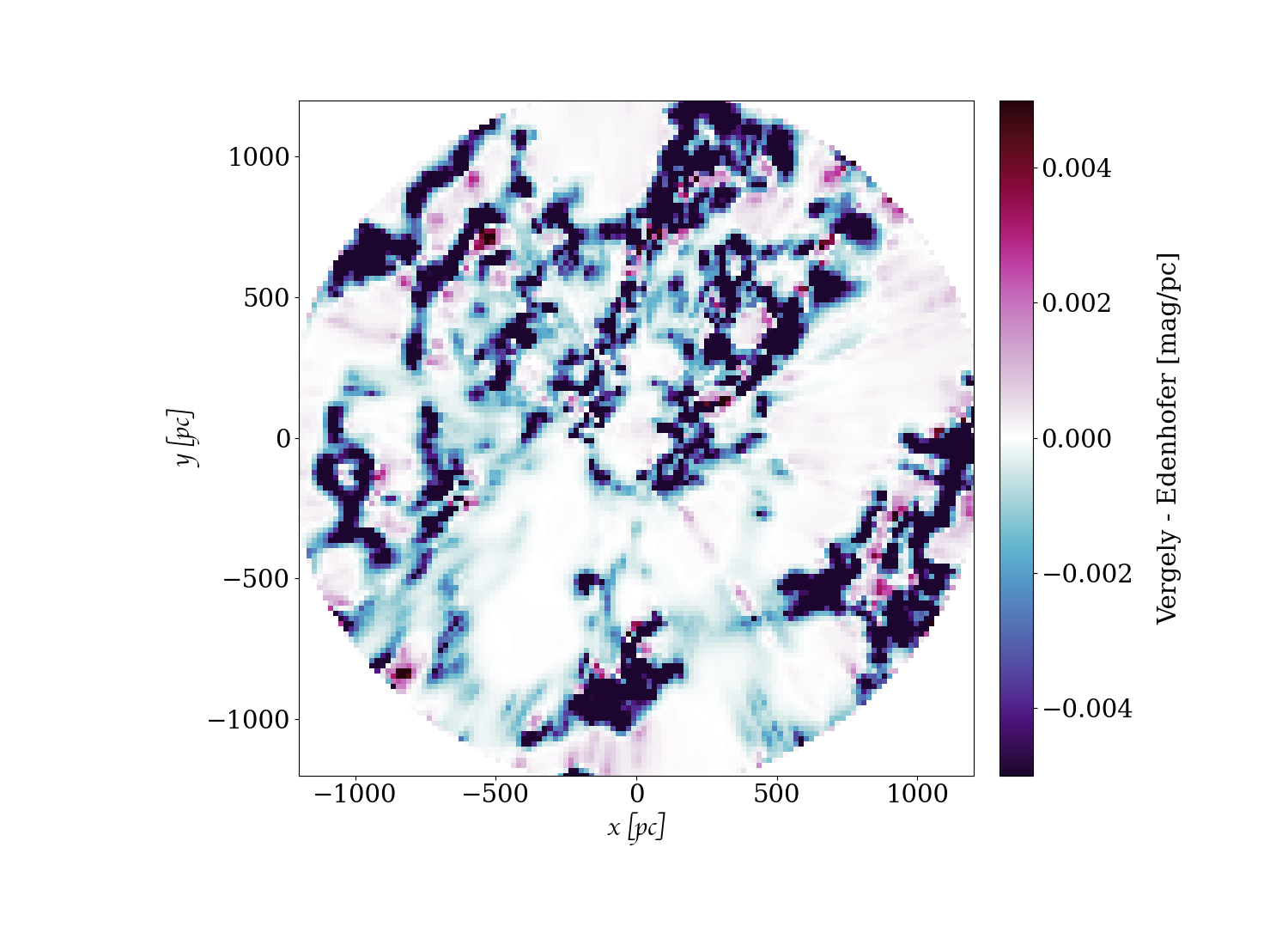}
    \end{subfigure}
\begin{subfigure}{0.49\textwidth}
    \includegraphics[width=\textwidth]%trim: left bottom right top
{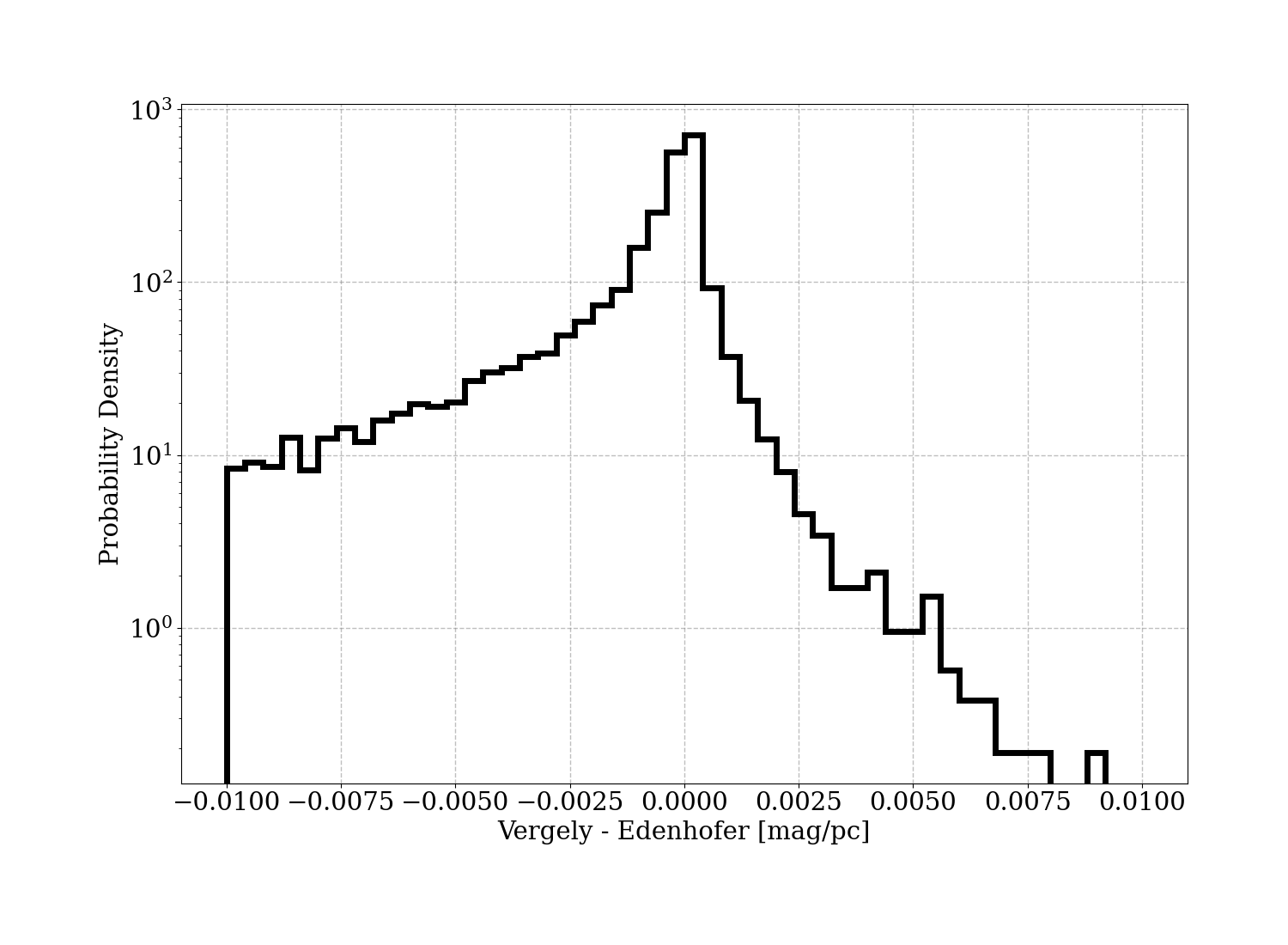}
    \end{subfigure}
    \caption{Comparison of this work with the dust density at $z=0$ from \citet{Vergely2022} and \citet{Edenhofer2023}. On the left side we show the subtraction of the the three different maps from each other. On the right side we show the histograms of subtraction of the the three different maps from each other.}
    \label{fig:xy_LitMapsComp_Sub}
\end{figure*}

\begin{figure*}
    \centering
\begin{subfigure}{0.49\textwidth}
    \includegraphics[width=\textwidth, trim=5cm 2cm 3cm 2cm, clip]%trim: left bottom right top
{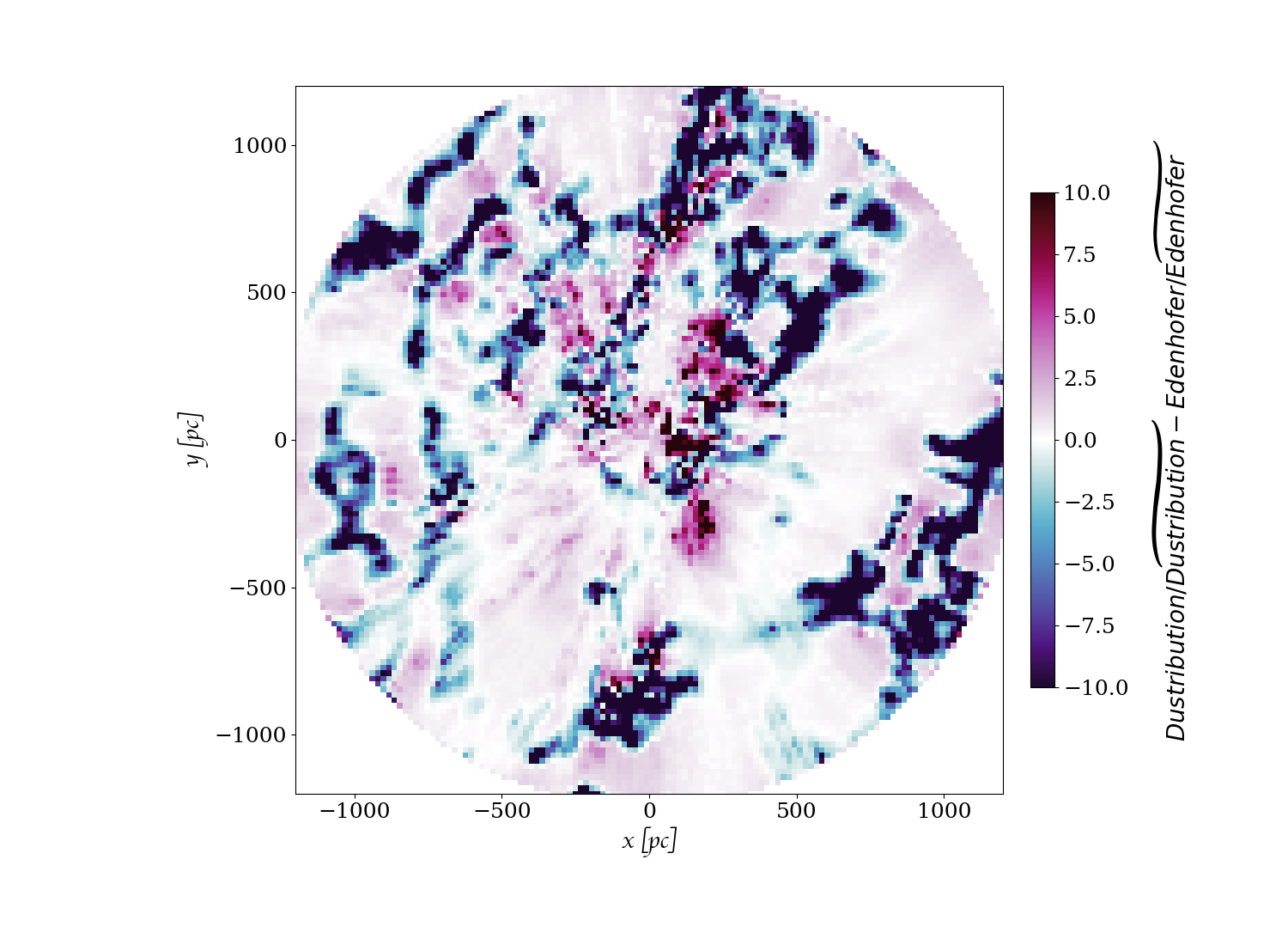}
    \end{subfigure}
\begin{subfigure}{0.49\textwidth}
    \includegraphics[width=\textwidth]%trim: left bottom right top
{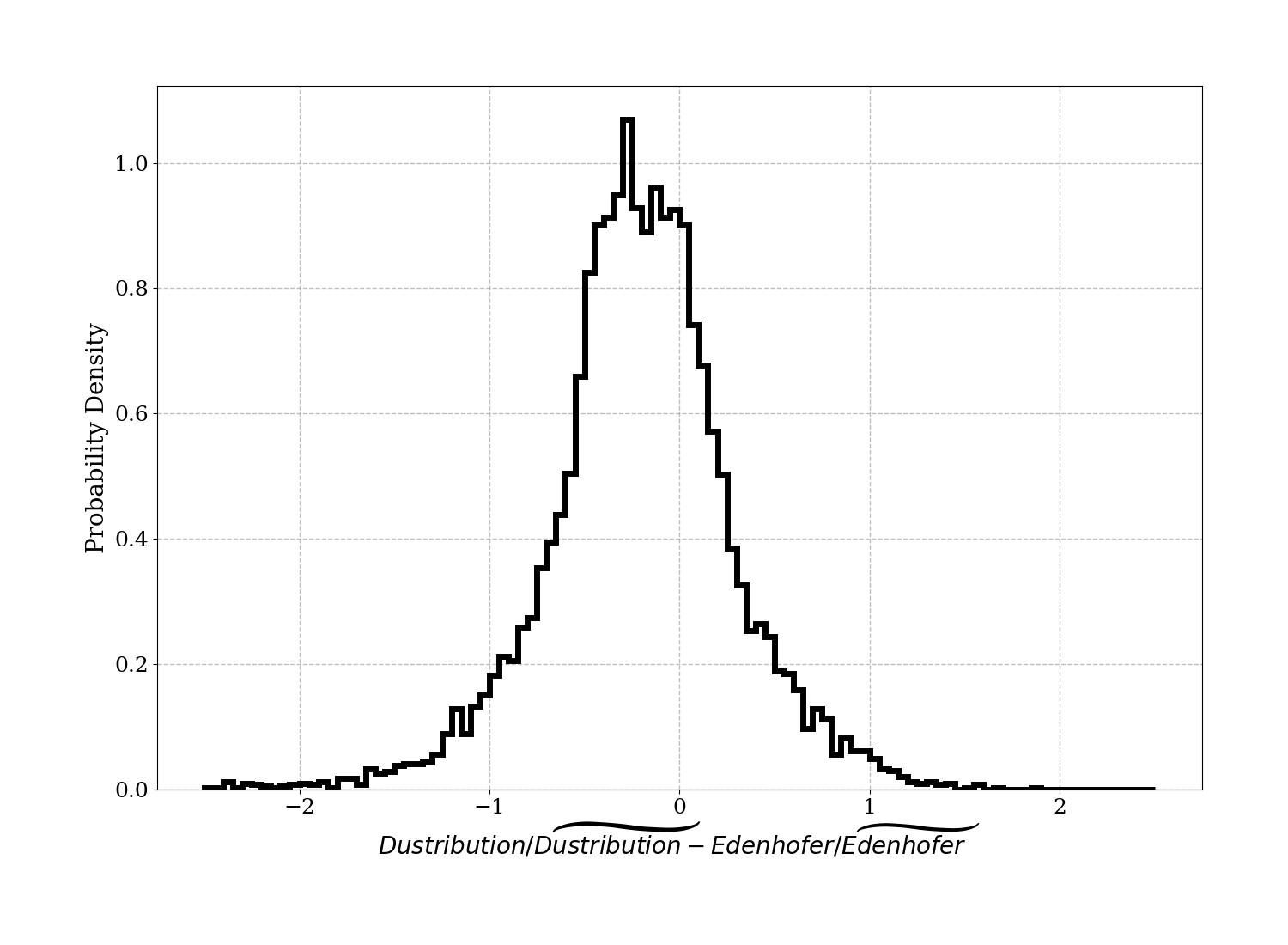}
    \end{subfigure}
\begin{subfigure}{0.49\textwidth}
    \includegraphics[width=\textwidth, trim=5cm 2cm 3cm 2cm, clip]%trim: left bottom right top
{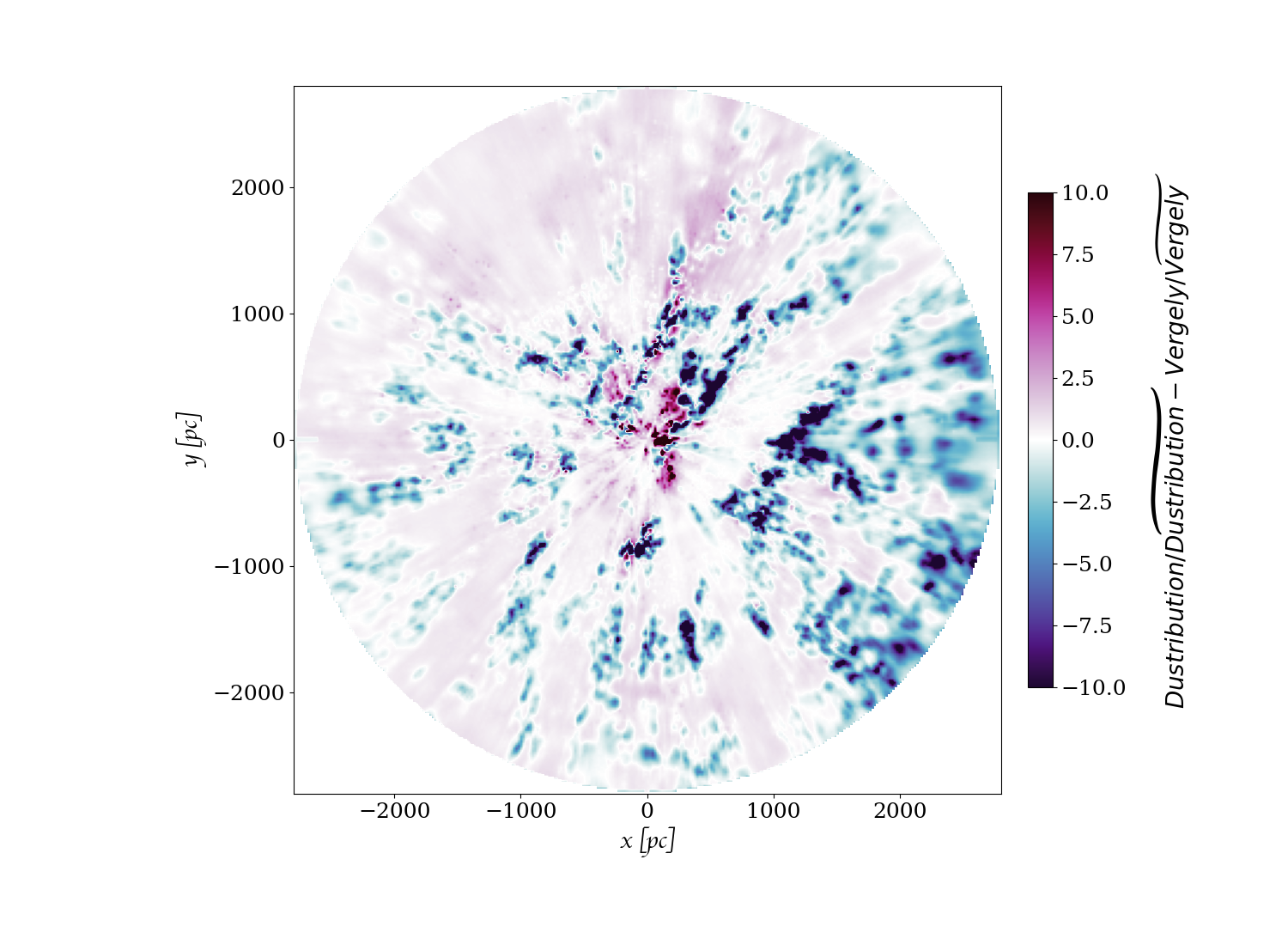}
    \end{subfigure}
\begin{subfigure}{0.49\textwidth}
    \includegraphics[width=\textwidth]%trim: left bottom right top
{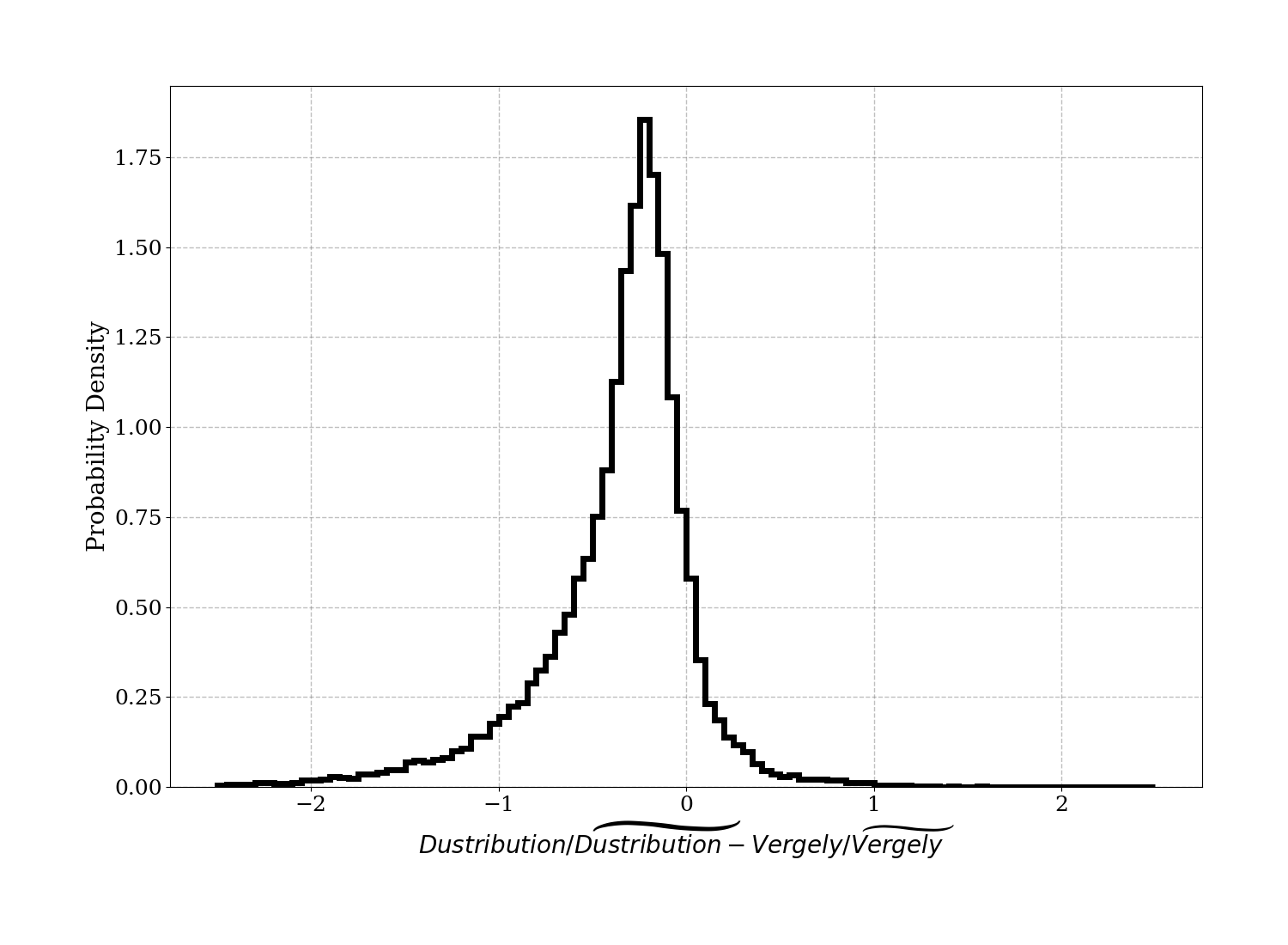}
    \end{subfigure}
\begin{subfigure}{0.49\textwidth}
    \includegraphics[width=\textwidth, trim=5cm 2cm 3cm 2cm, clip]%trim: left bottom right top
{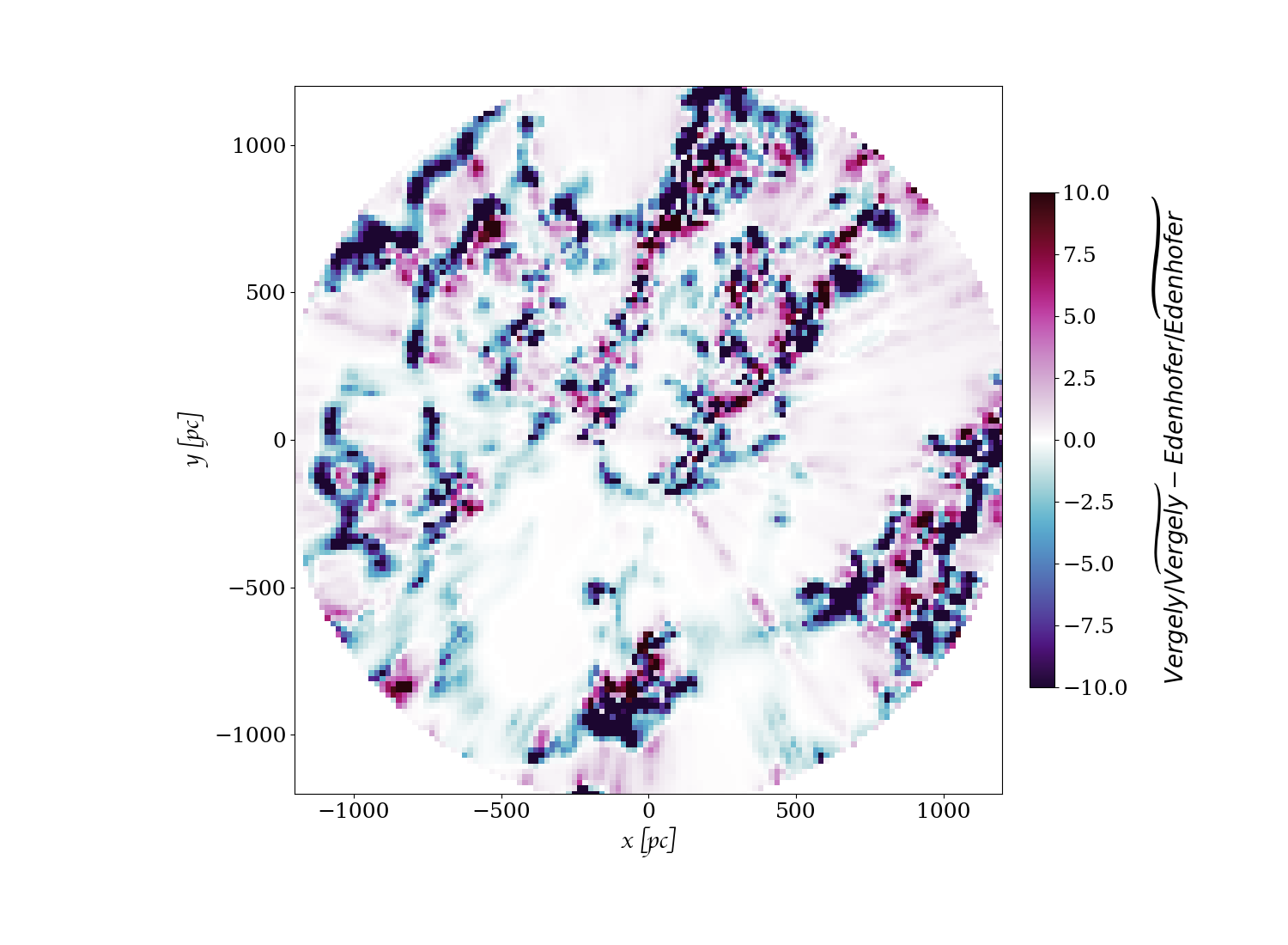}
    \end{subfigure}
\begin{subfigure}{0.49\textwidth}
    \includegraphics[width=\textwidth]%trim: left bottom right top
{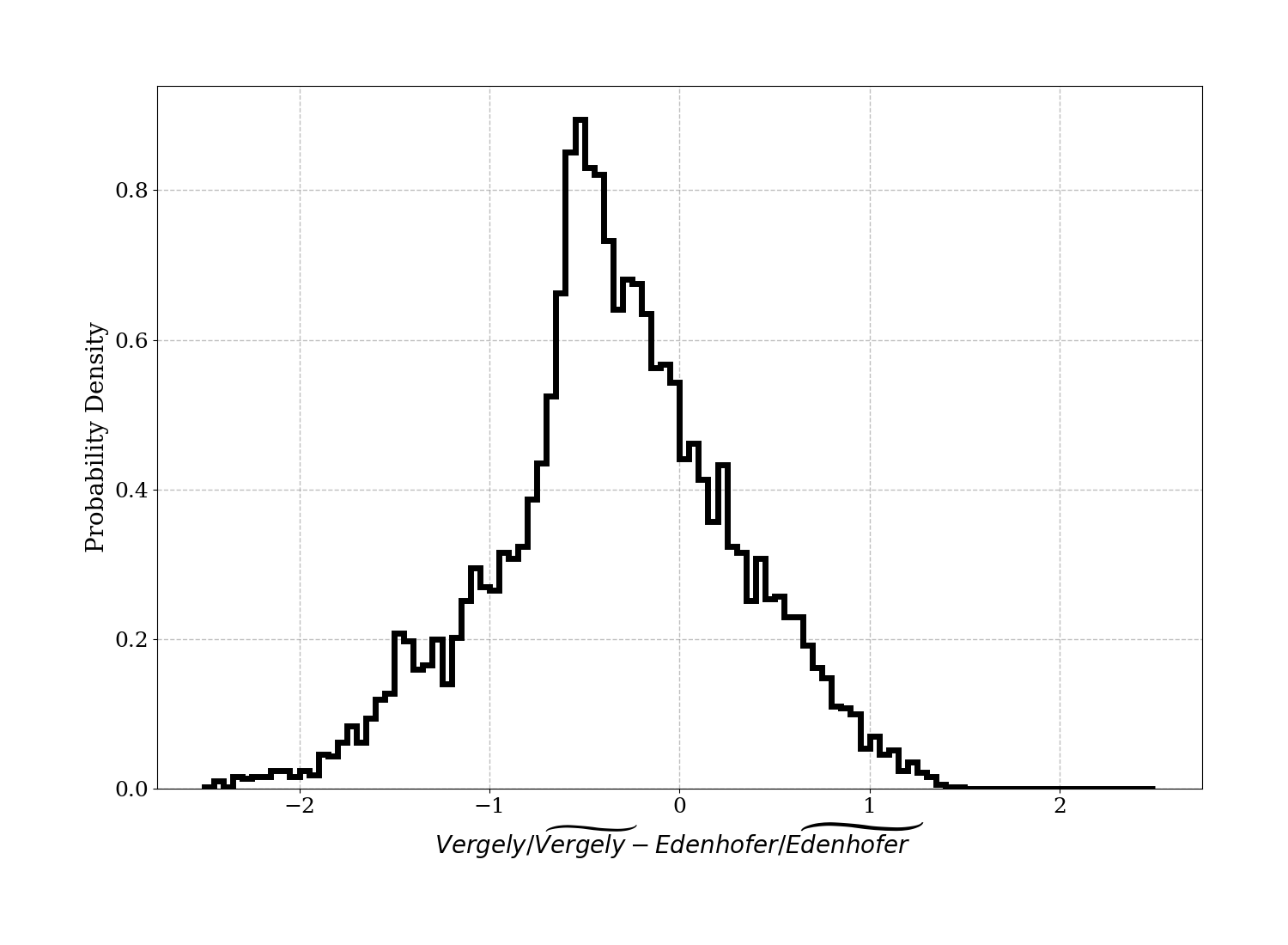}
    \end{subfigure}
    \caption{Comparison of this work with the dust density at $z=0$ from \citet{Vergely2022} and \citet{Edenhofer2023}. On the left side we show the subtraction of the the three different maps from each other after dividing by their respective median to account for any differences in zero point. On the right side we show the histograms of subtraction of the the three different maps from each other after dividing by their respective median.}
    \label{fig:xy_LitMapsComp_SubMed}
\end{figure*}

In Fig.~\ref{fig:xy_LitMapsComp} we compare the three maps by overlaying our dust density at $z=0$ on top of the two maps from literature. The overlaid pink contours show our dust density at $4\times10^{-4}$ mag pc$^{-1}$. In Fig.~\ref{fig:xy_LitMapsComp_Sub} we compare the three maps by subtracting the dust density at $z=0$ of each map from one other, while fig.~\ref{fig:xy_LitMapsComp_SubMed} compares the maps by first dividing each one by its respective median to normalise them and remove any differences in zero point, and then takes the difference. 

% First: Features are typically in the same places
% Second: Differences between the three maps are of the same order in all plots, and are less than or similar to the values.

There is a high degree of correspondence between the three maps. As seen in Fig.~\ref{fig:xy_LitMapsComp}, differences in the location of dense regions are small, typically similar to the scale length of our GP model (10~pc). Moreover, the difference images in Fig.~\ref{fig:xy_LitMapsComp_Sub} show that the differences in density at any given point are typically similar to or smaller than the densities. Similarly, in fig.~\ref{fig:xy_LitMapsComp_SubMed} the normalised maps and histograms show that the differences between the maps are smaller than the typical values of the maps. In high-density regions, where the differences are naturally largest, the three maps have a typical difference of less than $0.01$~mag~pc$^{-1}$, while all three maps agree at the $0.001$~mag~pc$^{-1}$ level in low-density regions. This is further emphasised by the sharp central peak of the histograms shown also in Fig.~\ref{fig:xy_LitMapsComp_Sub}, where the bin widths are 0.0005 mag/pc, showing that any differences in the zero point between the maps must be smaller than or comparable to this level. These differences are all comparable to or smaller than the typical point-wise uncertainty in our map. Since \citet{Vergely2022} do not publish the point-wise uncertainties on their map we cannot comprehensively account for the statistical noise in comparing the maps. Given the differences in input data and reconstruction approach between the three maps, this constitutes a good agreement. Comparatively our work has better agreement with \citet{Vergely2022} than either our work or \citet{Vergely2022} has with \citet{Edenhofer2023}. This is further highlighted by the difference in the skew and kurtosis of the histograms. While all three histograms fall of comparably fast in the positive direction, the histograms of the differences involving the Edenhofer map have much larger density in the negative tail.

All three maps nonetheless show features that are not recovered in the other two, and it is important to bear in mind that we have no ground truth to compare against. 
The \citet{Vergely2022} map shows slightly higher densities than ours in clouds associated with the Radcliffe wave or towards the Galactic Centre, while we find slightly higher densities overall close to the Sun and in the foreground of the Aquila Rift.
Both \citet{Vergely2022} and our map show nearby clouds inside the Local Bubble in similar locations, however they are generally slightly more extended and higher density in our map.
We see many of the same trends when comparing out map with that of \citet{Edenhofer2023} - we see higher densities in the foreground of the Aquila Rift in out map, and the highest-density clouds generally have slightly higher densities in \citet{Edenhofer2023}.
We also have a region of higher density close to the Sun in between the Vela and Circinus lines-of-sight.
The largest differences between \citet{Edenhofer2023} and our work seem to follow a wave-like pattern concentric around the Sun with a $\sim 300$~pc raidal spacing.
Similar quasi-concentric patterns are visible in the difference image between the \citet{Vergely2022} and \citet{Edenhofer2023} maps. This could be an effect of their construction as presented in \citet{Leike2022} and seen in Fig.D1 of \citet{Edenhofer2023}.

We also see interesting features in the Vela region, with the \citet{Vergely2022} map having higher densities on the near-side of the clouds and \citet{Edenhofer2023} higher on the far side, perhaps indicating a shift of the clouds along the line-of-sight between the two maps. In general, the \citet{Edenhofer2023} map shows the most compact features with the most substructure and the least line-of-sight extension of material. 

Many possible reasons exist for the differences observed  between the three maps. All three maps used different data sets. Although Gaia data was used in all three maps, parallaxes were chosen from different data releases, photometry were selected from different sources (e.g. DR2 vs DR3, Pan-STARRS, WISE) and in some cases spectra were used for part or all of the sources. and different methods to construct the maps. These choices come with different systematics and noise properties which could give rise to the differences observed. For example, as described above IR data allows one to probe deeper into dense dusty regions, thereby recovering more accurate extinction measurements for stars in these regions. Further, different catalogues assign different distances to the same stars; changing the distances alters how material is distributed along lines of sight. Nevertheless, the differences may also have a physical origin; by using different datasets, derived with different underlying extinction models, the different maps are liable to encode differences in dust properties as differences in the dust density instead. The \citet{Fouesneau2022_LBol} catalogue that we used treats the ratio of total-to-selective extinction ($R_V$) as a free parameter, allowing for changes in the extinction curve (and hence dust properties) between lines of sight. Since the input data to the \citet{Edenhofer2023} map do not include $R_V$ as a parameter, the choice of unitless extinction with the same extinction curve for all sources may shift how changes in the extinction curve are reflected in the density map. Meanwhile, the input data used by \citet{Vergely2022} do not report $R_V$, and assume fixed dust properties. However, it is beyond the scope of this paper to identify how and where these factors contribute to the differences between the maps, and it is likely that new and improved datasets will be required to explore this.

%Basically, the results are limited by the data, not the models! We see differences but they're not that big, so we need much better data to start disentangling everything.

%\subsection{Comparison to literature star and cloud catalogues}
%\label{sec:LitCats}
% \clearpage

\section{Conclusions}
\label{sec:Conclusions}

We have presented all-sky 3D dust density and integrated extinction maps of the Milky Way out to 2.8~kpc with a sampling of $l,b,d = 1^{\circ} \times1^{\circ} \times 1.7$~pc. The map is constructed by adapting the \DustT\ code \citep{Dharmawardena2022, Dharmawardena2023} to break large regions into chunks that were subsequently merged, allowing us to map the Milky Way on scales that are otherwise computationally infeasible. To mitigate edge effects, we adopted a straightforward weighting scheme, which smooths out edges in $l, b, d$ directions.

We recover both large-scale structures in both under- and -over-densities, such as the Split, Sagittarius spur, Vela Superbubble, and the gulf of Camelopardalis, as well as other smaller molecular clouds and the Local Bubble. The smallest-scale structures we recover are tens of parsecs across.

We find evidence for low density dust clouds within the Local Bubble, indicating it is not completely devoid of dust. We have also identified an under-density which could be an inter-arm/-spur region towards the Galactic centre, centred at $x,y \approx 0,750$~pc. 

In the absence of a ground-truth comparison, we compare our map to two of the most-recent 3D dust maps in the literature: \citet{Vergely2022} and \citet{Edenhofer2023}. Both qualitatively and quantitatively, there are few deviations between the density values in the maps. However, each map has its own unique features that the others do not recover. These differences may arise from the choices of data, distances and methods used in each map, or potentially from differences in assumed dust properties and the shape of the extinction curve.

Our 3D dust map can be applied to various use cases, e.g.\ determining the line-of-sight extinctions to arbitrary positions where a direct measurement is impossible, such as: young or evolved stars with thick circumstellar obscuration; estimating the Galactic foreground extinction towards high-latitude extragalactic sources; understanding the 3-dimensional structure of star-forming regions; understanding the wider environment and context of peculiar sources or clusters.

We will further refine this map in the future through two approaches. We currently use Gaia DR2 data combined with IR data from \citet{Fouesneau2022_LBol} for our maps. By simultaneously inferring distance, stellar properties and extinction using Gaia DR3 results and infrared photometry, significant improvements in the input catalogue -- in particular the individual extinction -- will allow much higher fidelity. The larger number of stars with high-precision parallaxes from using GDR3 will both improve the quality of the map within the current volume and allow a larger volume to be mapped. However, without using IR photometry, the precision of the extinction estimates -- which is a key limitation of the map quality -- from Gaia DR3 is no better than in datasets based on DR2. At the time of writing, no catalogues forward-modelling the combination of DR3 and IR data are available. 

%%%%%%%%%%%%%%%%%%%%%%%%%%%%%%%%%%

\section*{Acknowledgements}

We wish to thank the editor and anonymous referee for their constructive feedback which helped improve the manuscript. T.E.D wishes to thank Andreas Marek at Max Planck Computing and Data Facility and Geoff Pleiss at University of British Columbia for useful discussions of optimisations and updates to \DustT. T.E.D acknowledges support for this work provided by NASA through the NASA Hubble Fellowship Program grant No. HST-HF2-51529 awarded by the Space Telescope Science Institute, which is operated by the Association of Universities for Research in Astronomy, Inc., for NASA, under contract NAS5-26555. This project is funded by the Sonderforschungsbereich SFB 881 “The Milky Way System” of the German Research Foundation (DFG). 

% ADD SOFTWARE!!!
% % https://astrofrog.github.io/acknowledgment-generator/
% % tools and software
% This publication made extensive use of the online authoring Overleaf platform (\url{https://www.overleaf.com/}).\\
% %
% %\orange{The data processing and analysis made use of the software A, B\\}
% The data processing and analysis made use of
%     matplotlib \citep{Hunter:2007},
%     NumPy \citep{harris2020array},
%     the IPython package \citep{PER-GRA:2007},
%     Vaex \citep{Breddels2018},
%     Astropy \\
%     TOPCAT \citep{Taylor2005}\\

%%%%%%%%%%%%%%%%%%%%%%%%%%%%%%%%%%%%%%%%%%%%%%%%%%
\section*{Data Availability}

Our code is available at \url{https://github.com/Thavisha/Dustribution}, and our results are available interactively at \url{www.mwdust.com}. All predicted 3D density and extinction data can be also downloaded from Zenodo via \url{https://doi.org/10.5281/zenodo.11448780}.
%The inclusion of a Data Availability Statement is a requirement for articles published in MNRAS. Data Availability Statements provide a standardised format for readers to understand the availability of data underlying the research results described in the article. The statement may refer to original data generated in the course of the study or to third-party data analysed in the article. The statement should describe and provide means of access, where possible, by linking to the data or providing the required accession numbers for the relevant databases or DOIs.

%%%%%%%%%%%%%%%%%%%% REFERENCES %%%%%%%%%%%%%%%%%%

% The best way to enter references is to use BibTeX:

\bibliographystyle{mnras}
\bibliography{Bib_MWExt}

%%%%%%%%%%%%%%%%%%%%%%%%%%%%%%%%%%%%%%%%%%%%%%%%%%

%%%%%%%%%%%%%%%%% APPENDICES %%%%%%%%%%%%%%%%%%%%%
\clearpage
\appendix

\section{Further description of the merge method and validating the weighting scheme}
\label{sec:app:merge_validation}

We use the python package \texttt{xarray} \citep{Hoyer2017_xarrayMain, Hoyer_xarrayV17} to derive the weighted quantiles using the weighting functions presented in equations \ref{eqn:lbweight} and \ref{eqn:weightingDist}. Given some input array of values, $v_i$, and an array of corresponding weights, $w_i$, \texttt{xarray} normalises the weights so that their sum is unity. Then, it sorts the values and calculates the cumulative weights, $w_c$, where $w_c(i) = \sum\limits_{0}^{i}w_i$. Once $v$ is sorted and cumulative weights are calculated, it will find the first entry where $w_c \geq 0.5$ and extract the value from $v$ associated with it - this is reported as the weighted median. The same applies for the 16$^{\rm th}$ and 84$^{\rm th}$ percentiles with $0.5$ being replaced with $0.16$ and $0.84$. 

% Fig~\ref{fig:ExtLos} shows lines-of-sight extinctions to multiple lines of sight drawn across the MW to show the smooth transition between the maps and no sudden large jumps.

\begin{figure*}
\centering
\begin{subfigure}{0.3\textwidth}
  \centering
 \includegraphics[width=\textwidth, trim=2cm 2cm 3cm 2cm, clip]{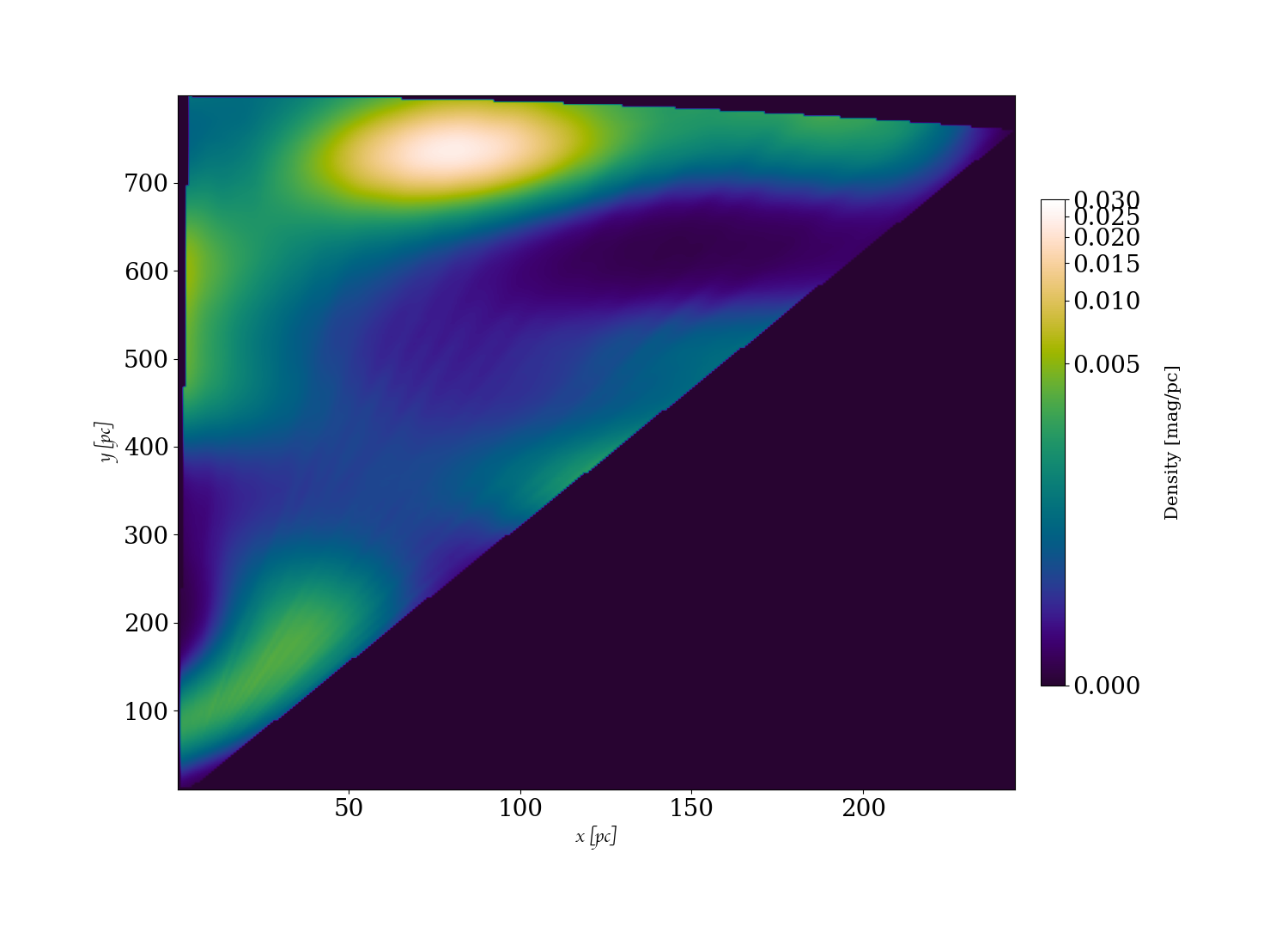}
  \end{subfigure}  
\begin{subfigure}{0.3\textwidth}
  \centering
   \includegraphics[width=\textwidth, trim=2cm 2cm 3cm 2cm, clip] {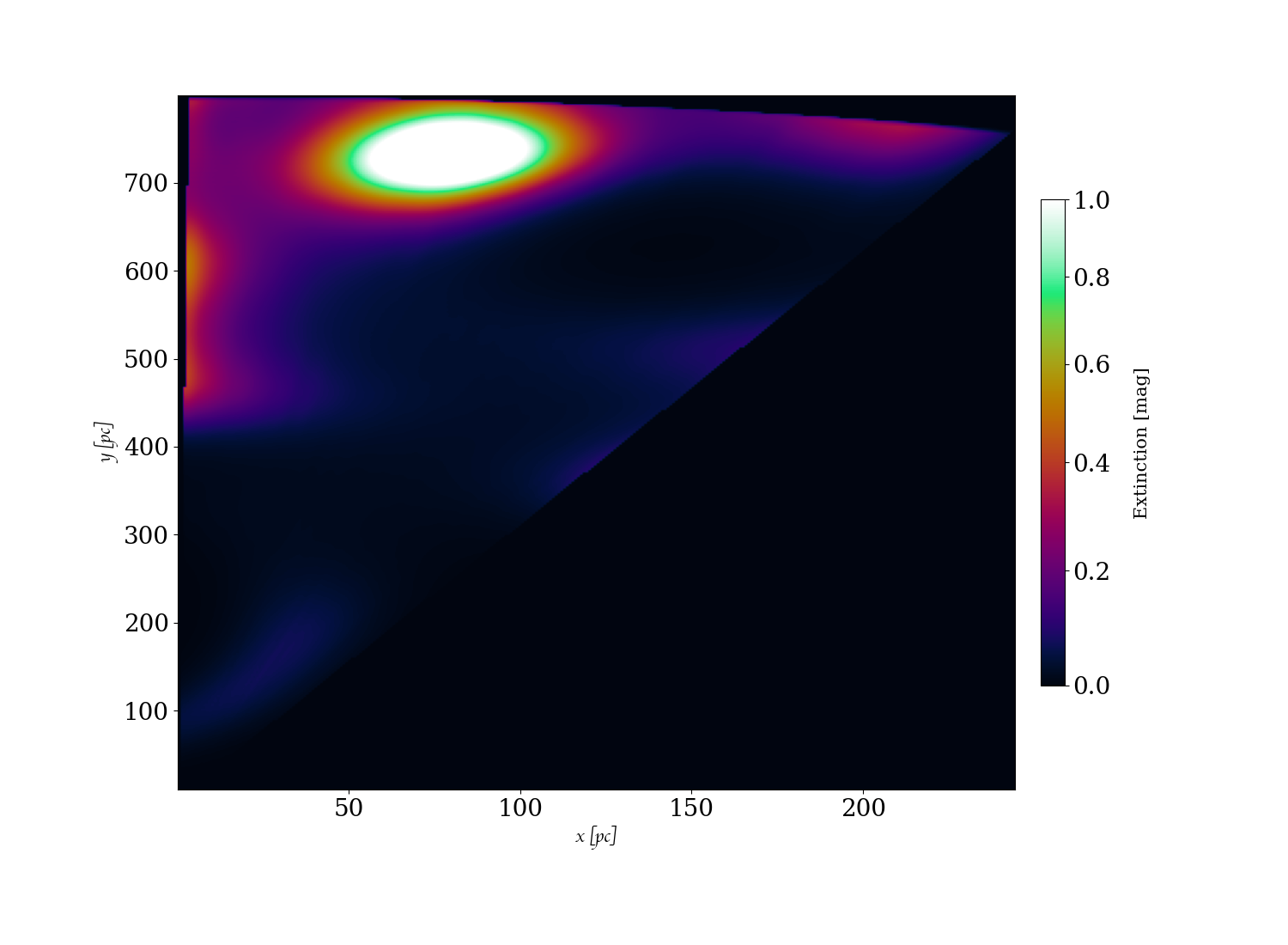}
  \end{subfigure}
\begin{subfigure}{0.3\textwidth}
\centering
\includegraphics[width=\textwidth, trim=2cm 2cm 3cm 2cm, clip] {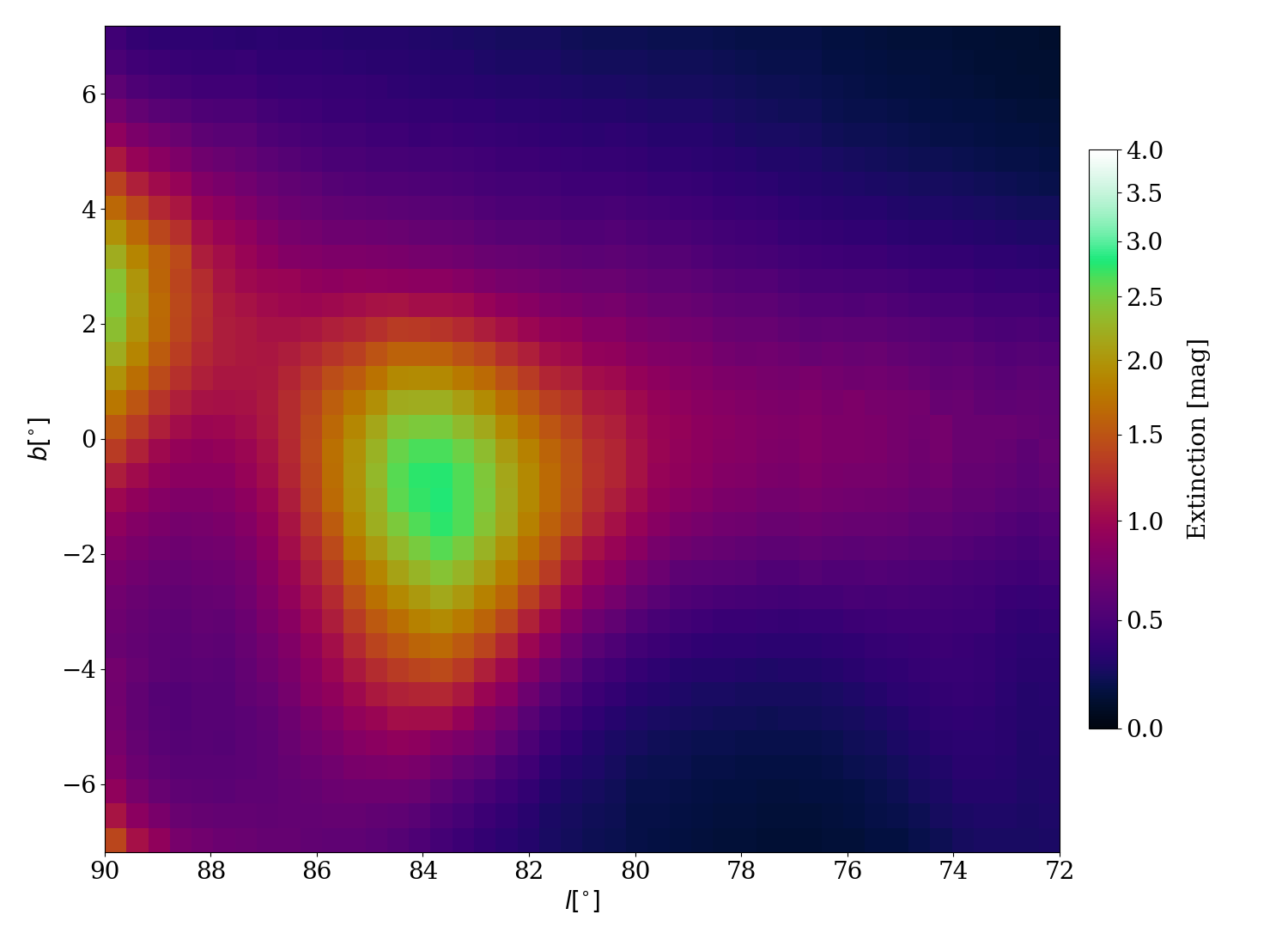}
\end{subfigure}

\begin{subfigure}{0.3\textwidth}
  \centering
 \includegraphics[width=\textwidth, trim=2cm 2cm 3cm 2cm, clip]{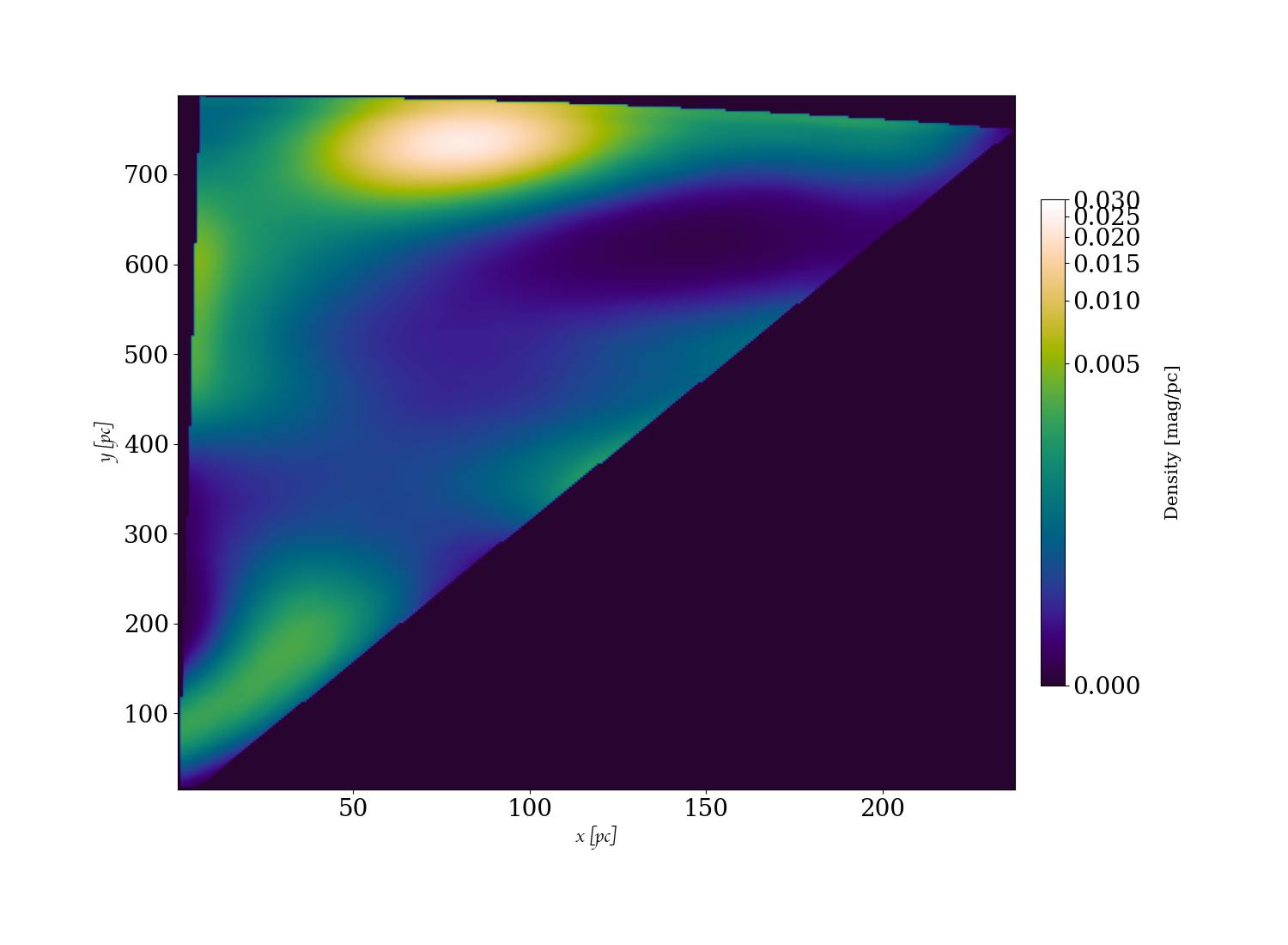}
  \end{subfigure}  
\begin{subfigure}{0.3\textwidth}
  \centering
   \includegraphics[width=\textwidth, trim=2cm 2cm 3cm 2cm, clip] {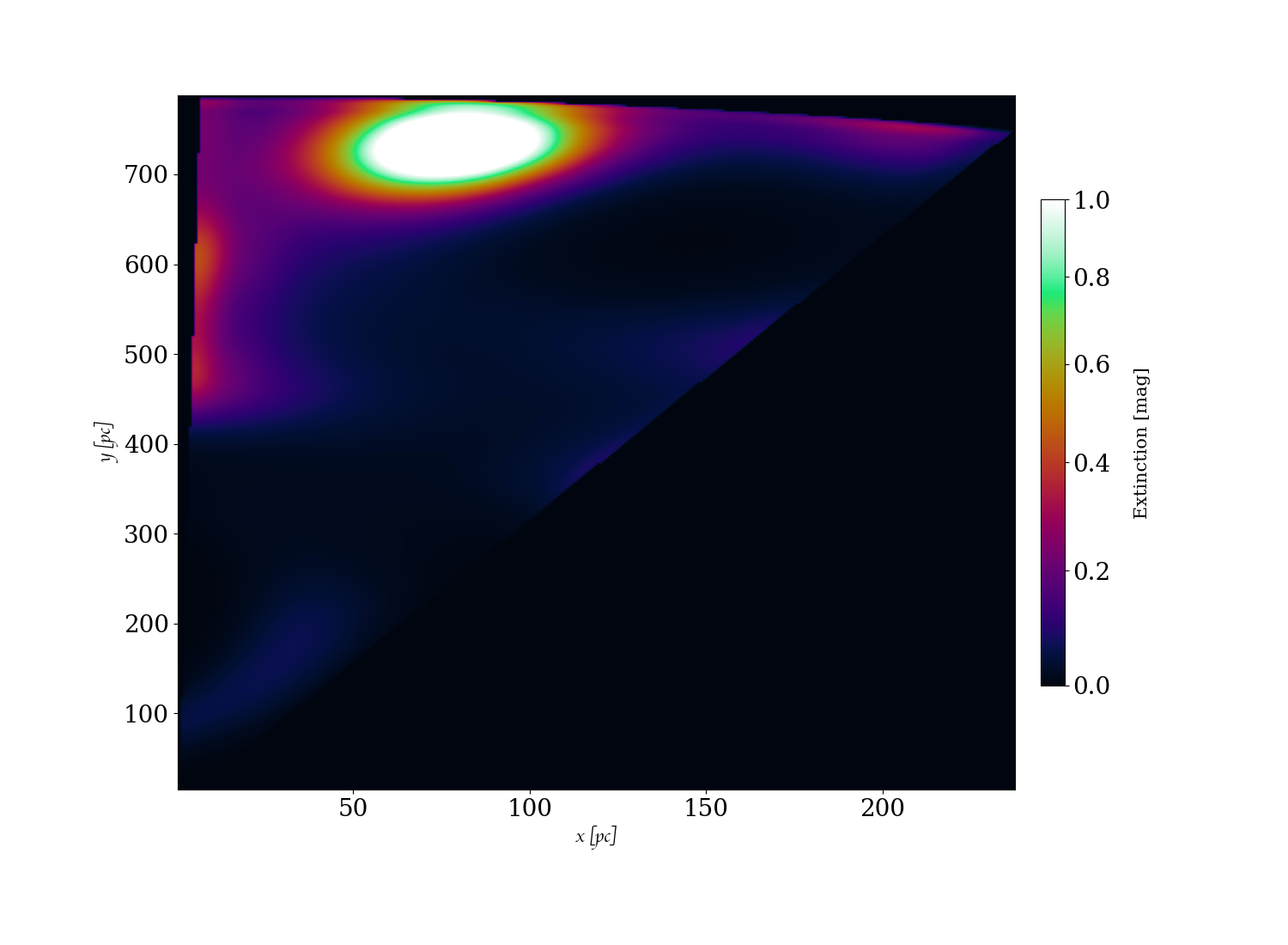}
  \end{subfigure}
\begin{subfigure}{0.3\textwidth}
\centering
\includegraphics[width=\textwidth, trim=2cm 2cm 3cm 2cm, clip] {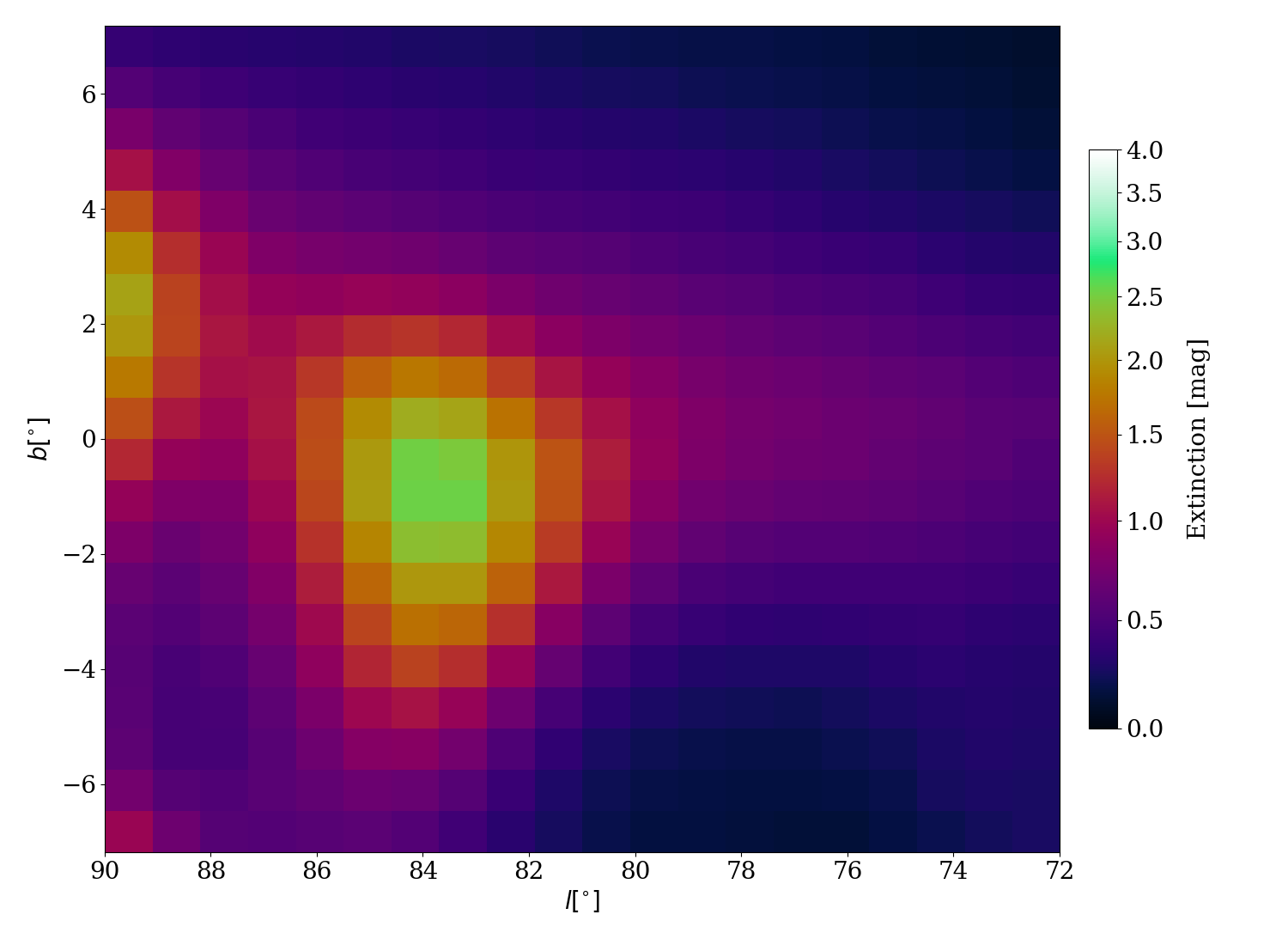}
\end{subfigure}

 \caption{Merging method test. Top row: \DustT\ predicted results for the single main chunk used in the Milky Way merging; Bottom row: chunk merged together using our merging method using the repredicted overlapping chunks from the main chunk; Left column: dust density at $z=0$ as seen above the Galactic plane; Middle column: Integrated dust extinction along z direction as seen from above the Galactic plane; Right column: Cumulative extinction as seen from the Sun integrated out to 800 pc.}
    \label{fig:MergeTest}
\end{figure*}

\begin{figure*}
    \centering
    \includegraphics[width=0.8\textwidth, trim=5cm 2cm 3cm 2cm, clip]%trim: left bottom right top
{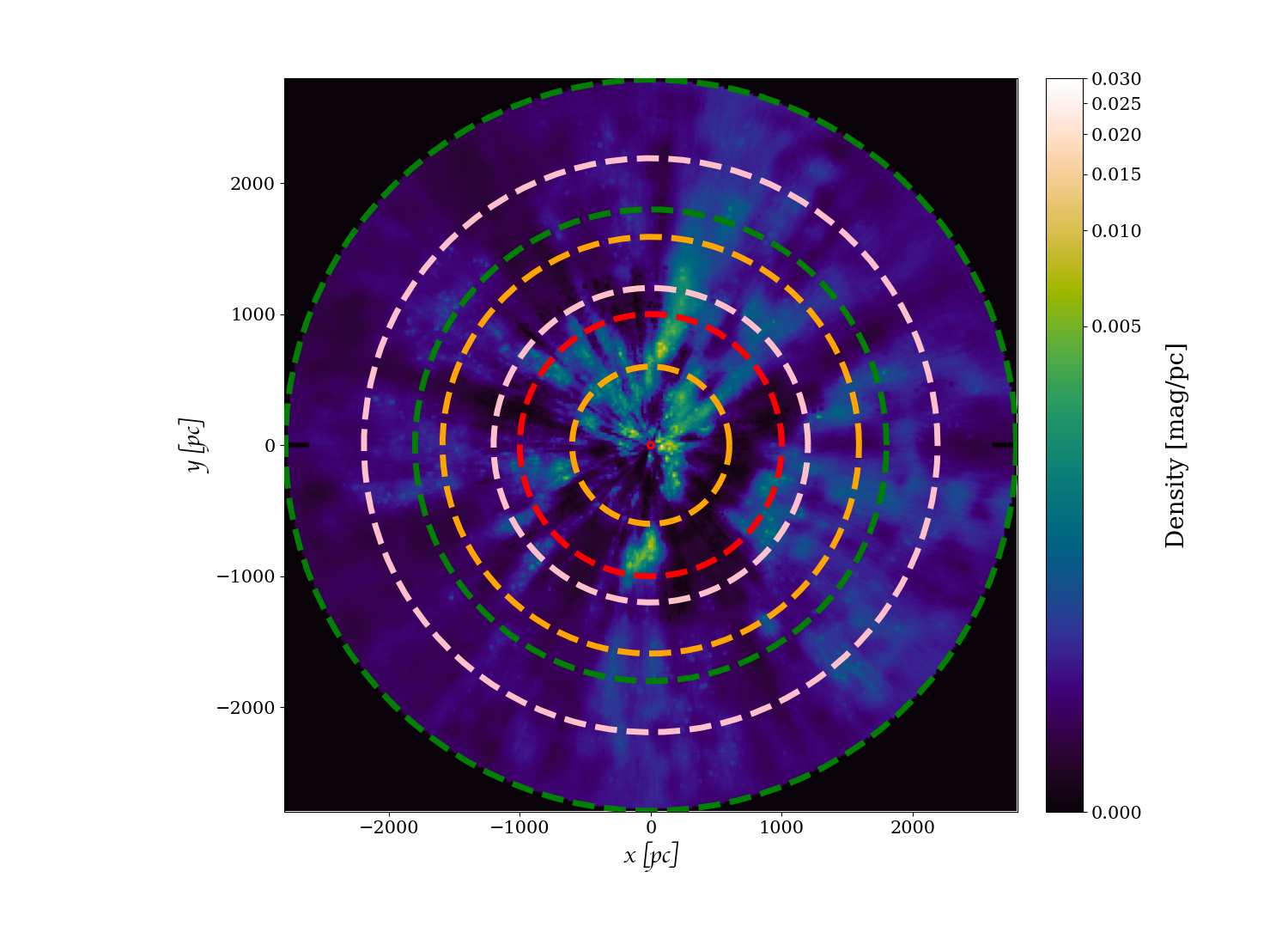}
    \caption{dust density at $z=0$ over plotted with the boundaries of each chunk. Red: 10-1000~pc; Orange: 600-1590~pc; Pink: 1200-2190~pc; Green: 1800-2790~pc.}
    \label{fig:MergeBoundaries}
\end{figure*}

\begin{figure*}
    \centering
    \includegraphics[width=0.49\textwidth]{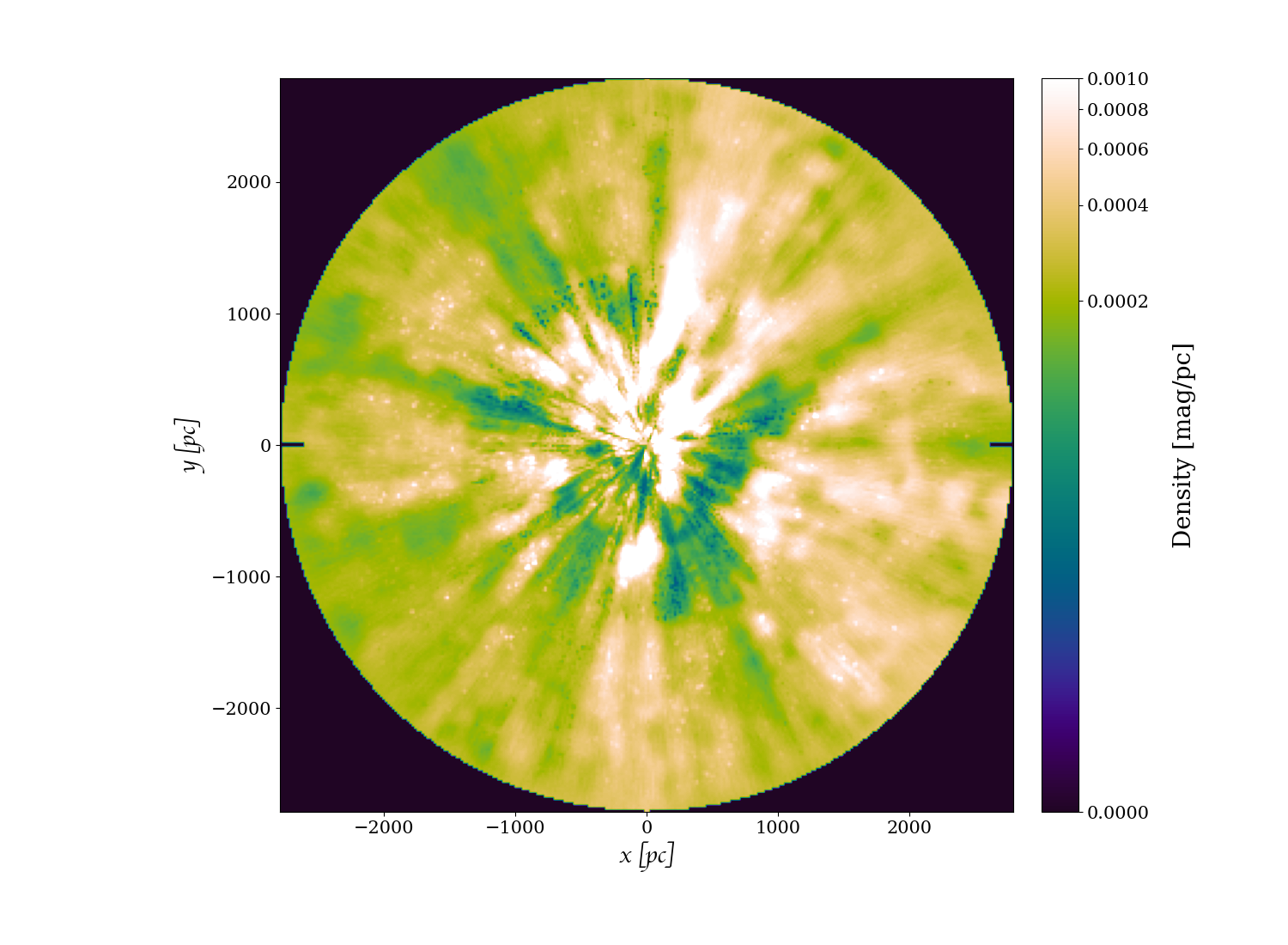}
    \includegraphics[width=0.49\textwidth]{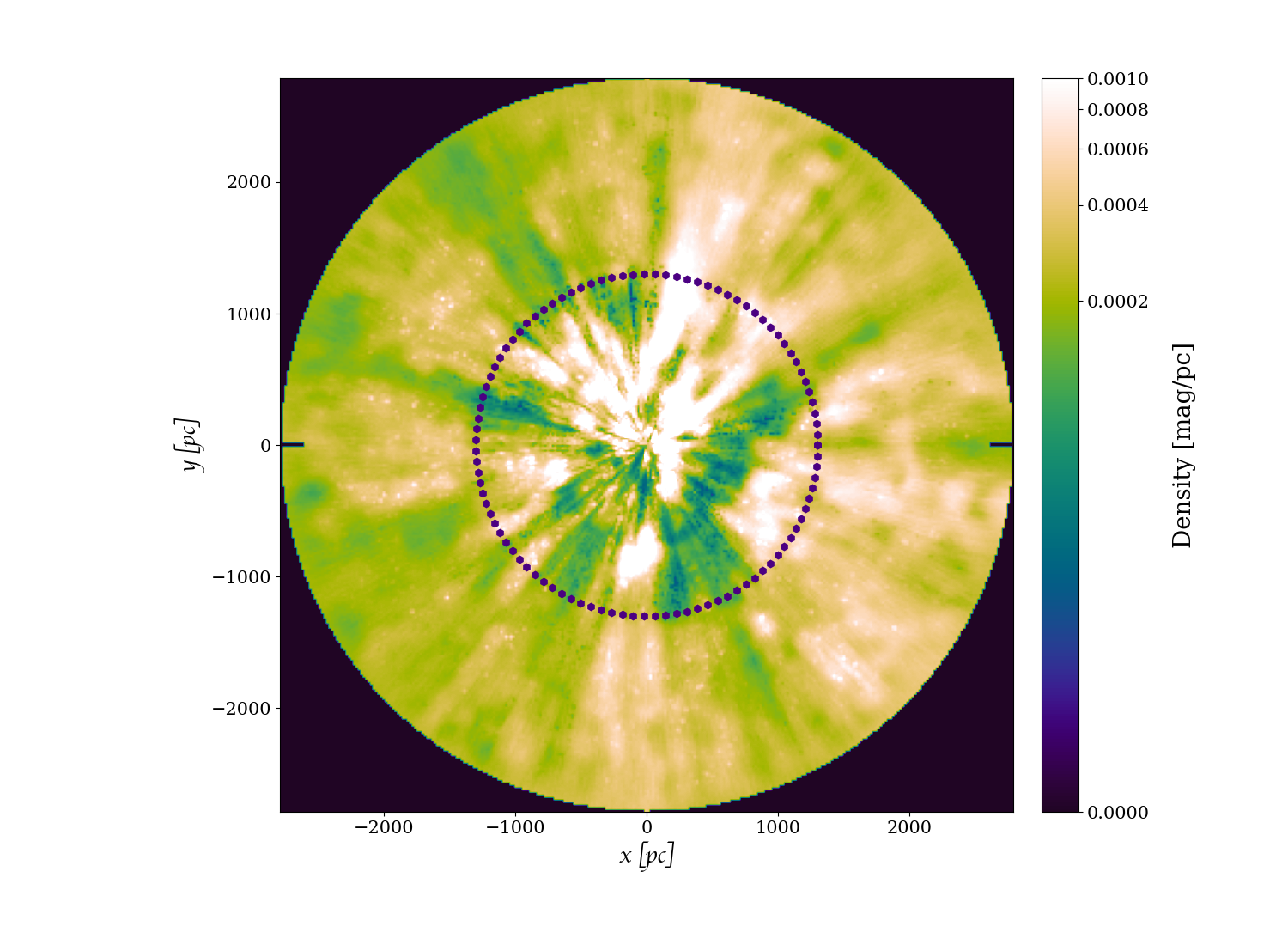}
    \caption{{\it Left:} Density in the Galactic Plane with the colour scale truncated by a factor of 30 compared to fig.~\ref{fig:xy_dustdense}. {\it Right:} As left, but with purple hexagonal symbols drawn at $d=1300$~pc highlighting the possible location of an artefact in the density.}
    \label{fig:sat_dens}
\end{figure*}

\begin{figure*}
    \centering
    \includegraphics[width=0.49\textwidth]{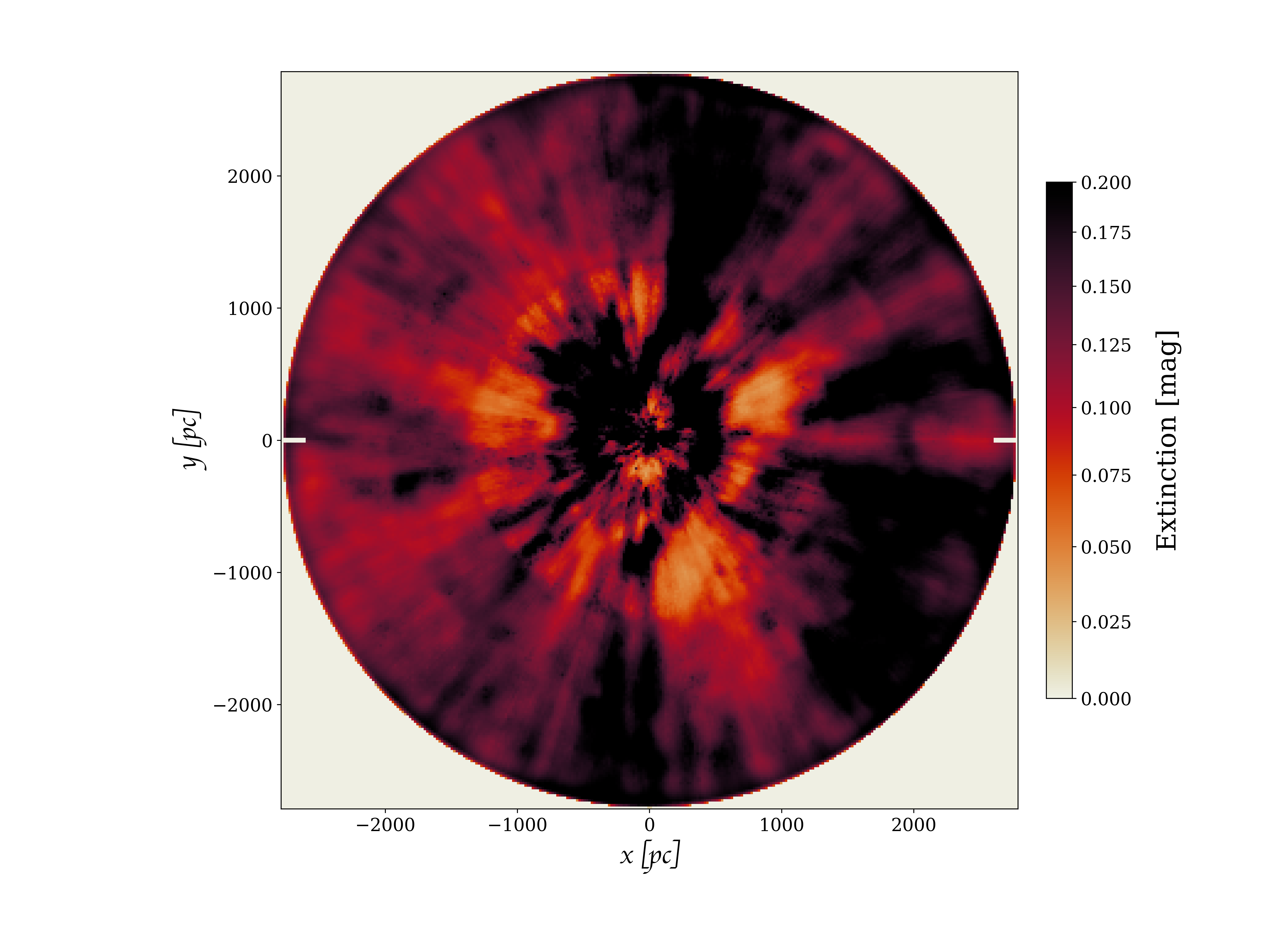}
    \includegraphics[width=0.49\textwidth]{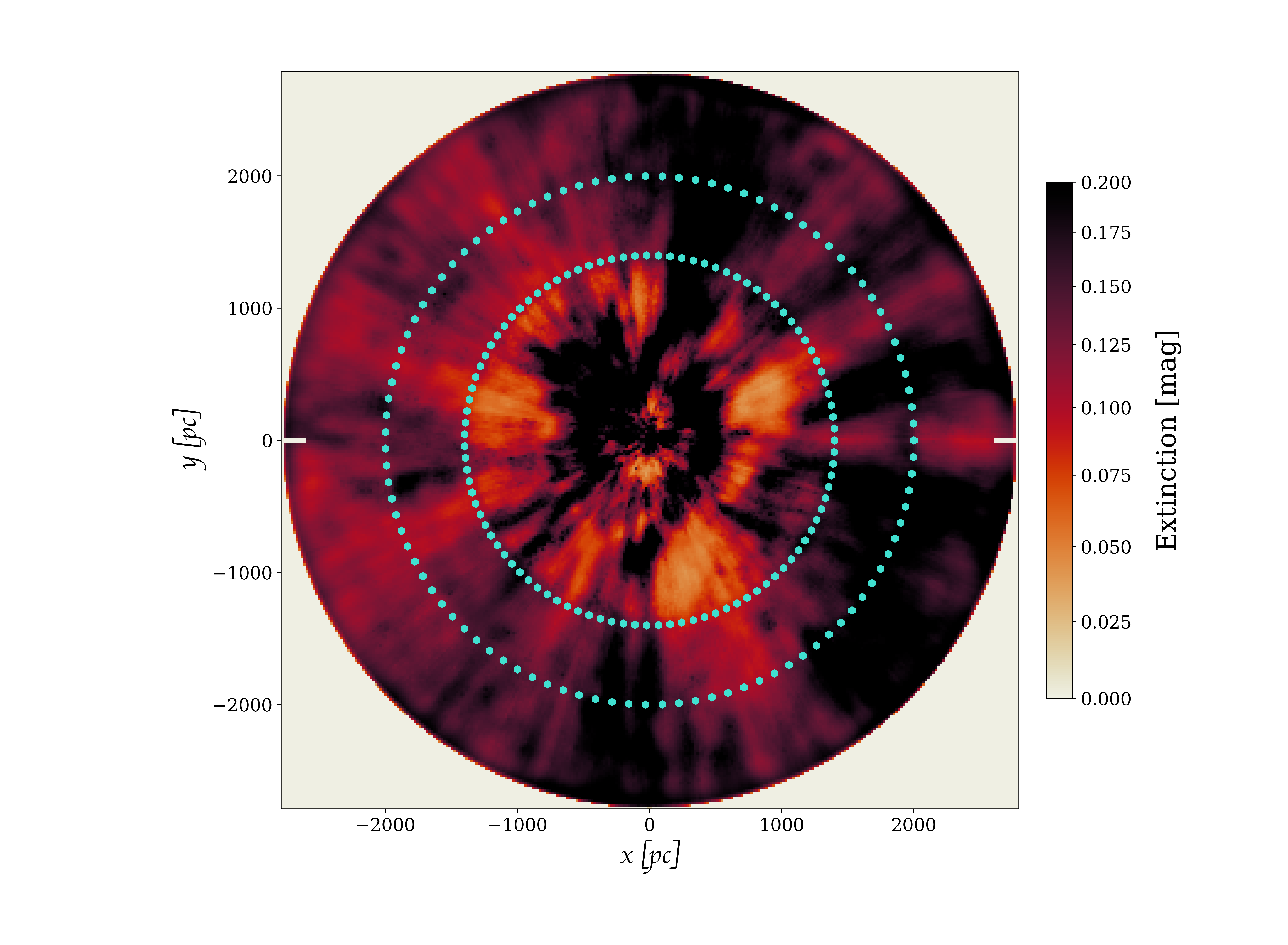}
    \caption{{\it Left:} Integrated density over the -500 $< z <$ 500 pc range with the colour scale truncated by a factor of 5 compared to fig.~\ref{fig:xy_dustInteg}. {\it Right:} As left, but with cyan hexagonal symbols drawn at $d$=1400 and 2000~pc highlighting the possible location of artefacts.}
    \label{fig:sat_ext}
\end{figure*}

\begin{figure*}
    \centering
    \includegraphics[width=0.27\textwidth]{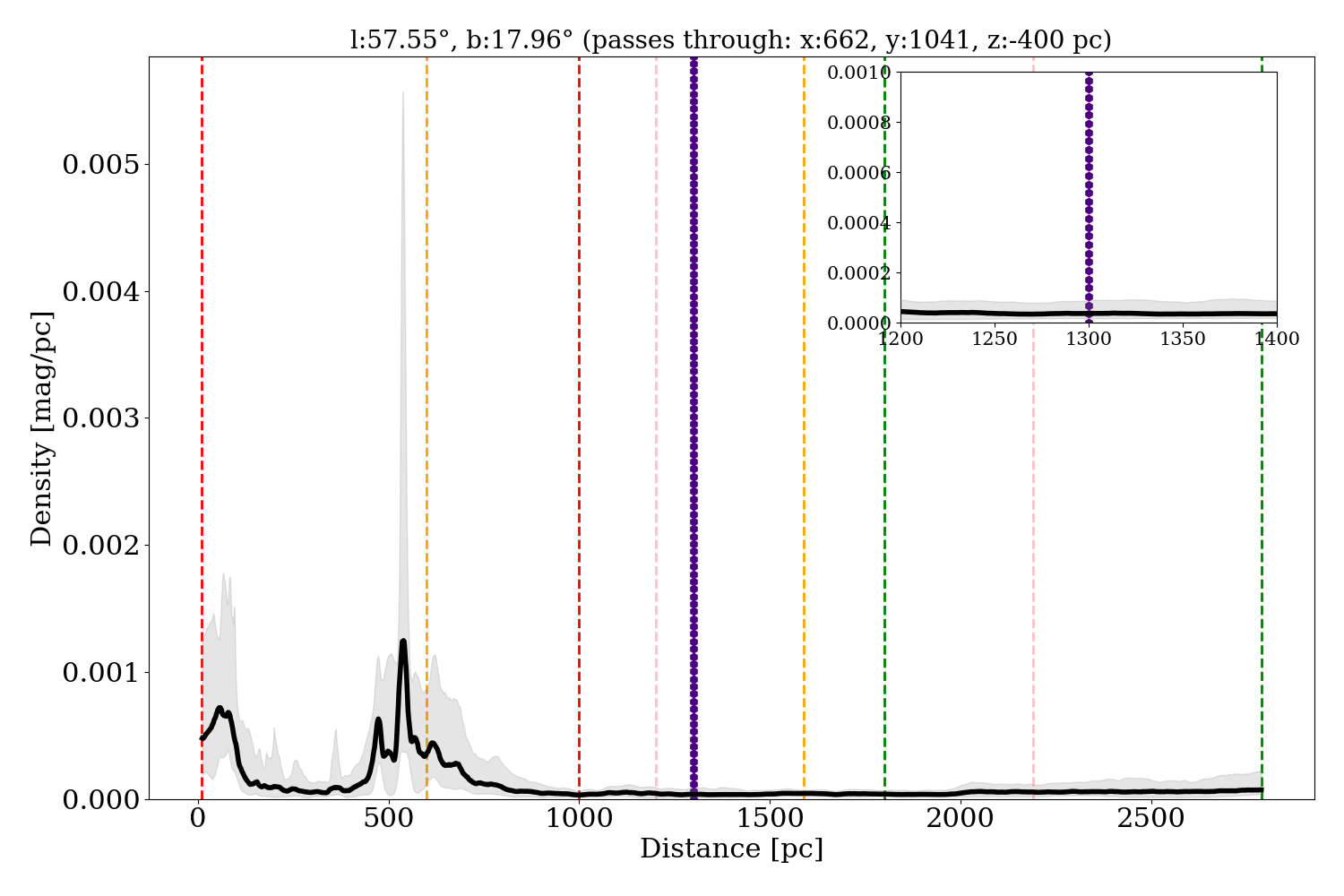}
    \includegraphics[width=0.27\textwidth]{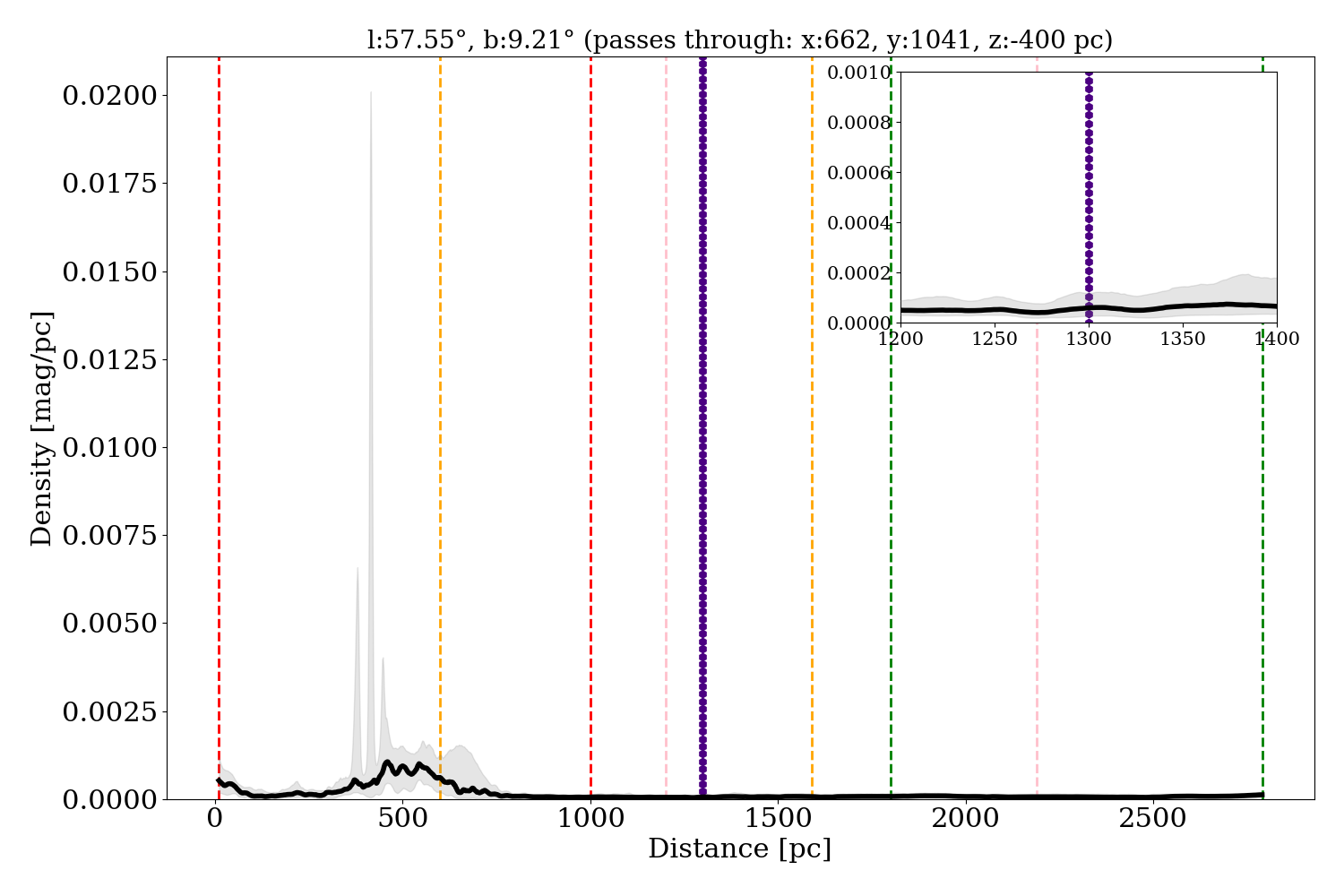}
    \includegraphics[width=0.27\textwidth]{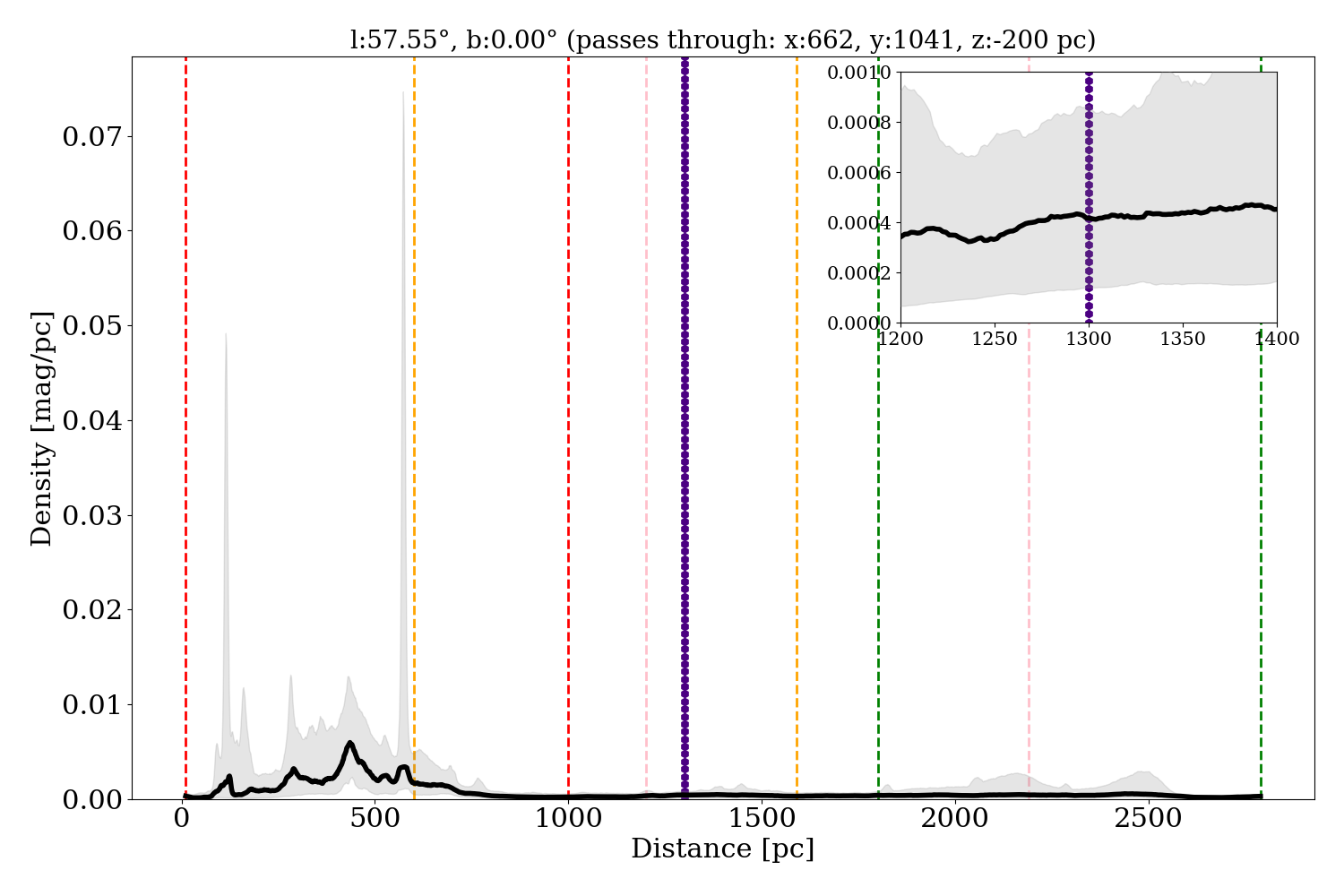}

    \includegraphics[width=0.27\textwidth]{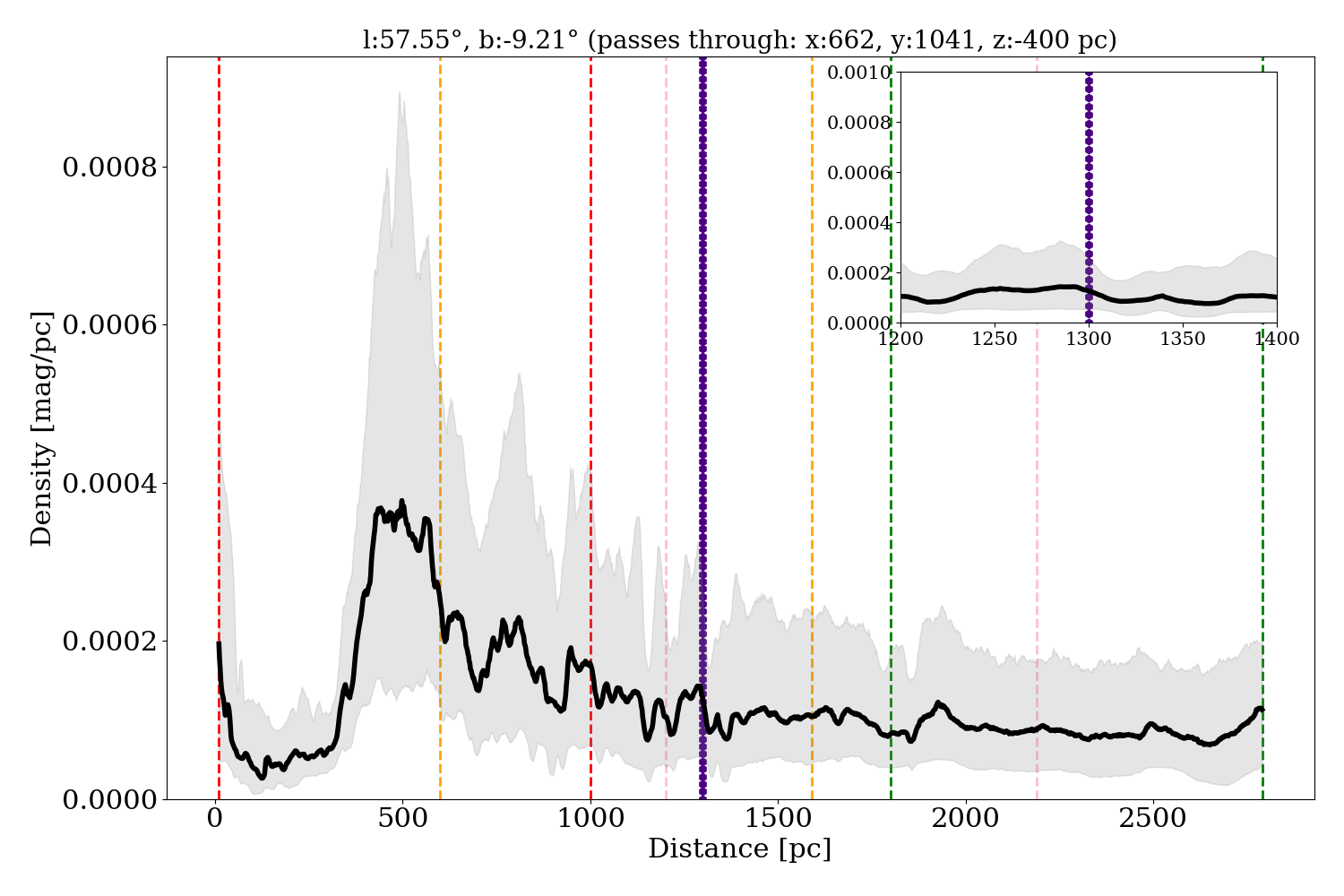}
    \includegraphics[width=0.27\textwidth]{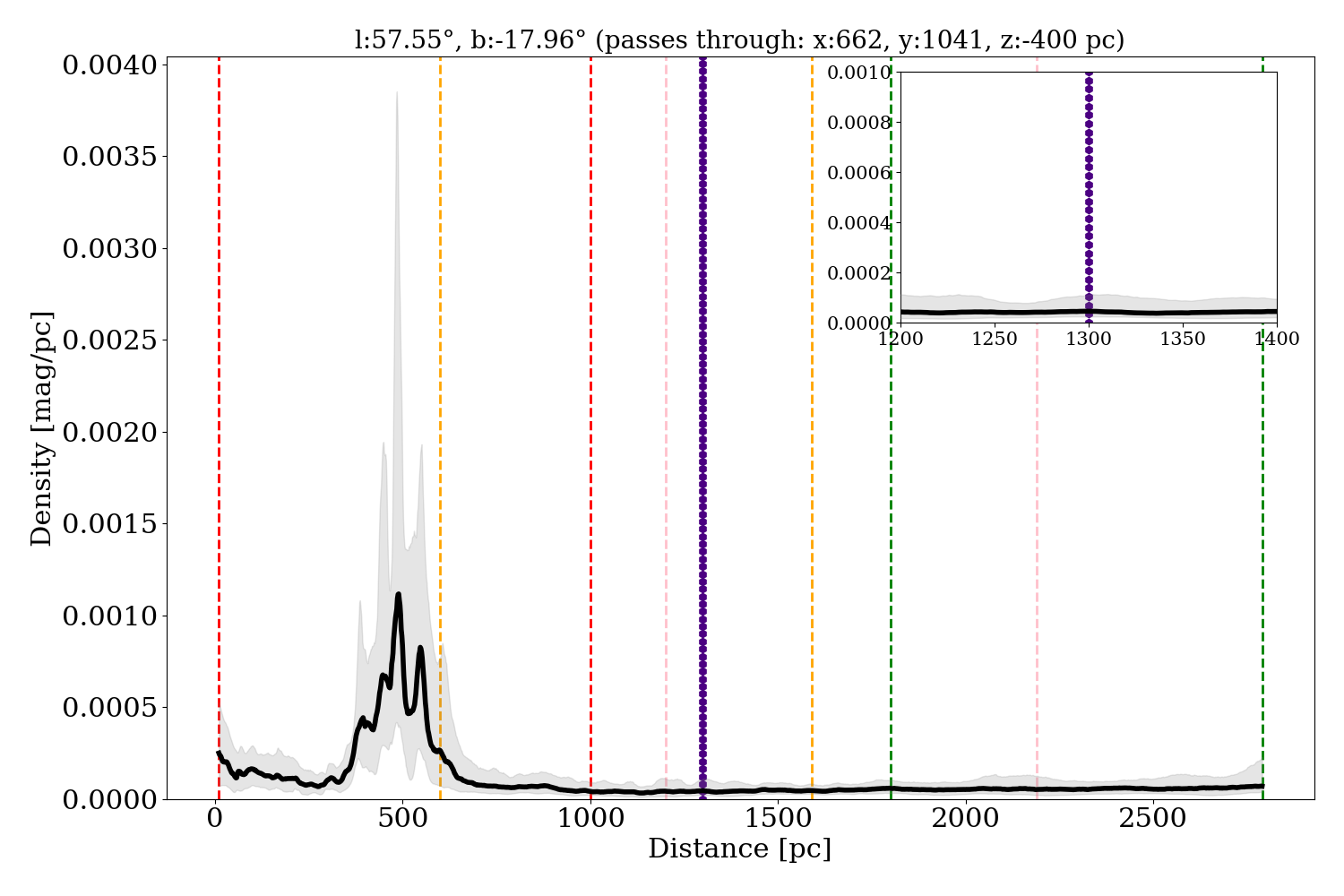}
    \includegraphics[width=0.27\textwidth]{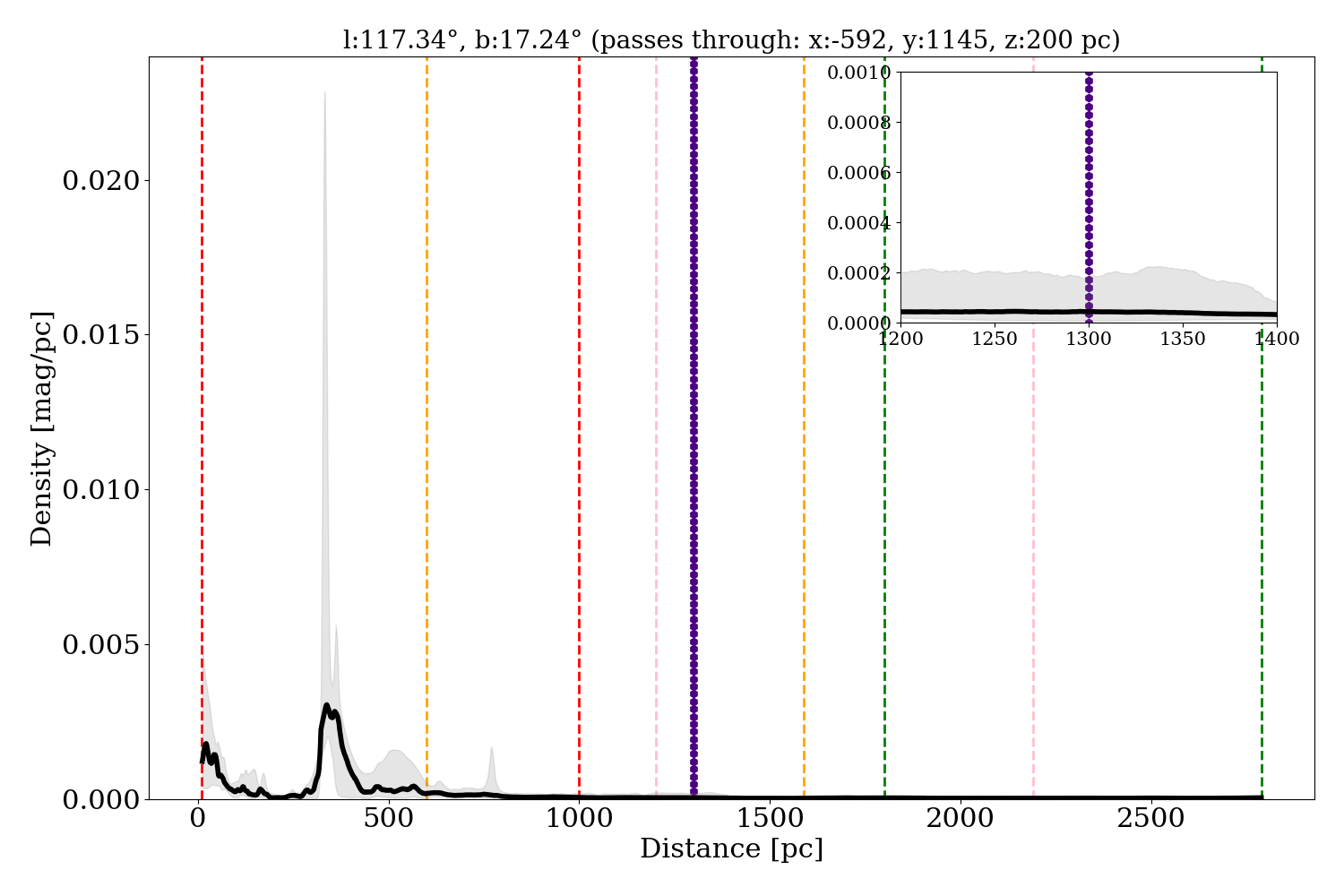}

    \includegraphics[width=0.27\textwidth]{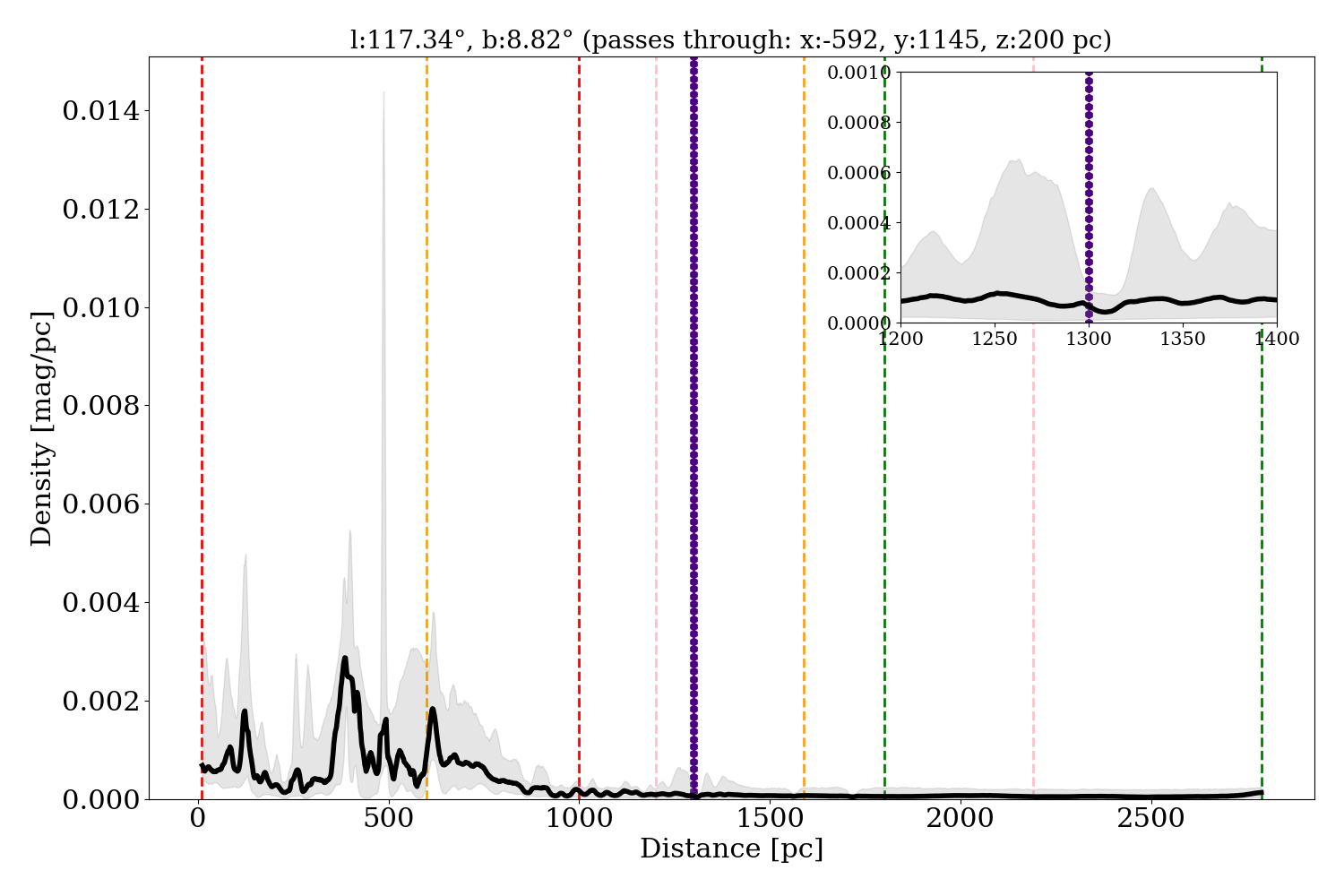}
    \includegraphics[width=0.27\textwidth]{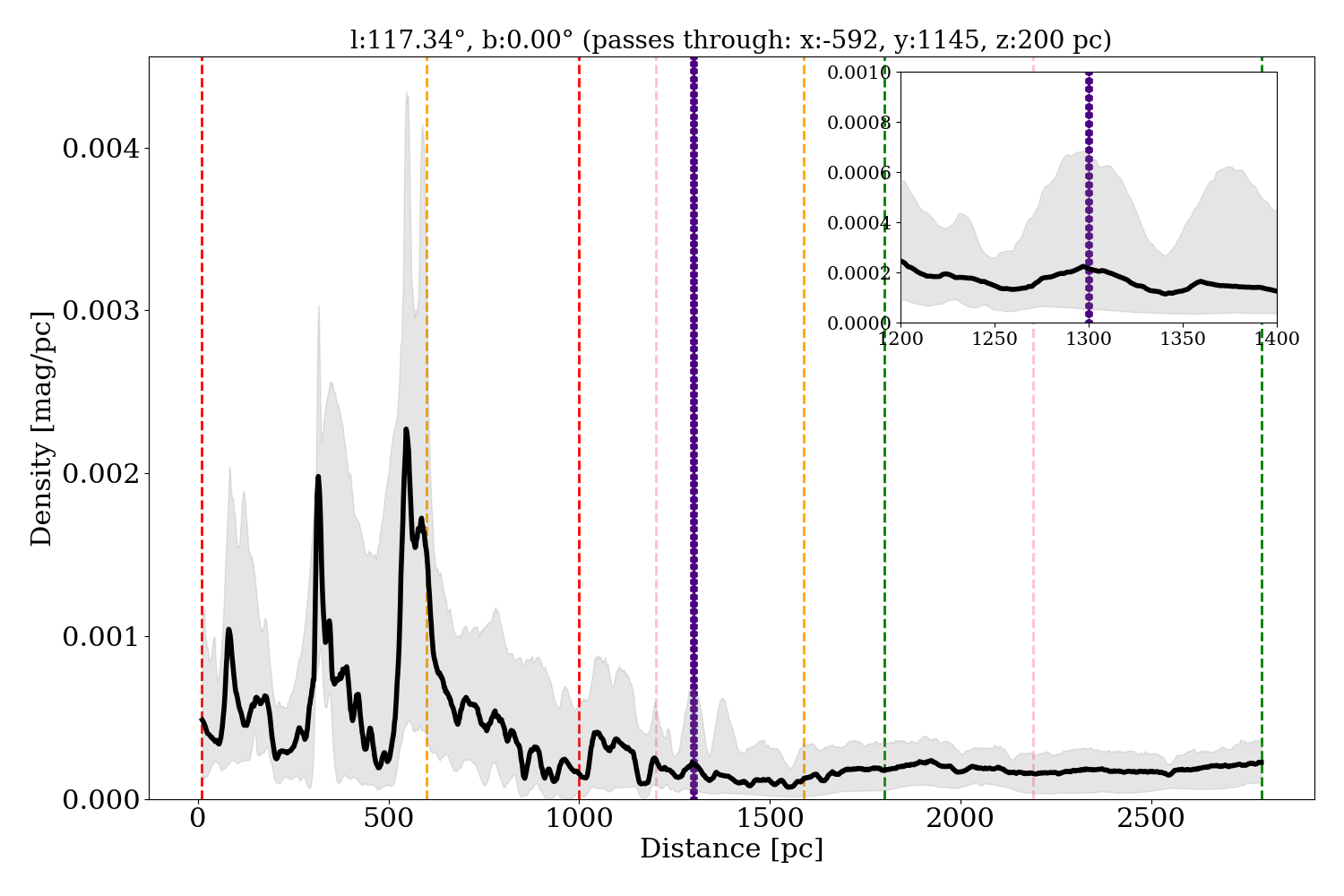}
    \includegraphics[width=0.27\textwidth]{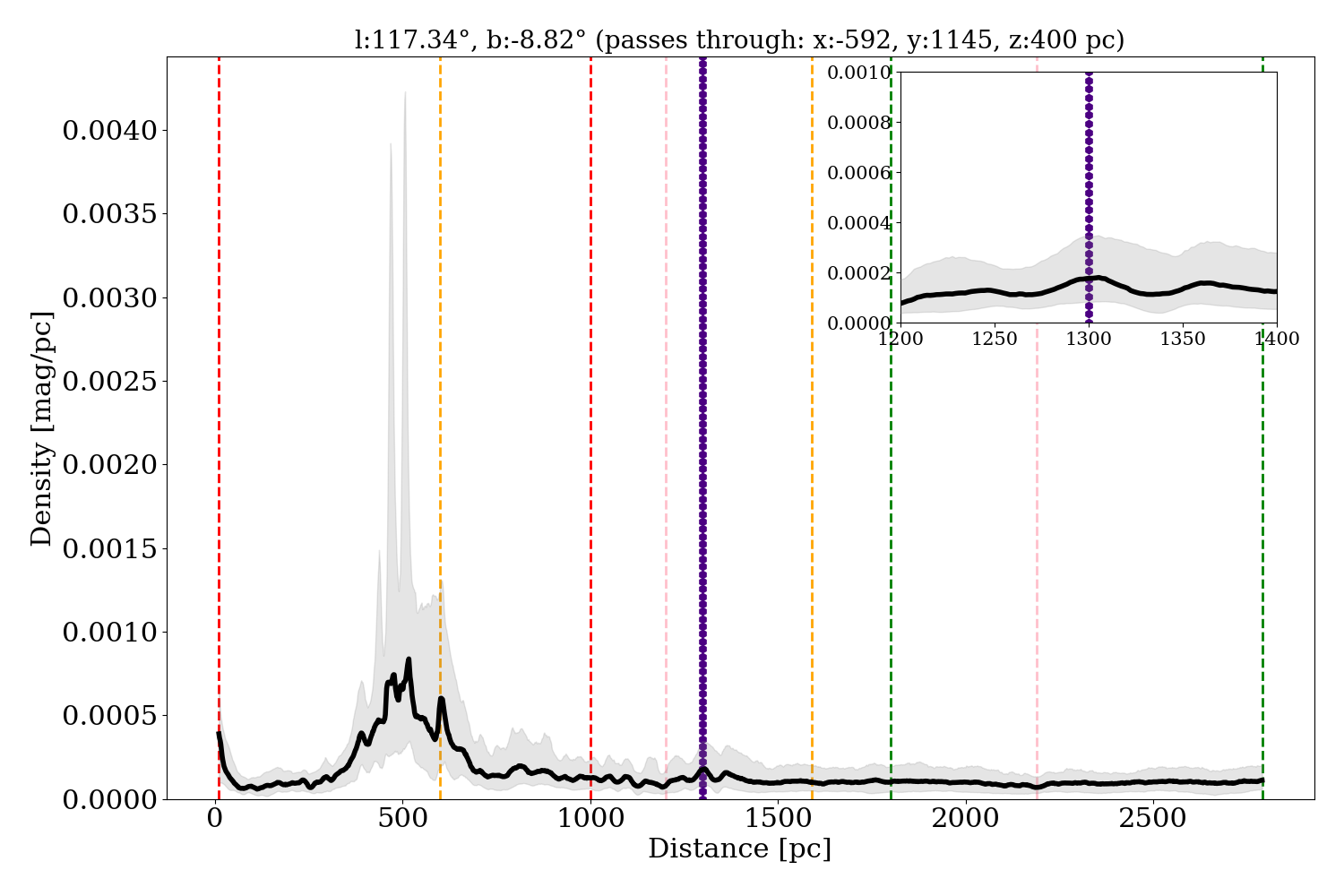}

    \includegraphics[width=0.27\textwidth]{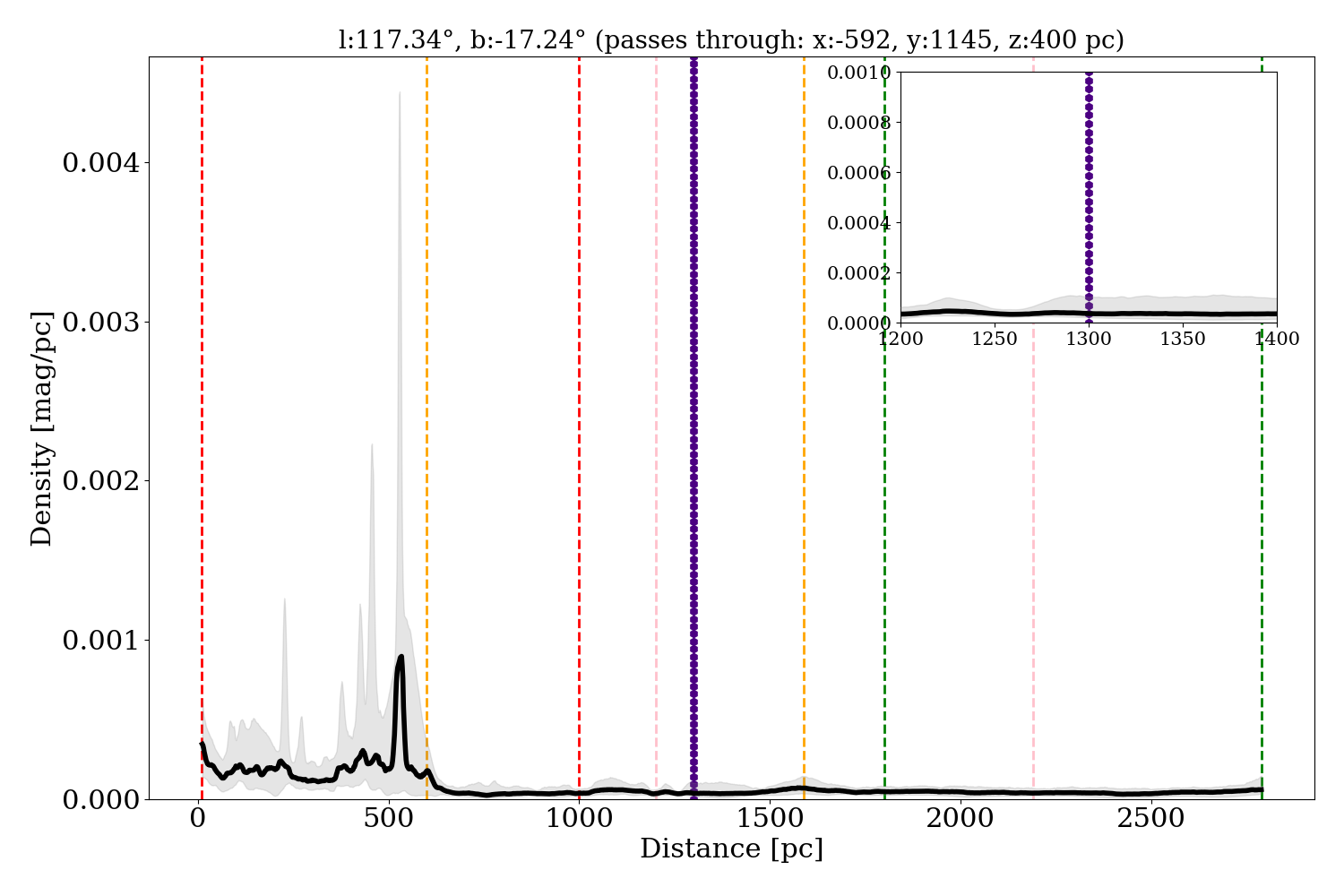}
    \includegraphics[width=0.27\textwidth]{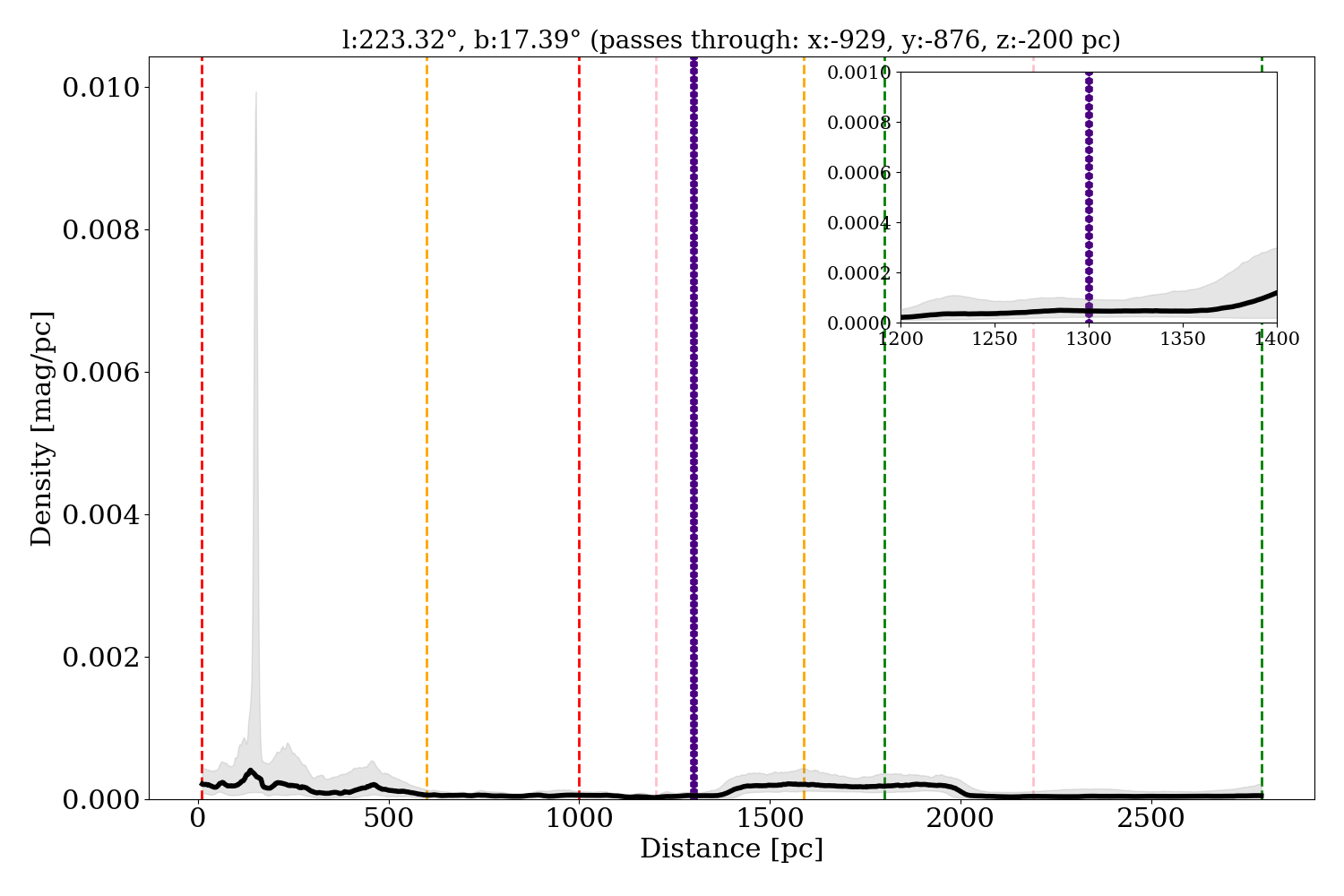}
    \includegraphics[width=0.27\textwidth]{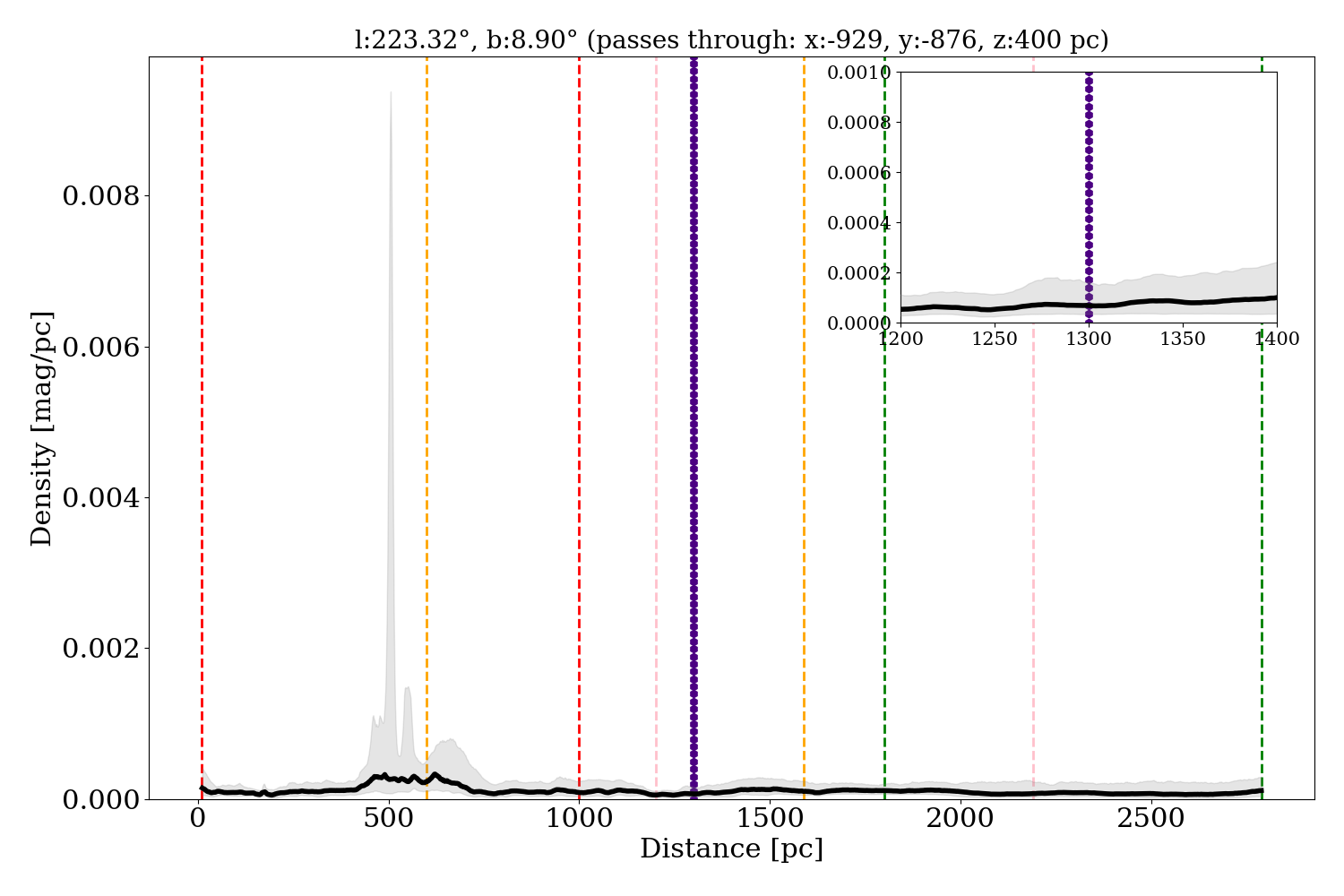}

    \includegraphics[width=0.27\textwidth]{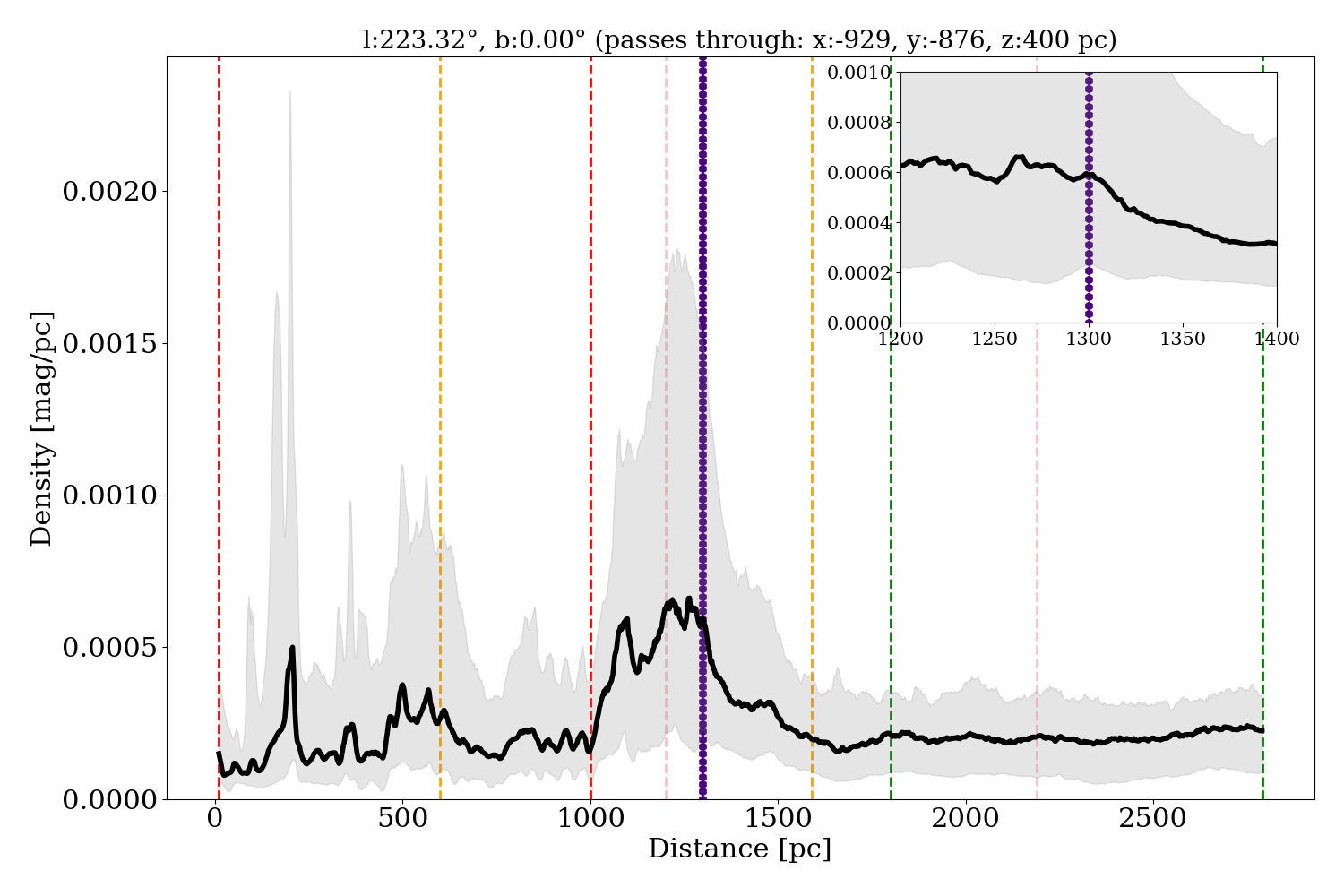}
    \includegraphics[width=0.27\textwidth]{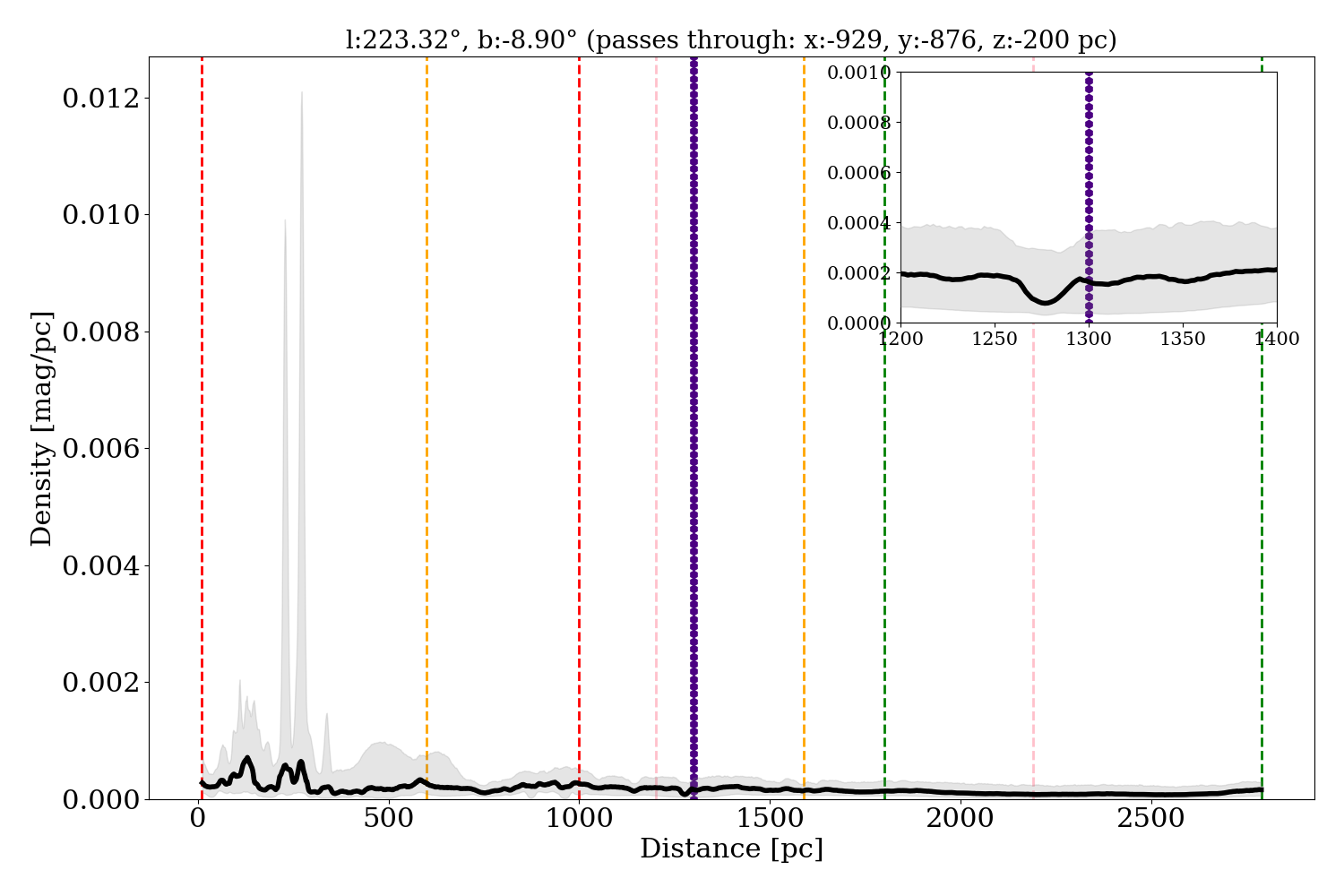}
    \includegraphics[width=0.27\textwidth]{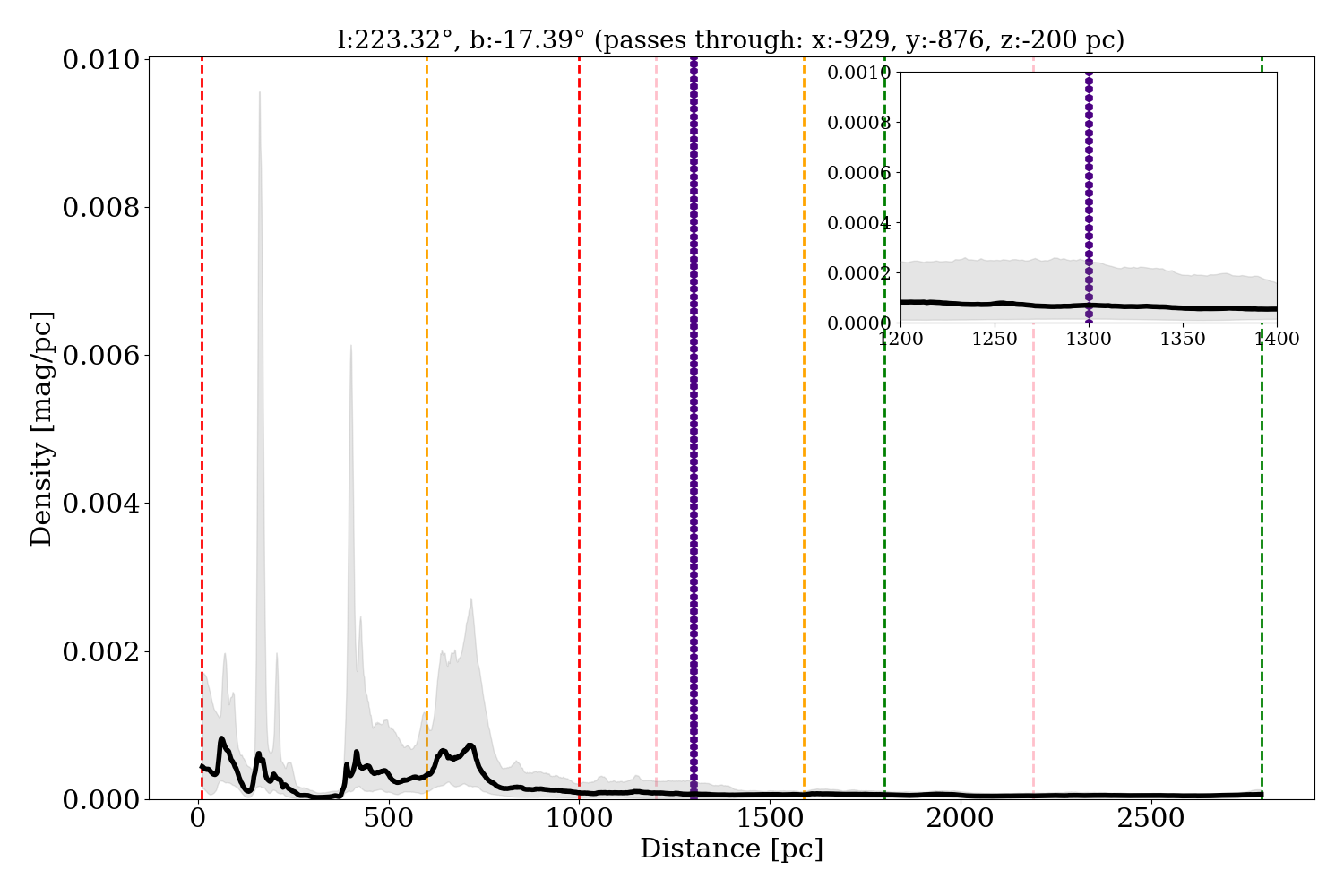}

    \includegraphics[width=0.27\textwidth]{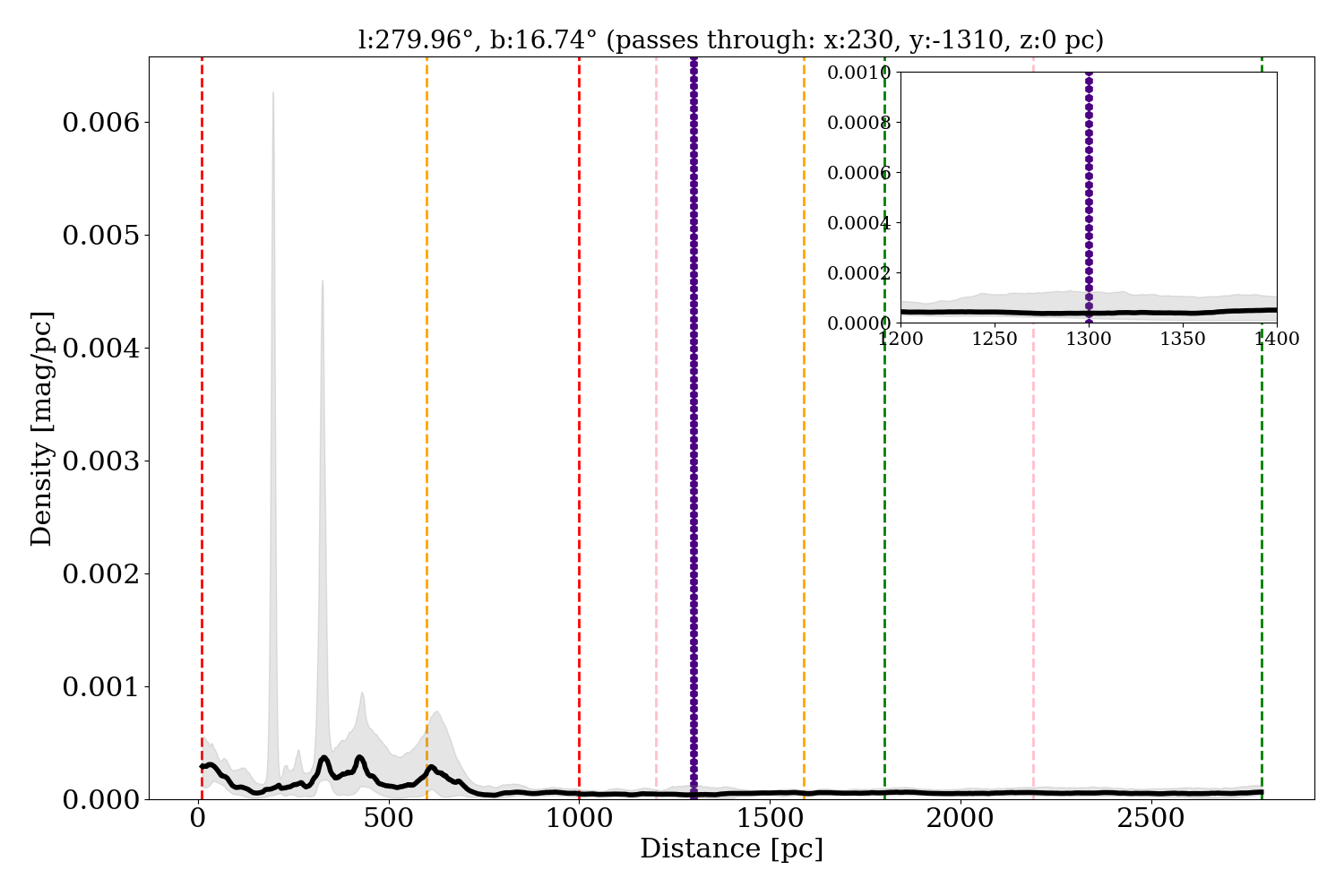}
    \includegraphics[width=0.27\textwidth]{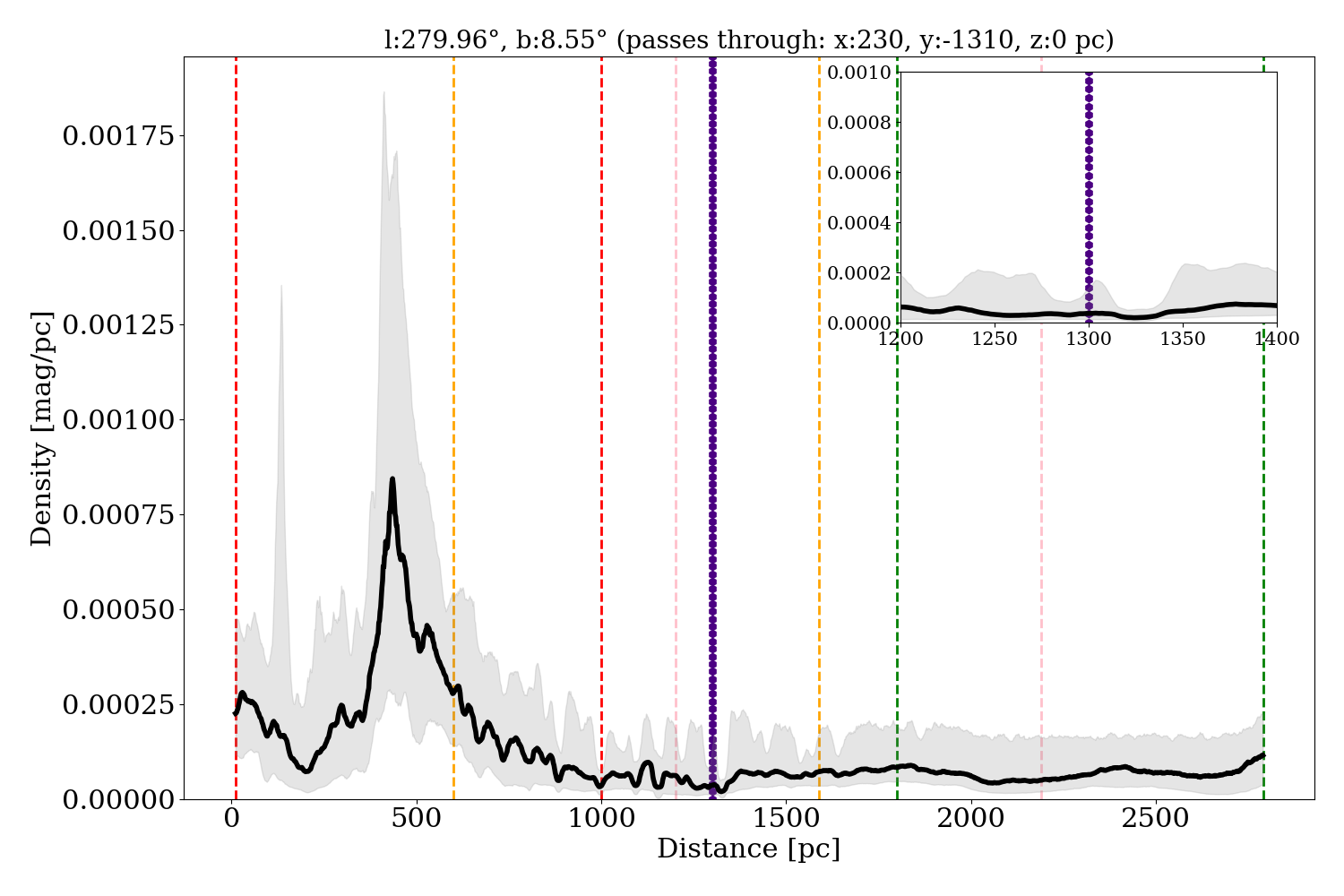}
    \includegraphics[width=0.27\textwidth]{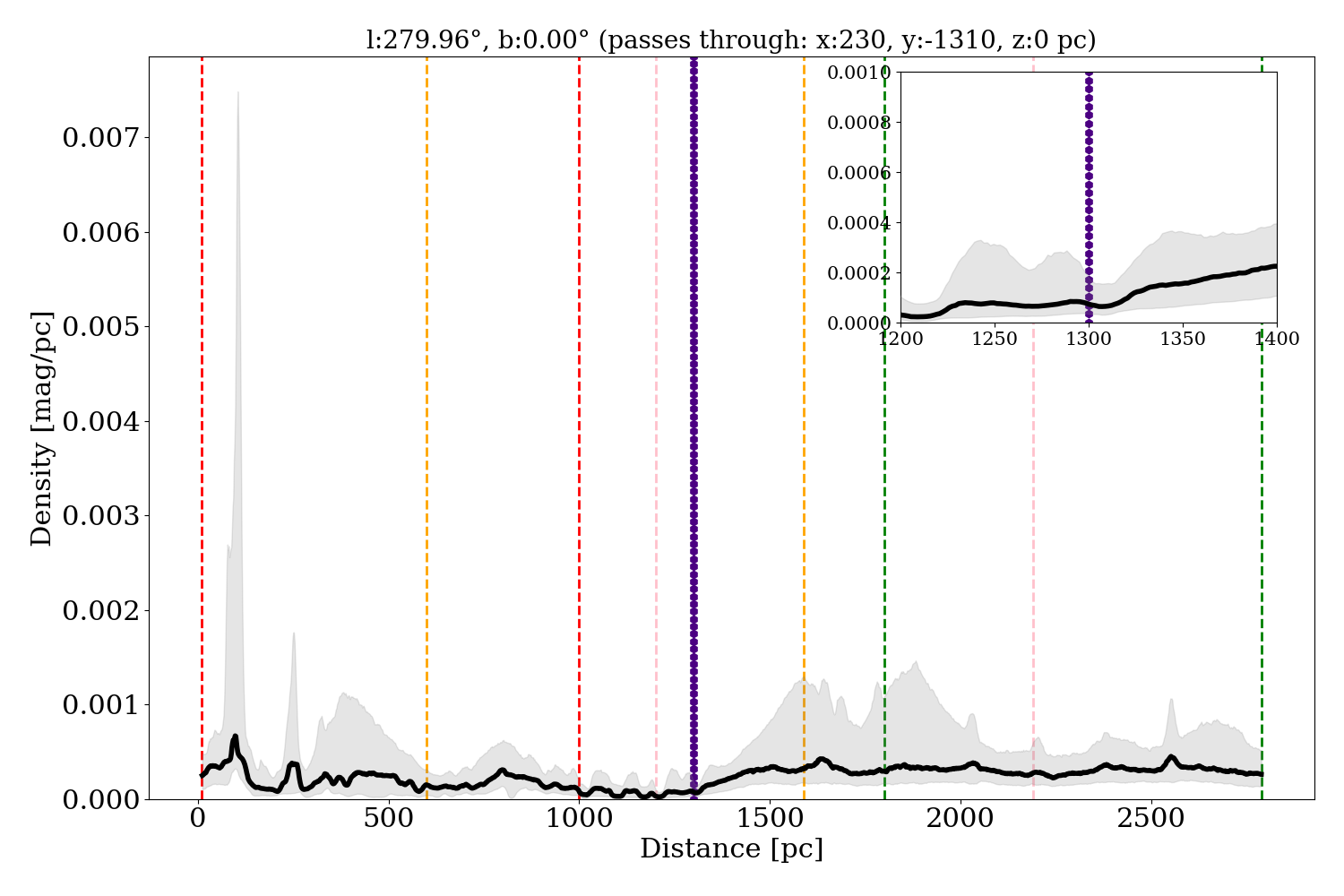}
    
    \includegraphics[width=0.27\textwidth]{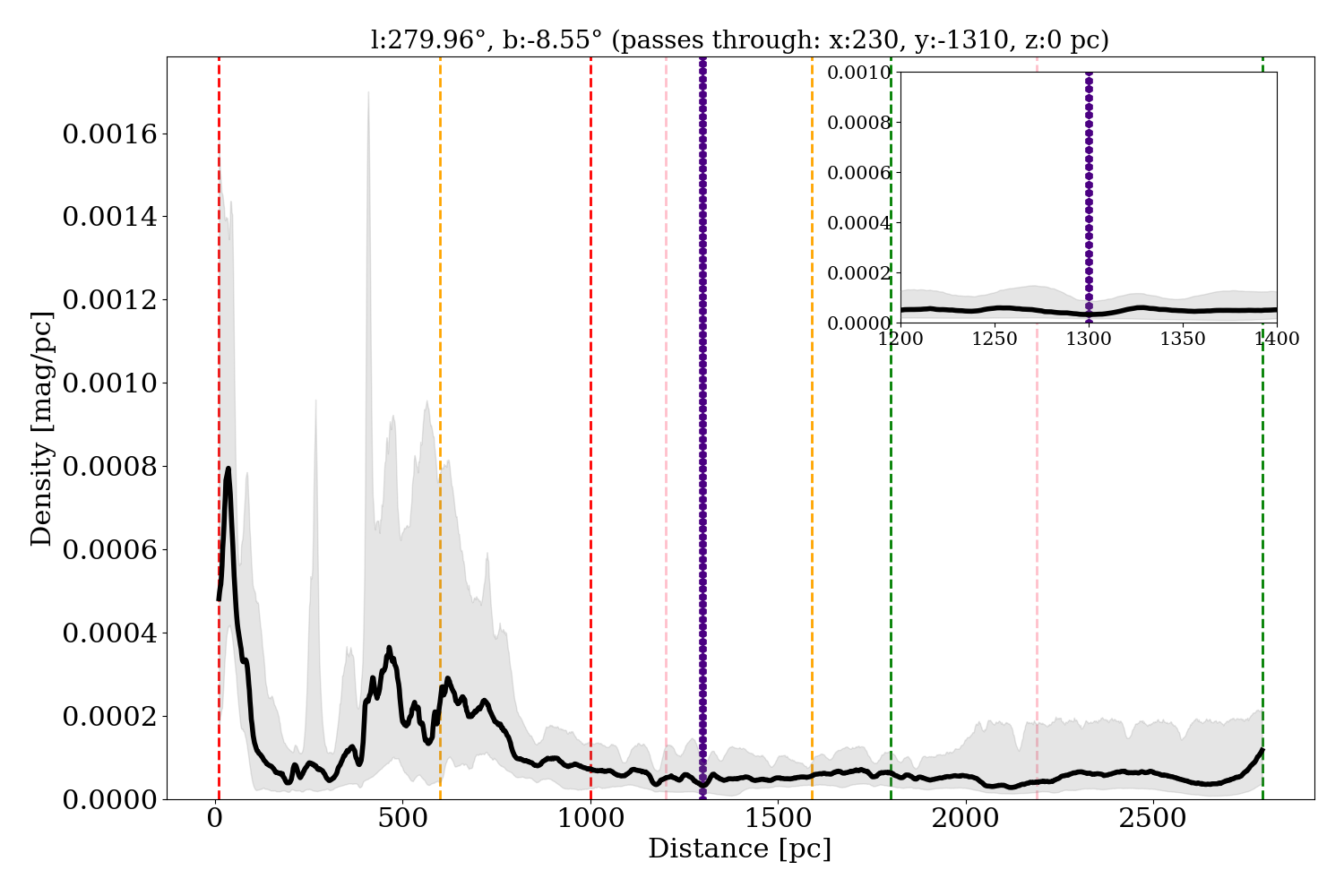}
    \includegraphics[width=0.27\textwidth]{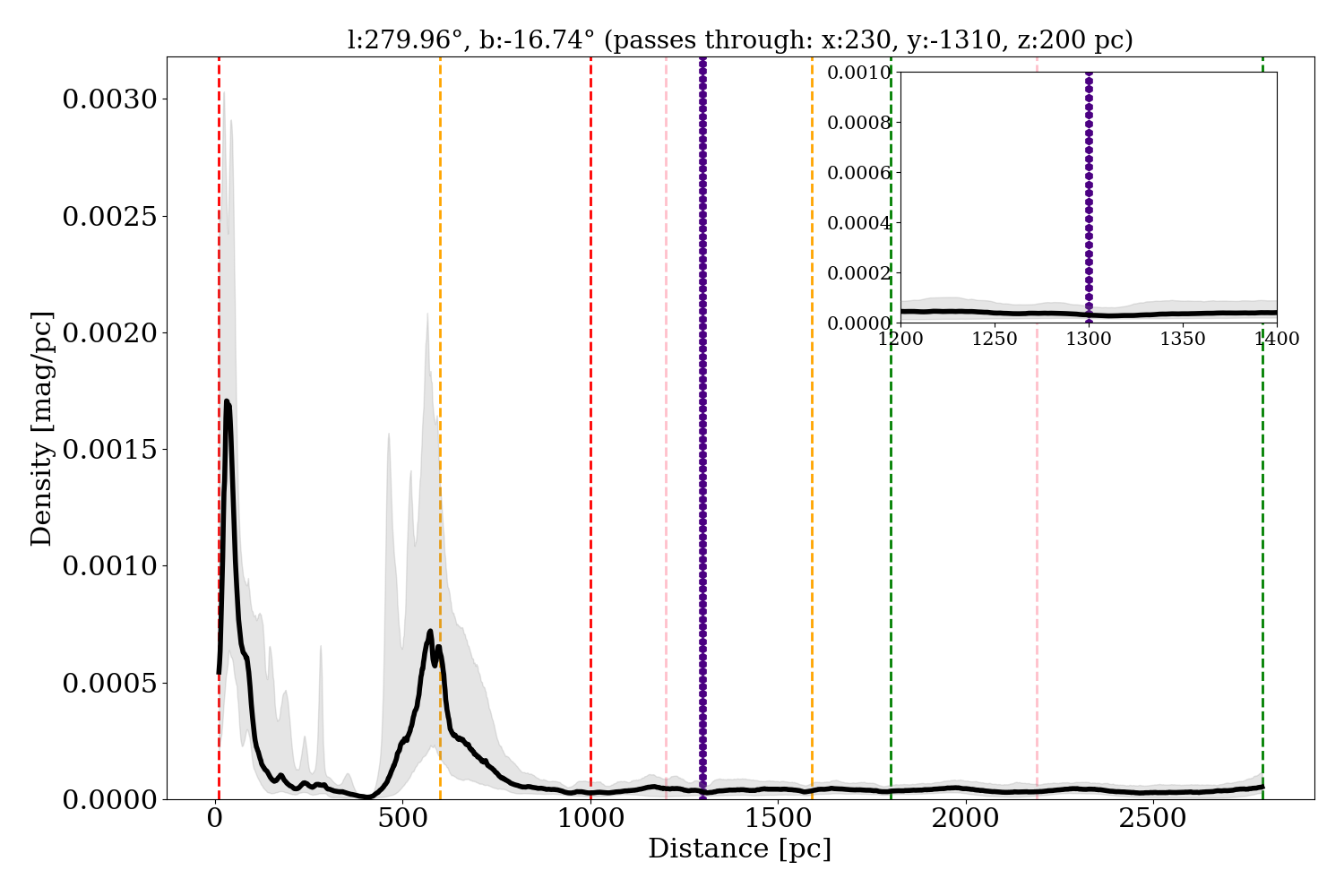}
    \includegraphics[width=0.27\textwidth]{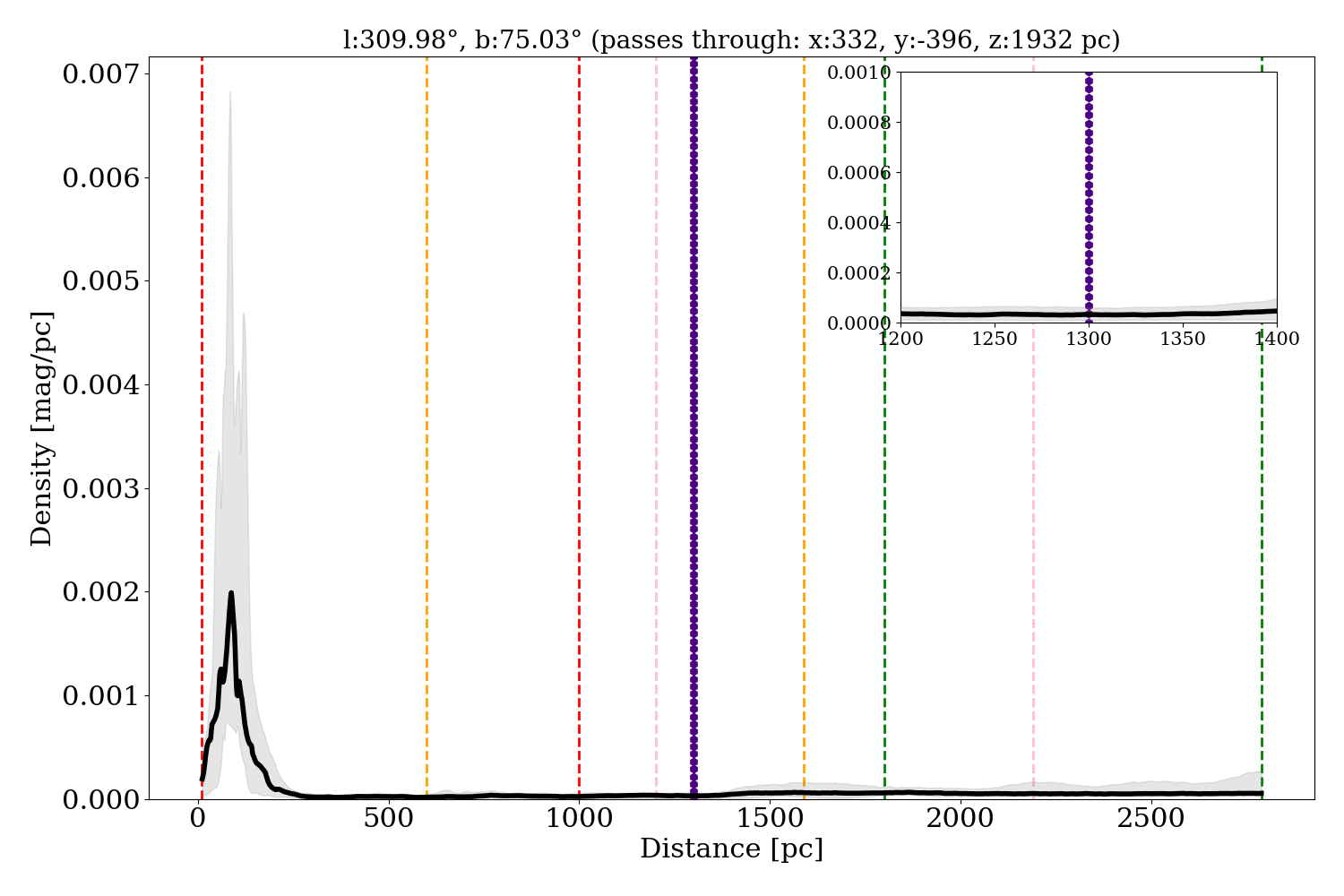}

    \caption{Density along lines of sight highlighting that any changes in the density at the location of the artefact are small compared to the uncertainty on the density. The purple hexagonal line drawn at $d=1300$ pc highlights the possible location of artefact as shown in Fig~\ref{fig:sat_dens}. The other coloured dashed lines represent the chunk boundaries as shown in Fig.~\ref{fig:MergeBoundaries}.}
    \label{fig:Dens_LOS}
\end{figure*}

\begin{figure*}
    \centering
    \includegraphics[width=0.27\textwidth]{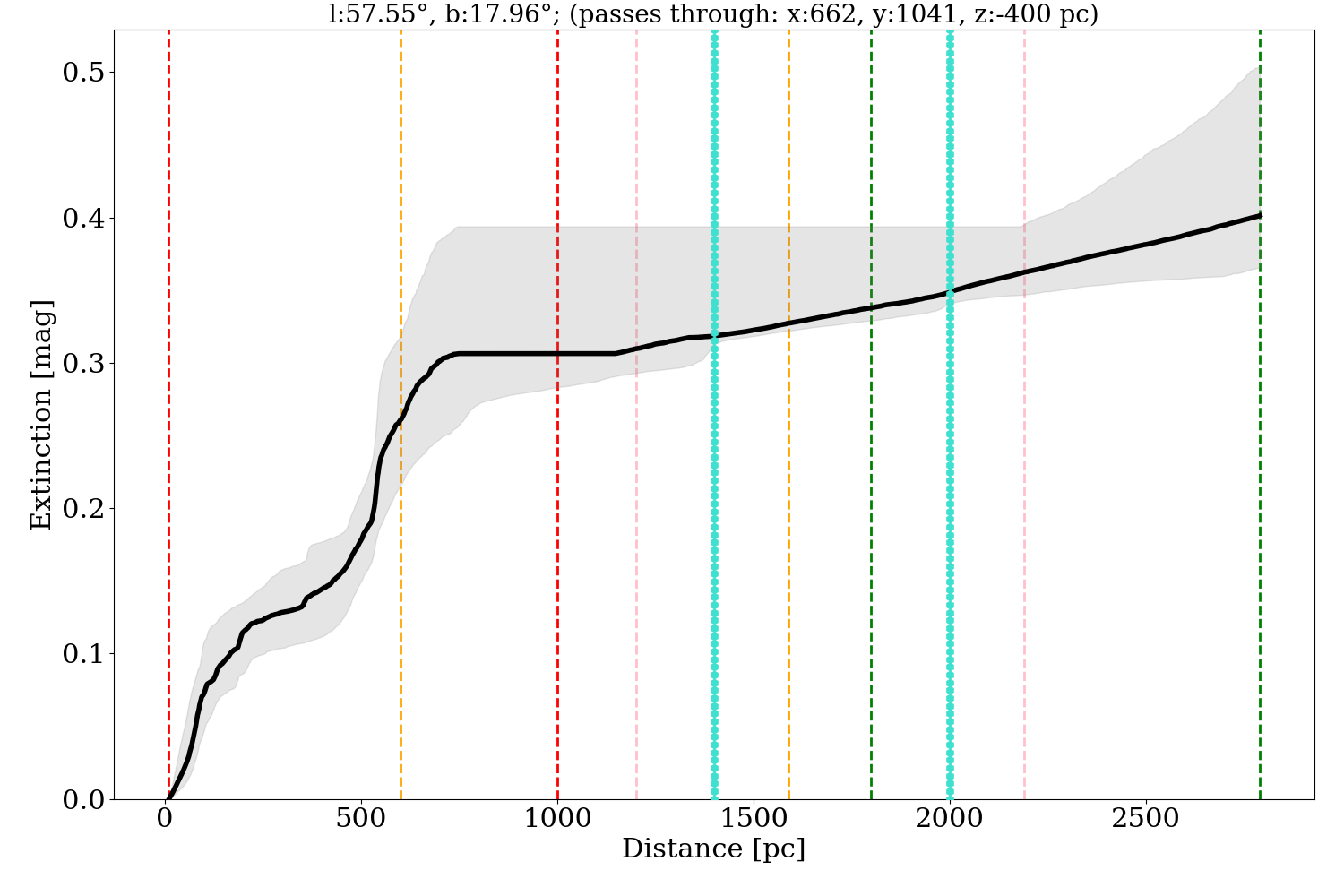}
    \includegraphics[width=0.27\textwidth]{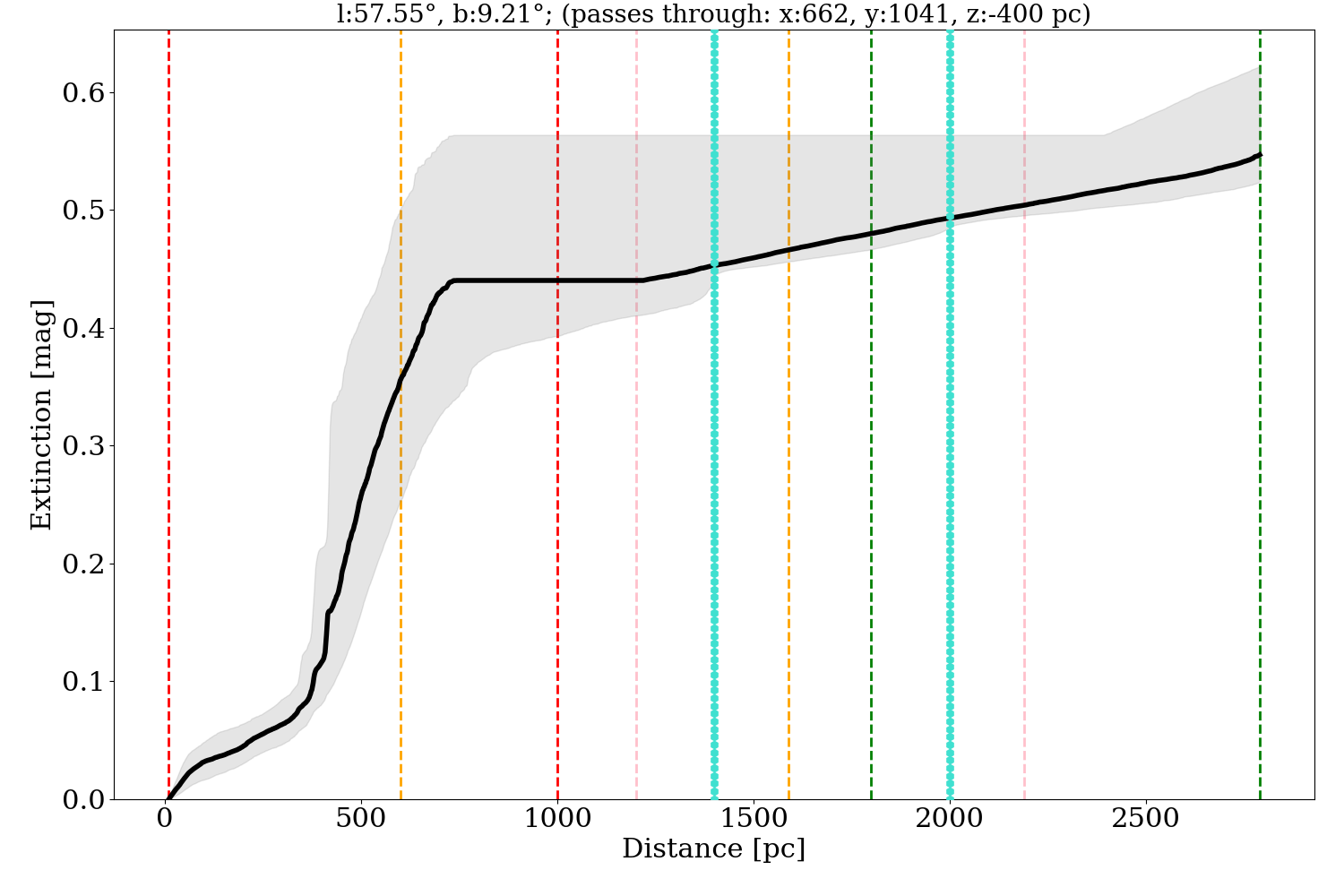}
    \includegraphics[width=0.27\textwidth]{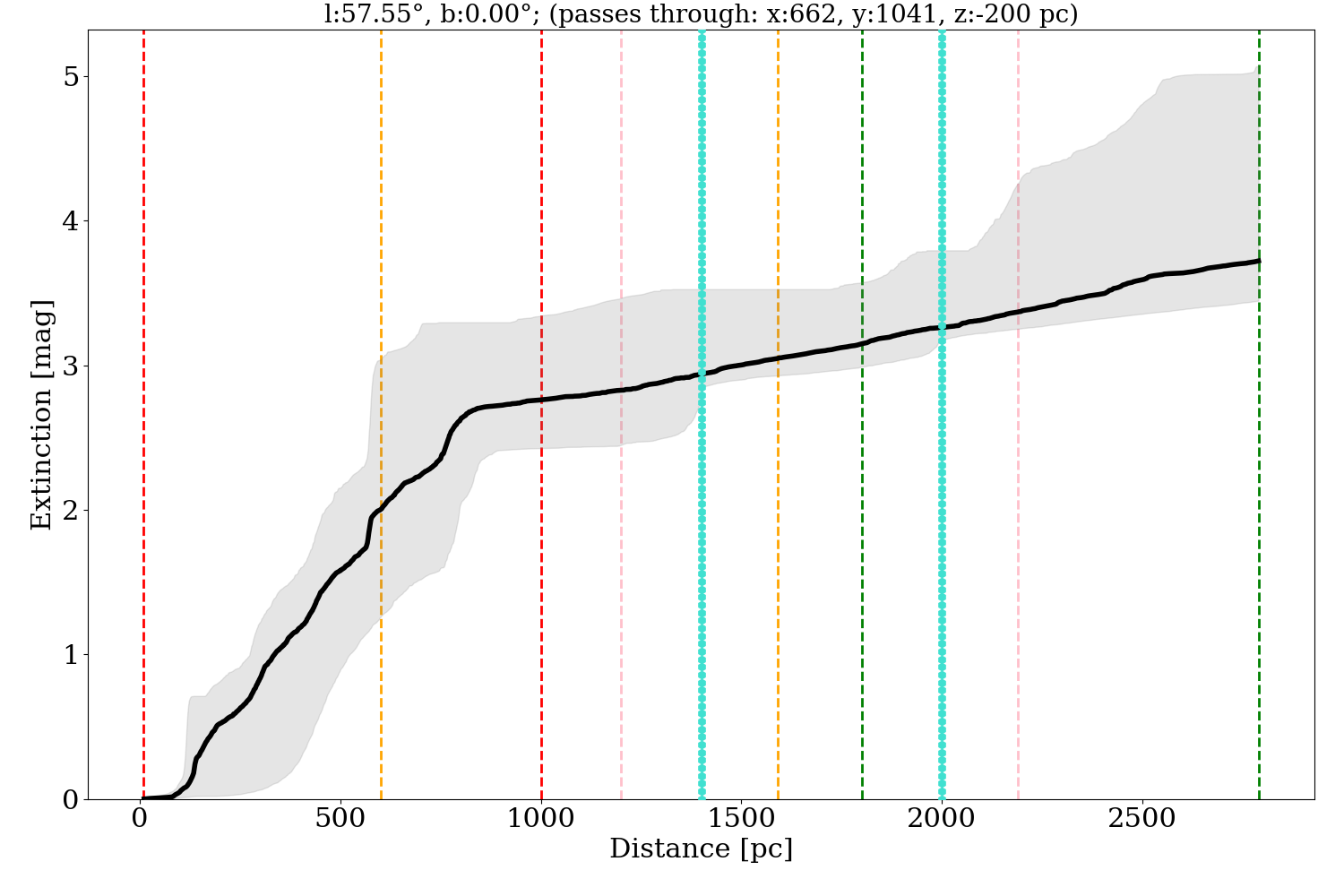}

    \includegraphics[width=0.27\textwidth]{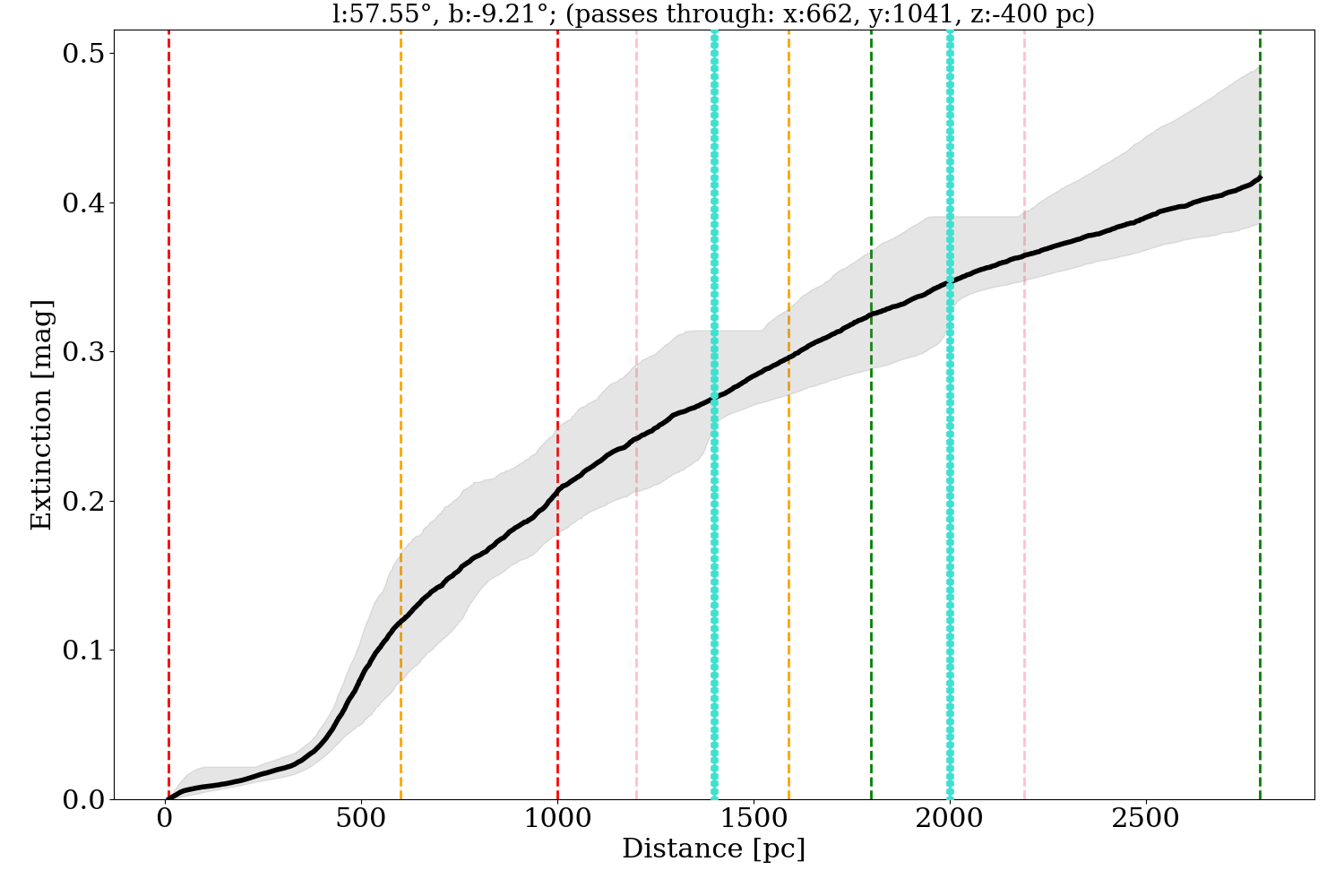}
    \includegraphics[width=0.27\textwidth]{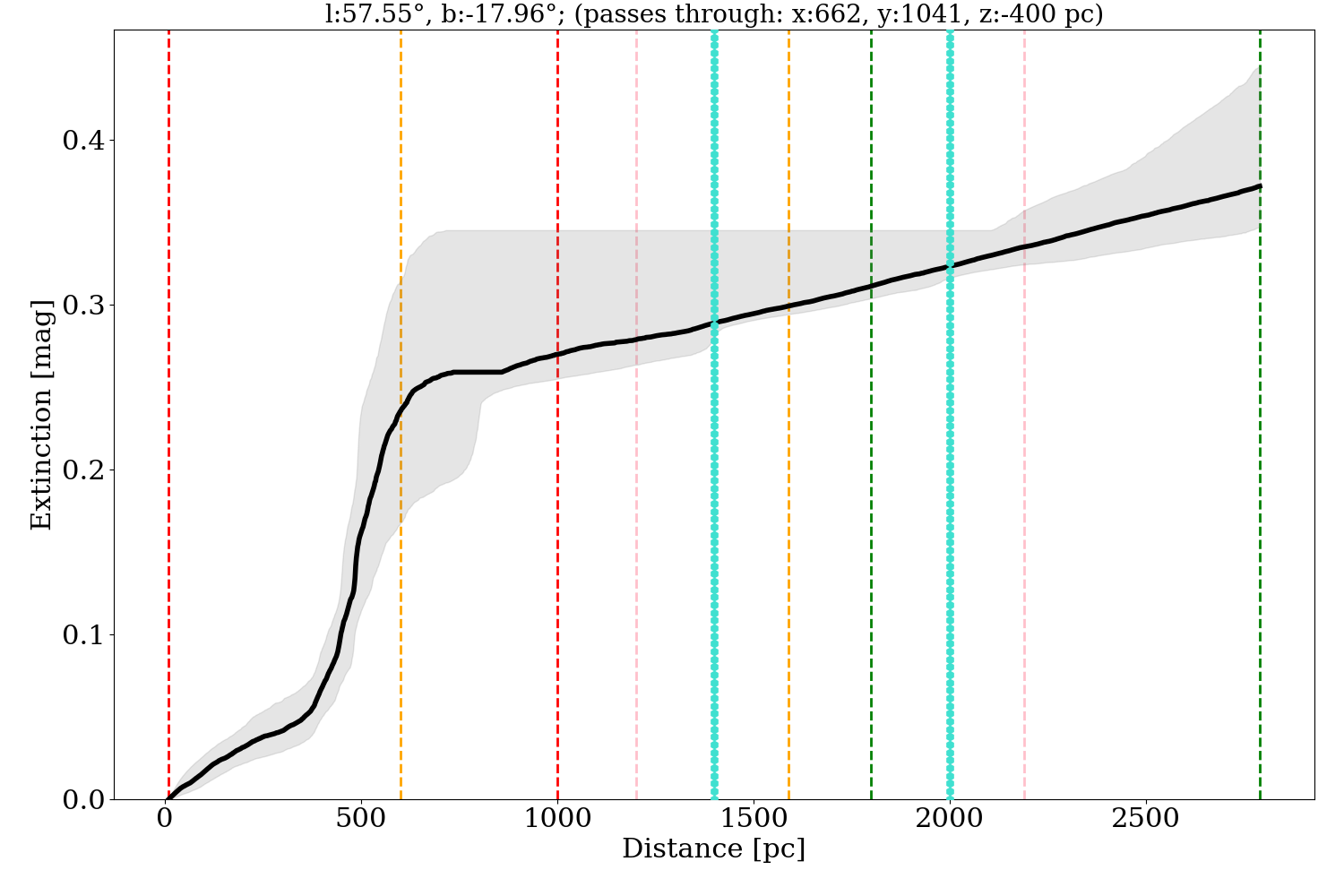}
    \includegraphics[width=0.27\textwidth]{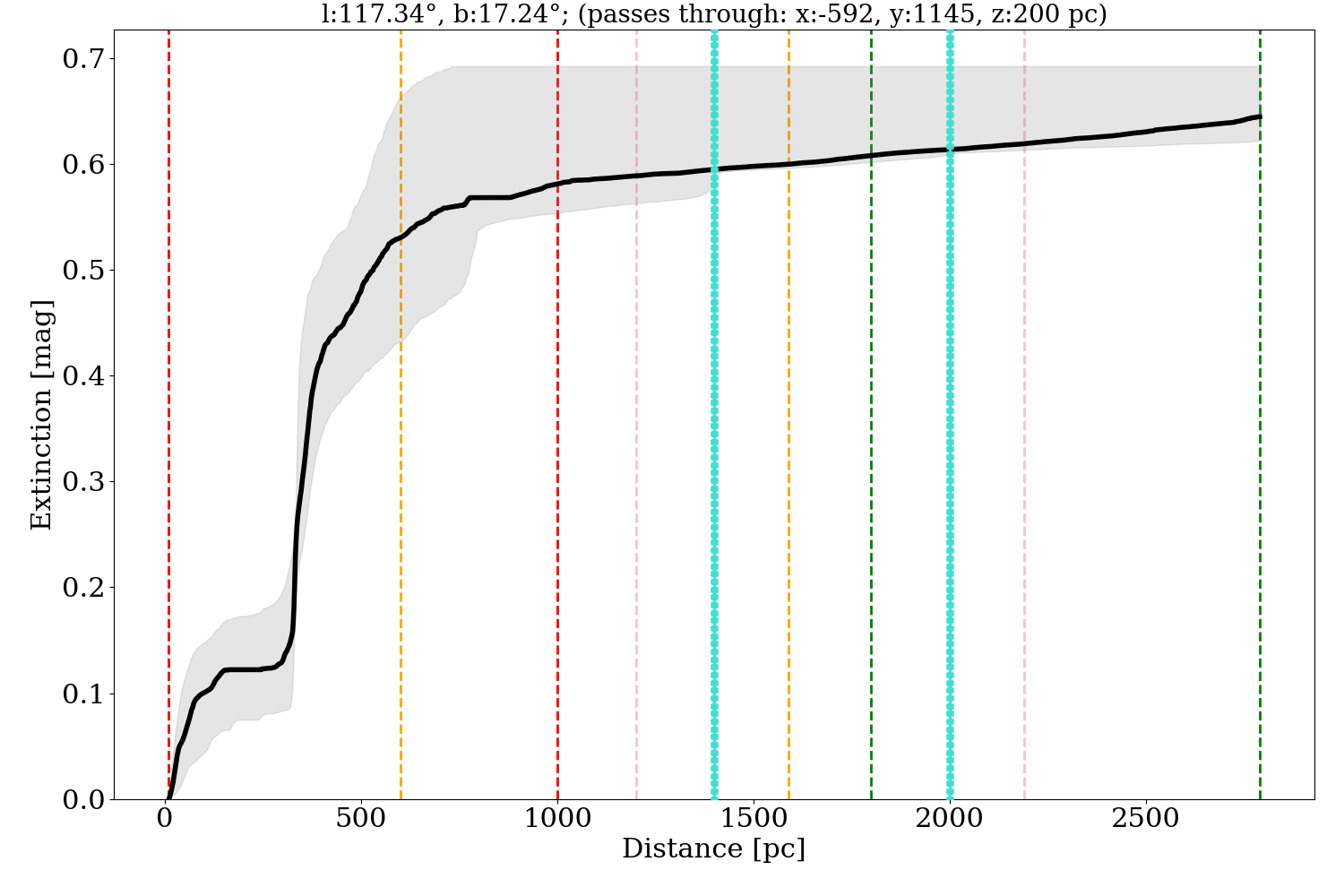}

    \includegraphics[width=0.27\textwidth]{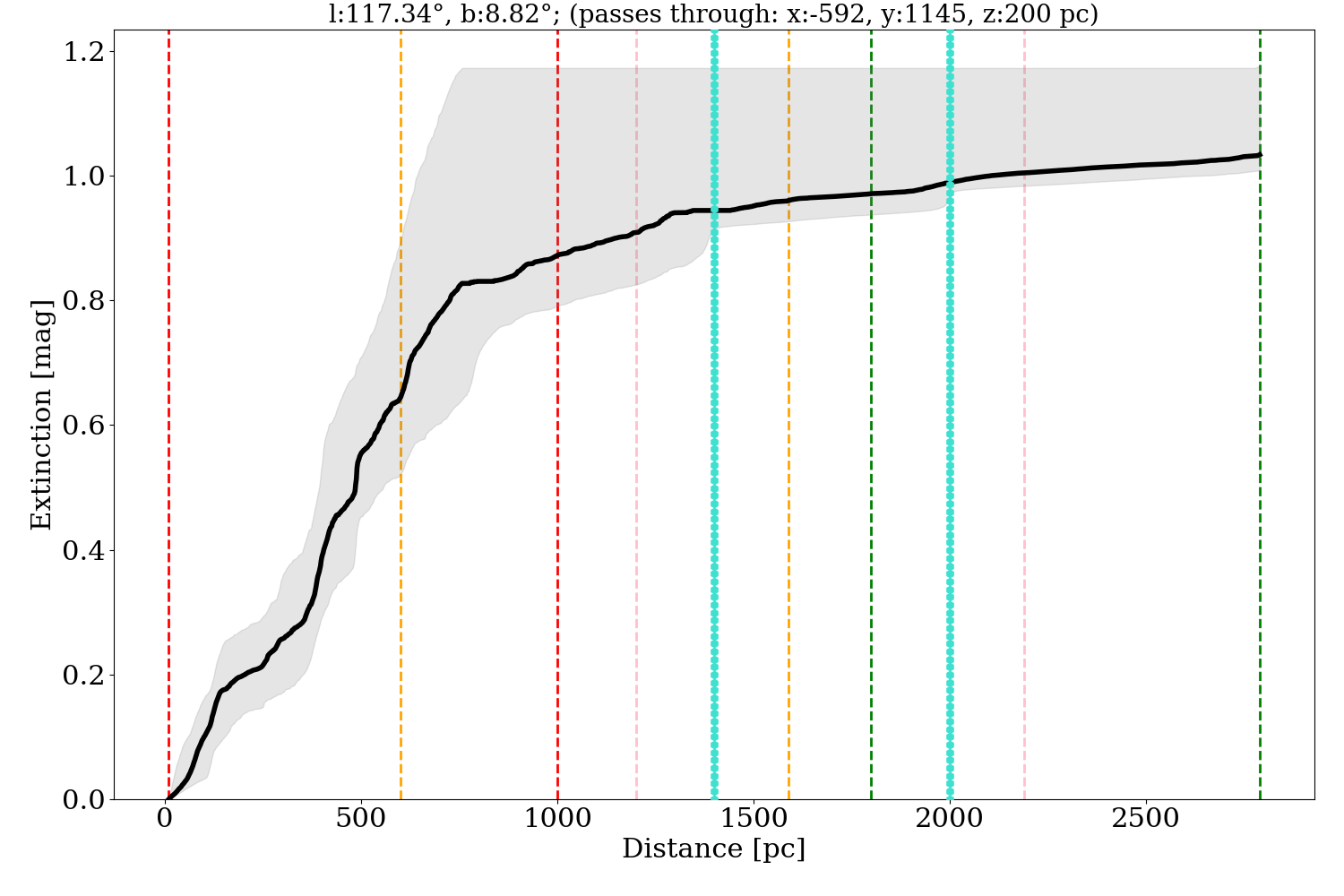}
    \includegraphics[width=0.27\textwidth]{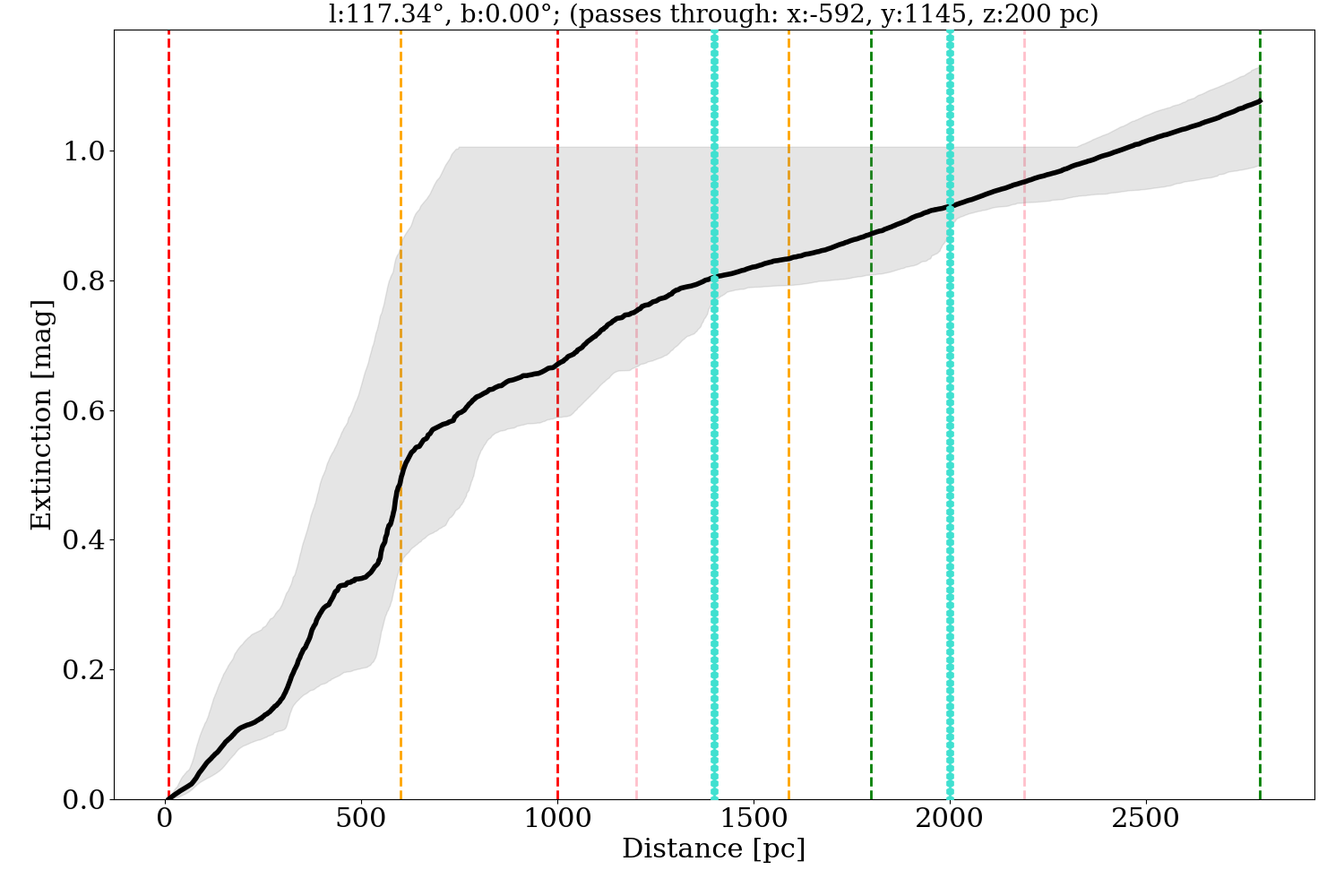}
    \includegraphics[width=0.27\textwidth]{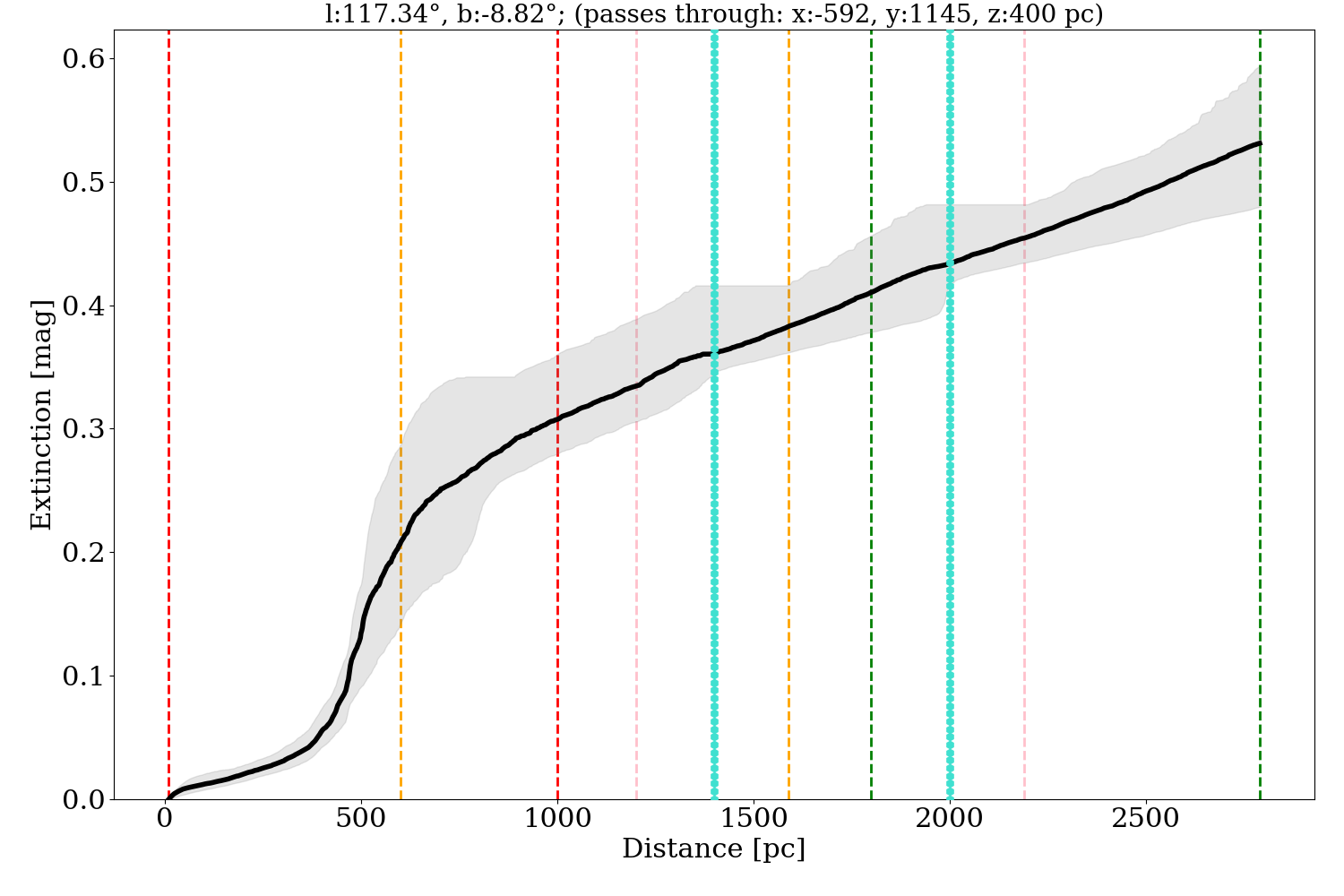}

    \includegraphics[width=0.27\textwidth]{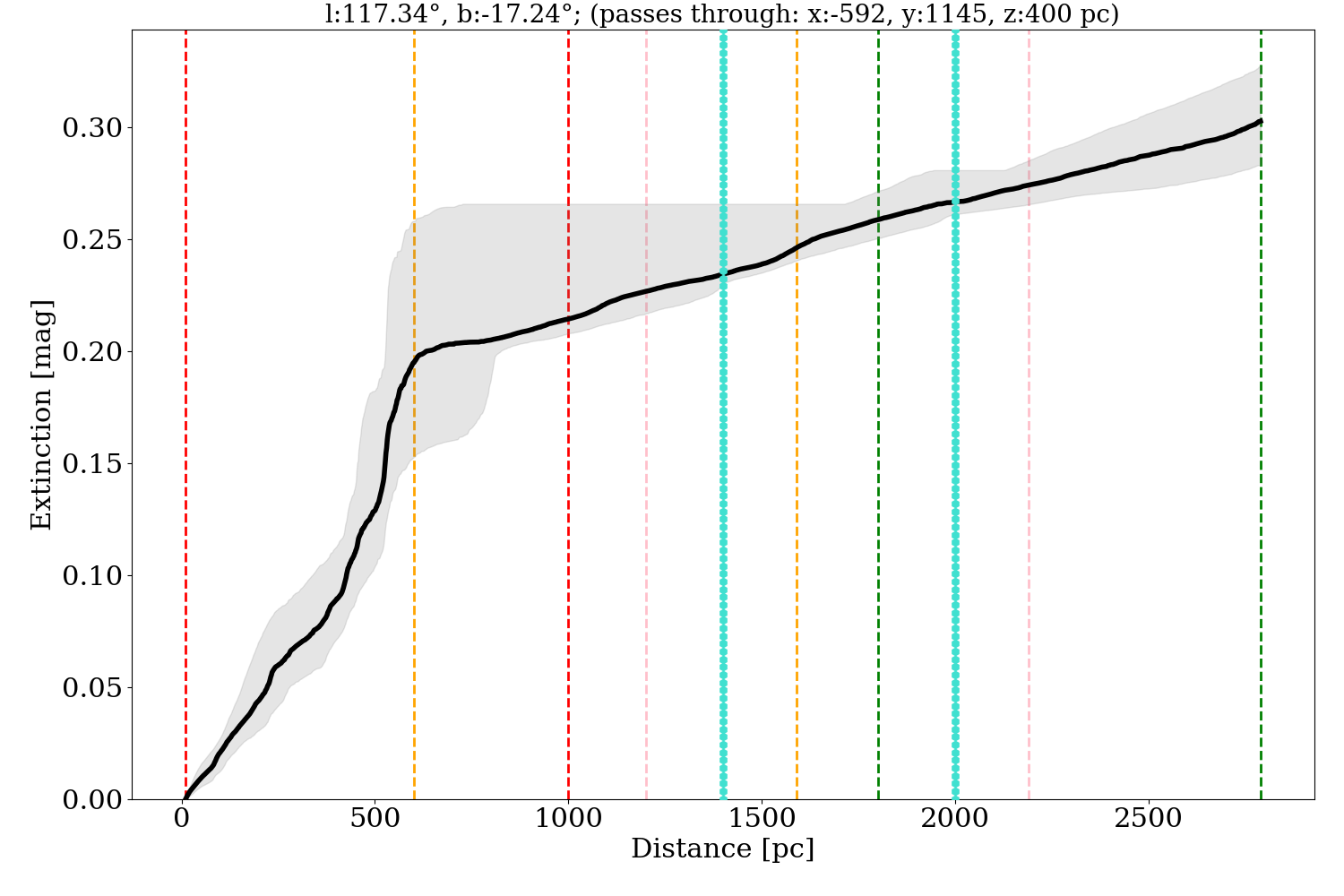}
    \includegraphics[width=0.27\textwidth]{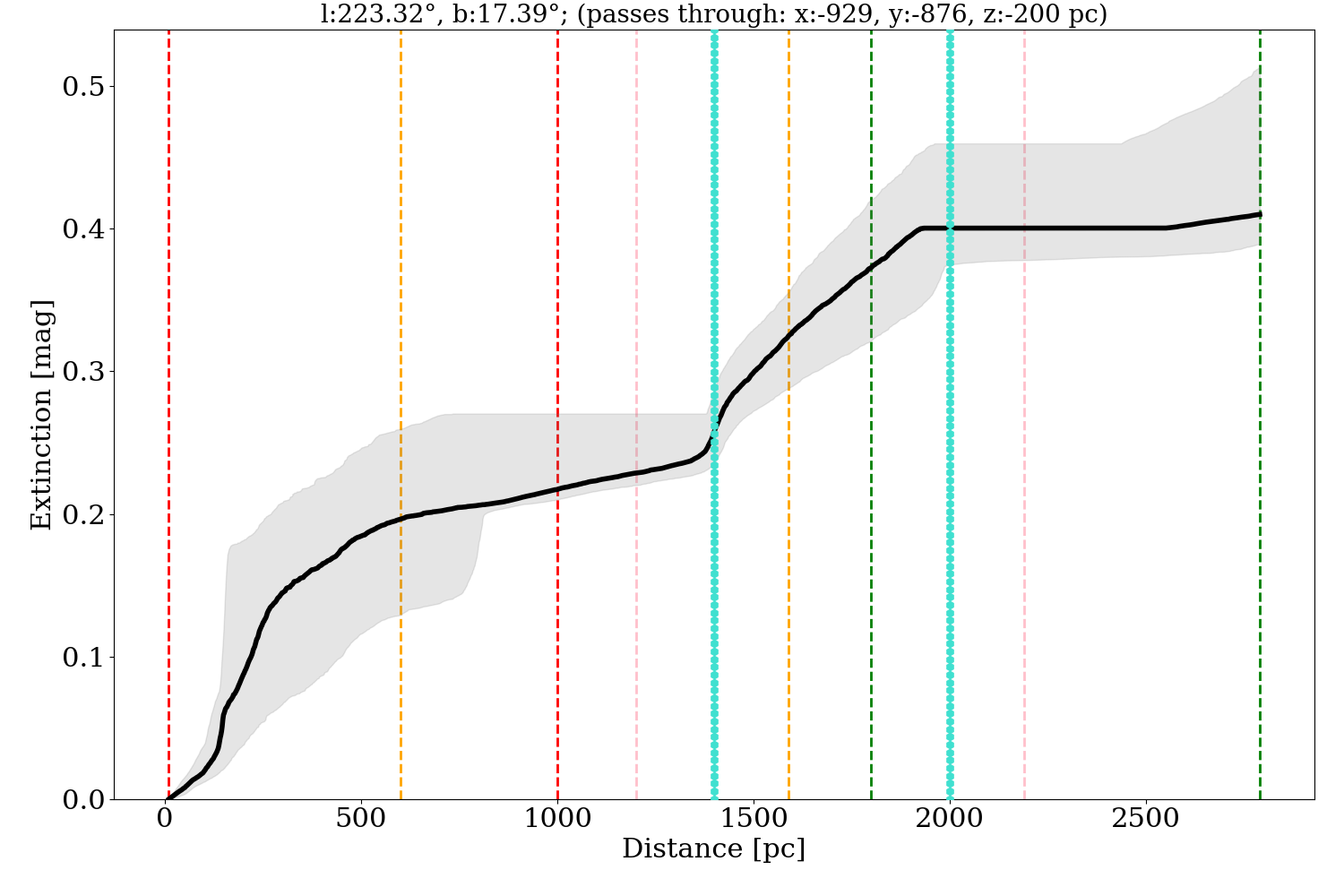}
    \includegraphics[width=0.27\textwidth]{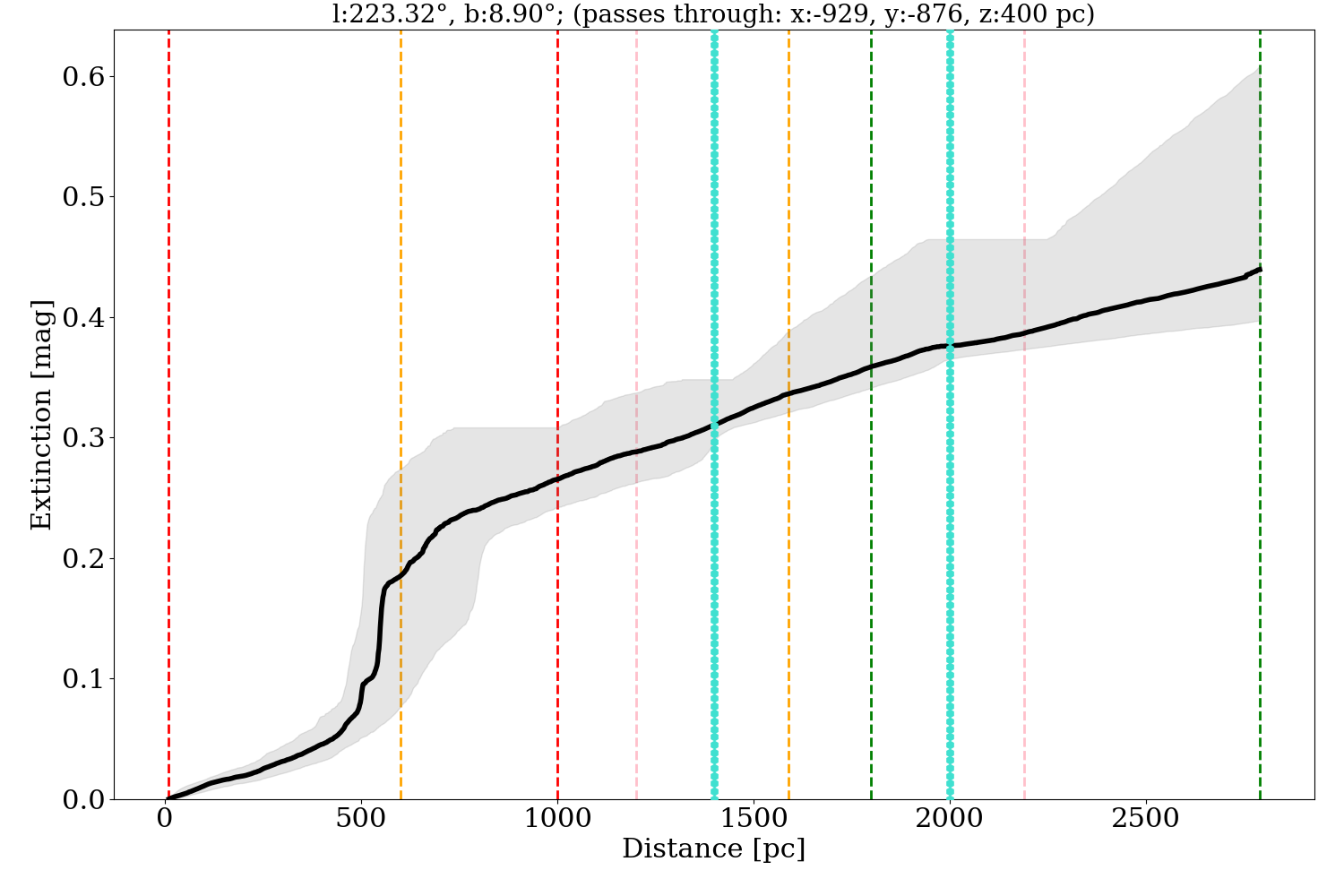}

    \includegraphics[width=0.27\textwidth]{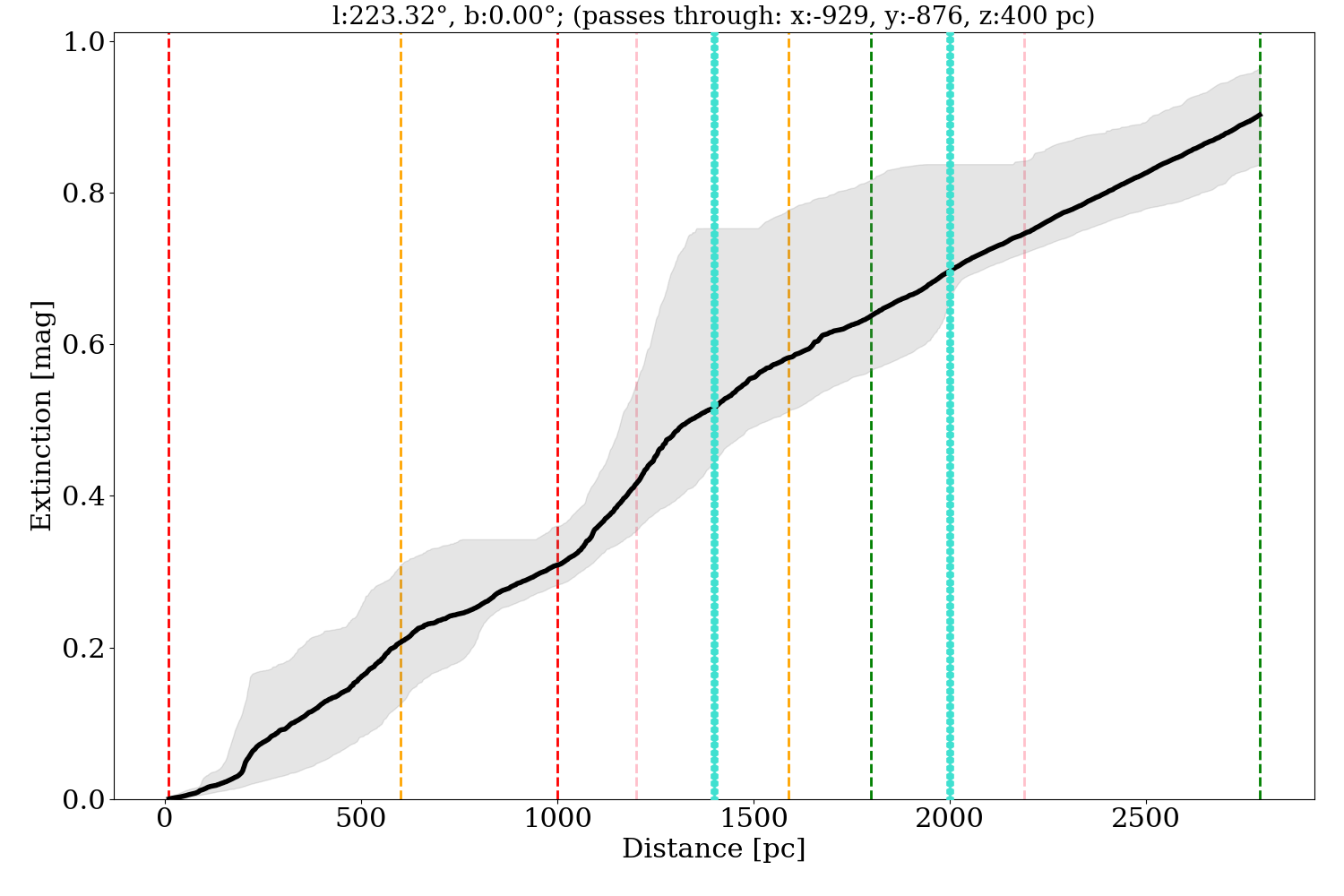}
    \includegraphics[width=0.27\textwidth]{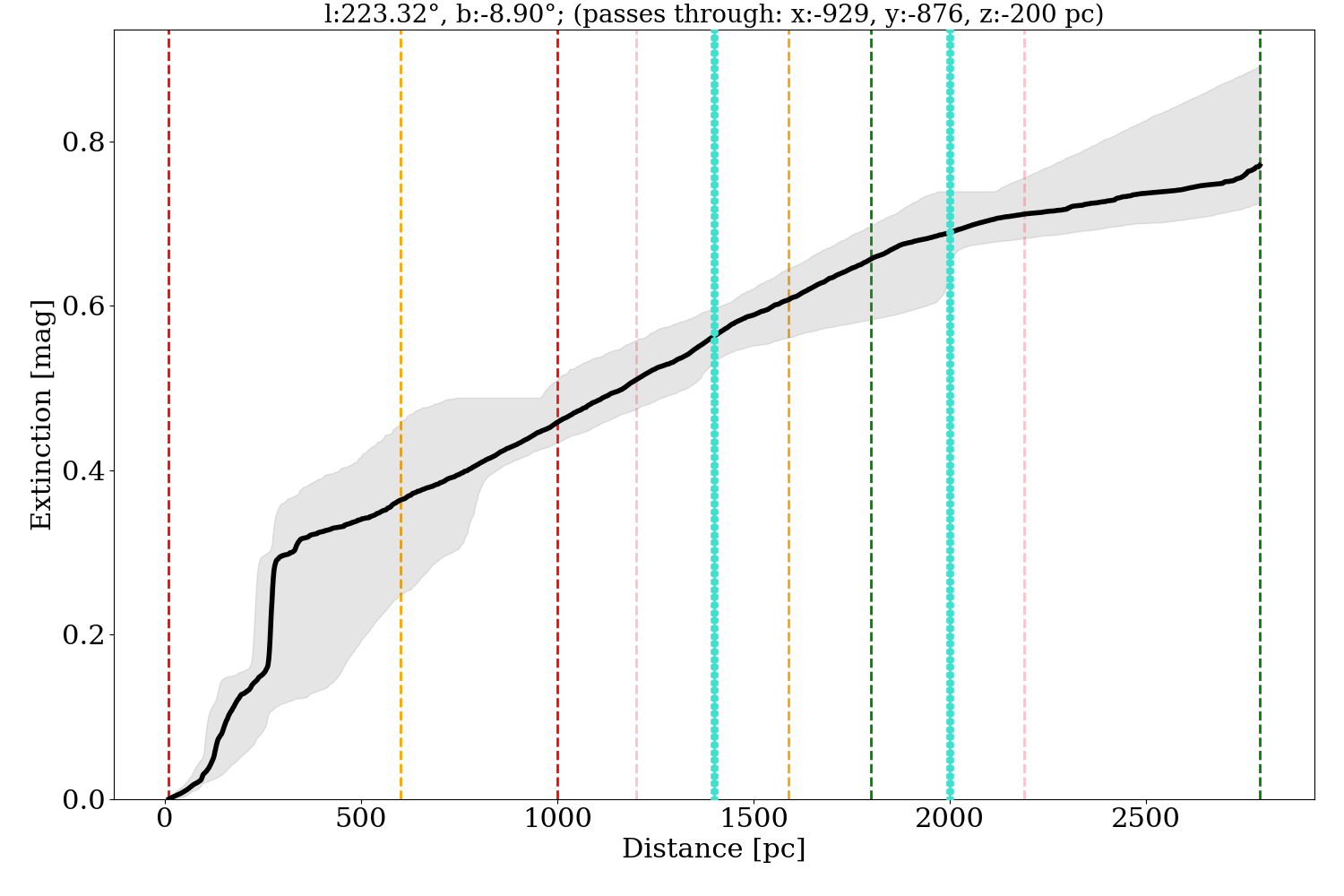}
    \includegraphics[width=0.27\textwidth]{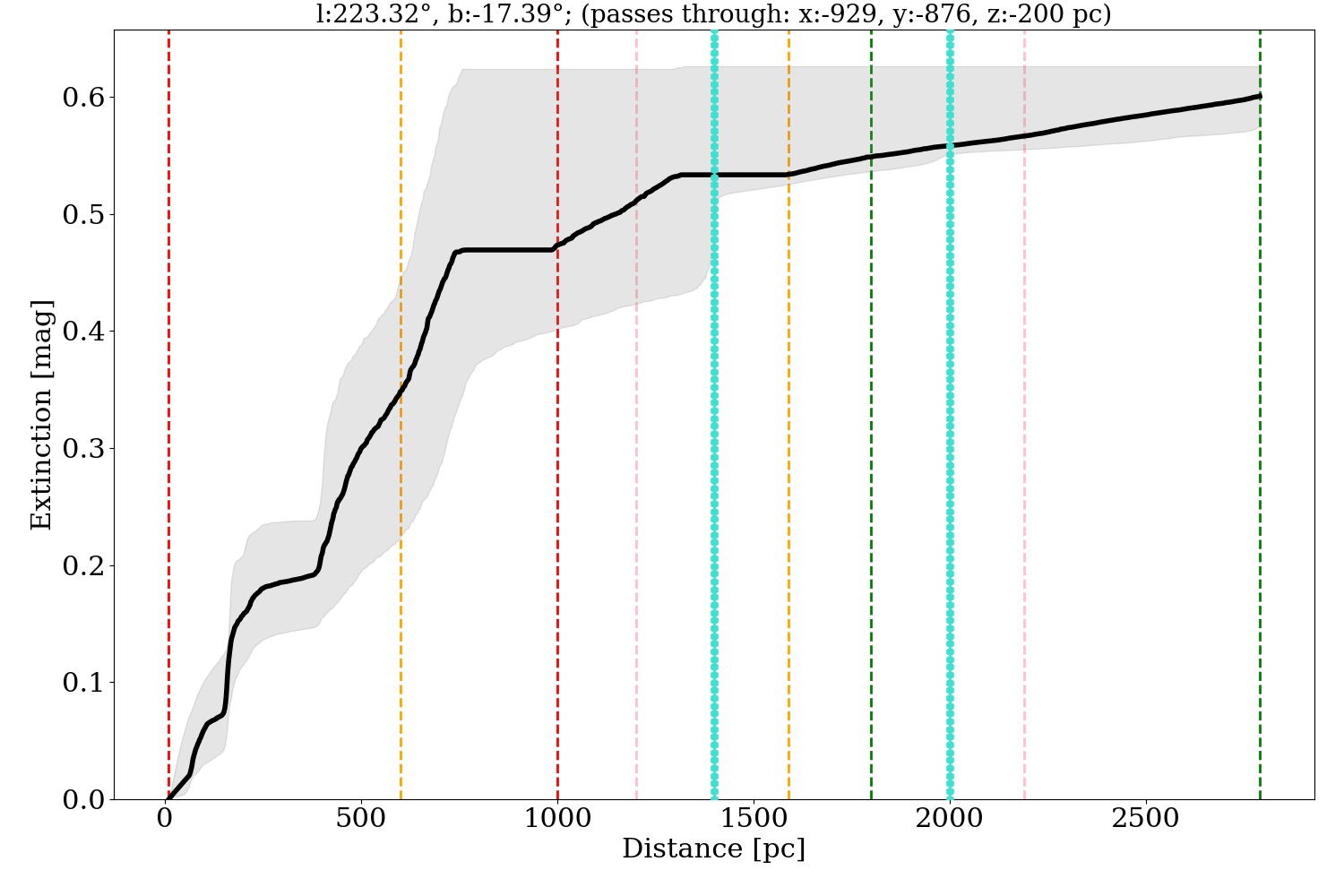}

    \includegraphics[width=0.27\textwidth]{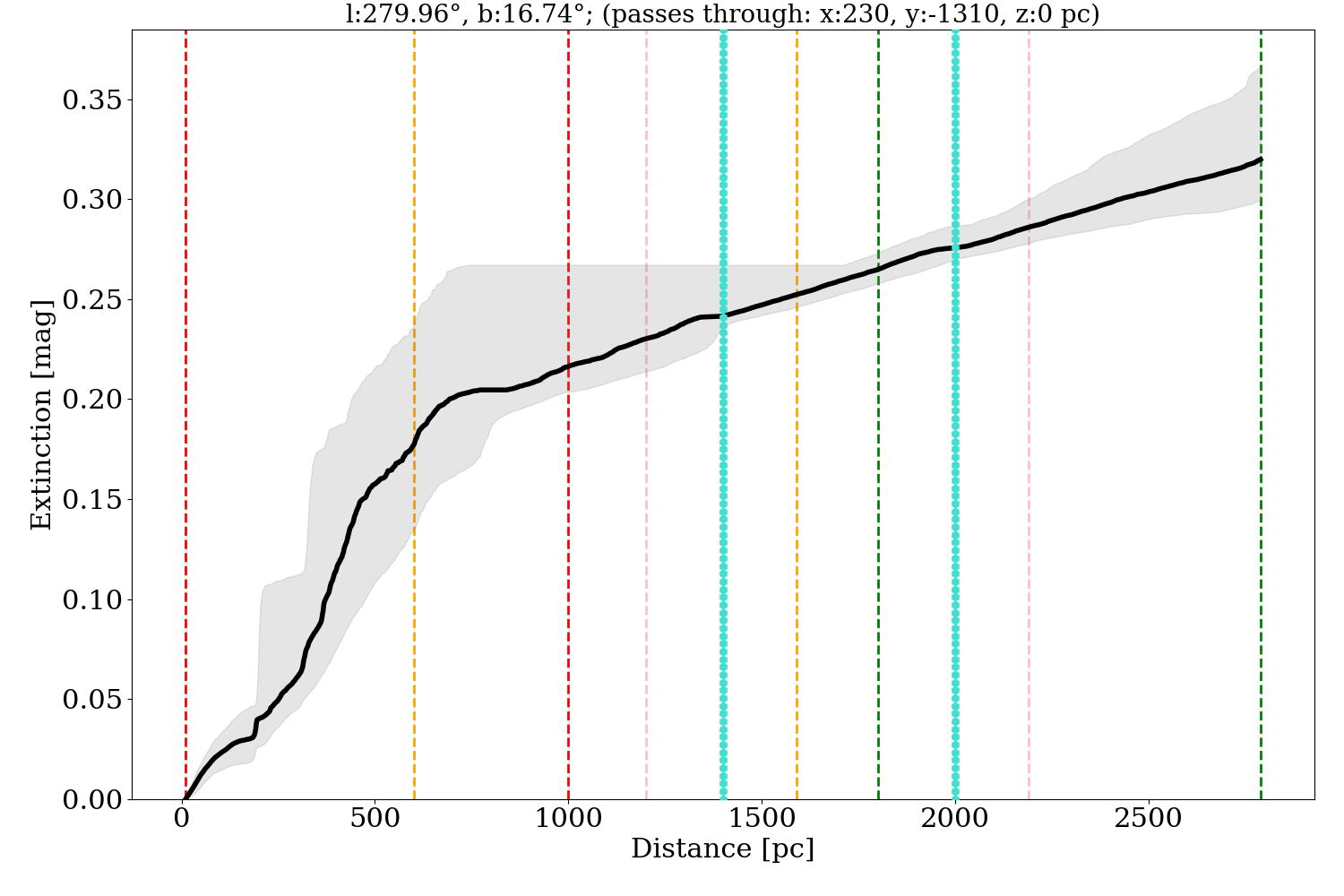}
    \includegraphics[width=0.27\textwidth]{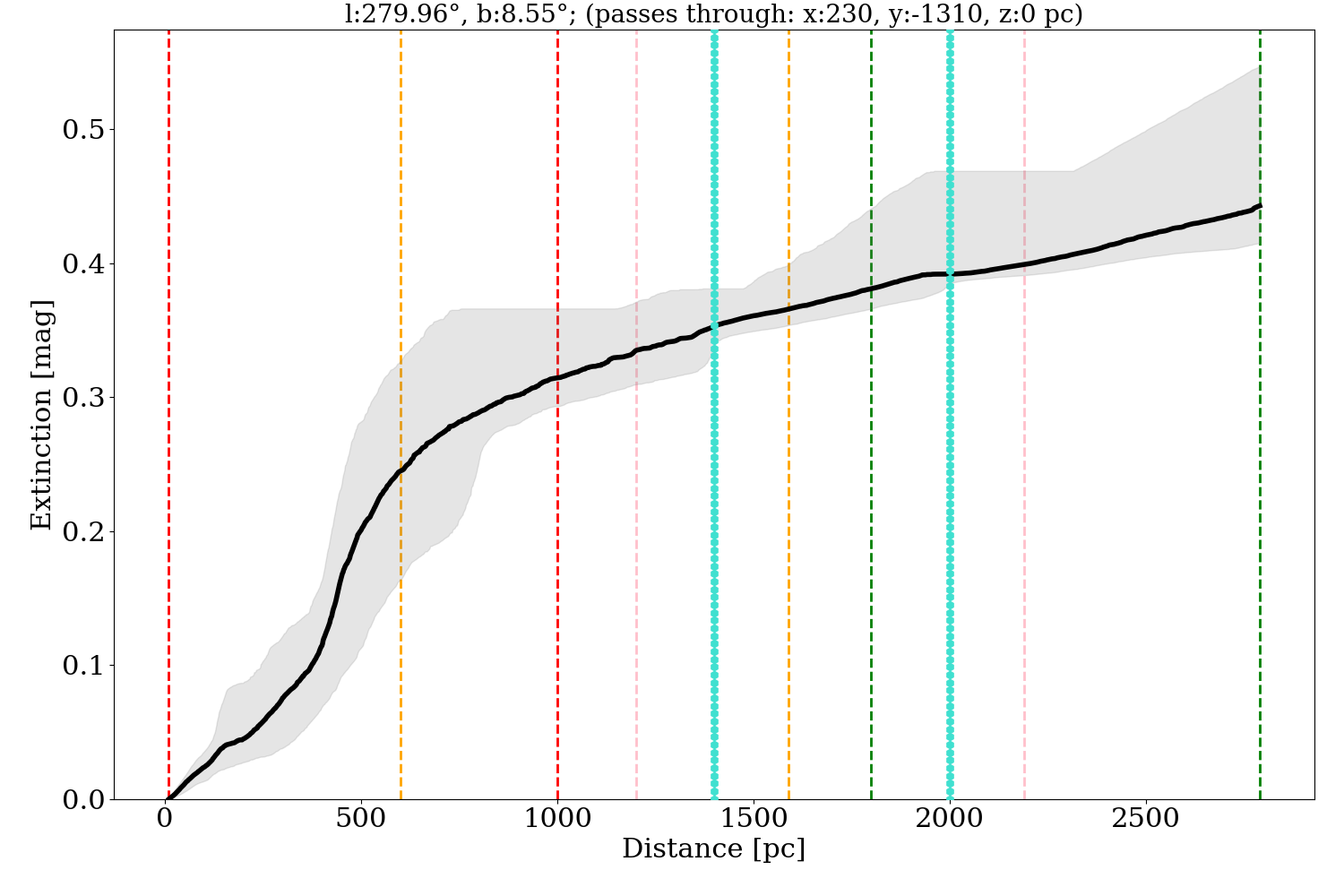}
    \includegraphics[width=0.27\textwidth]{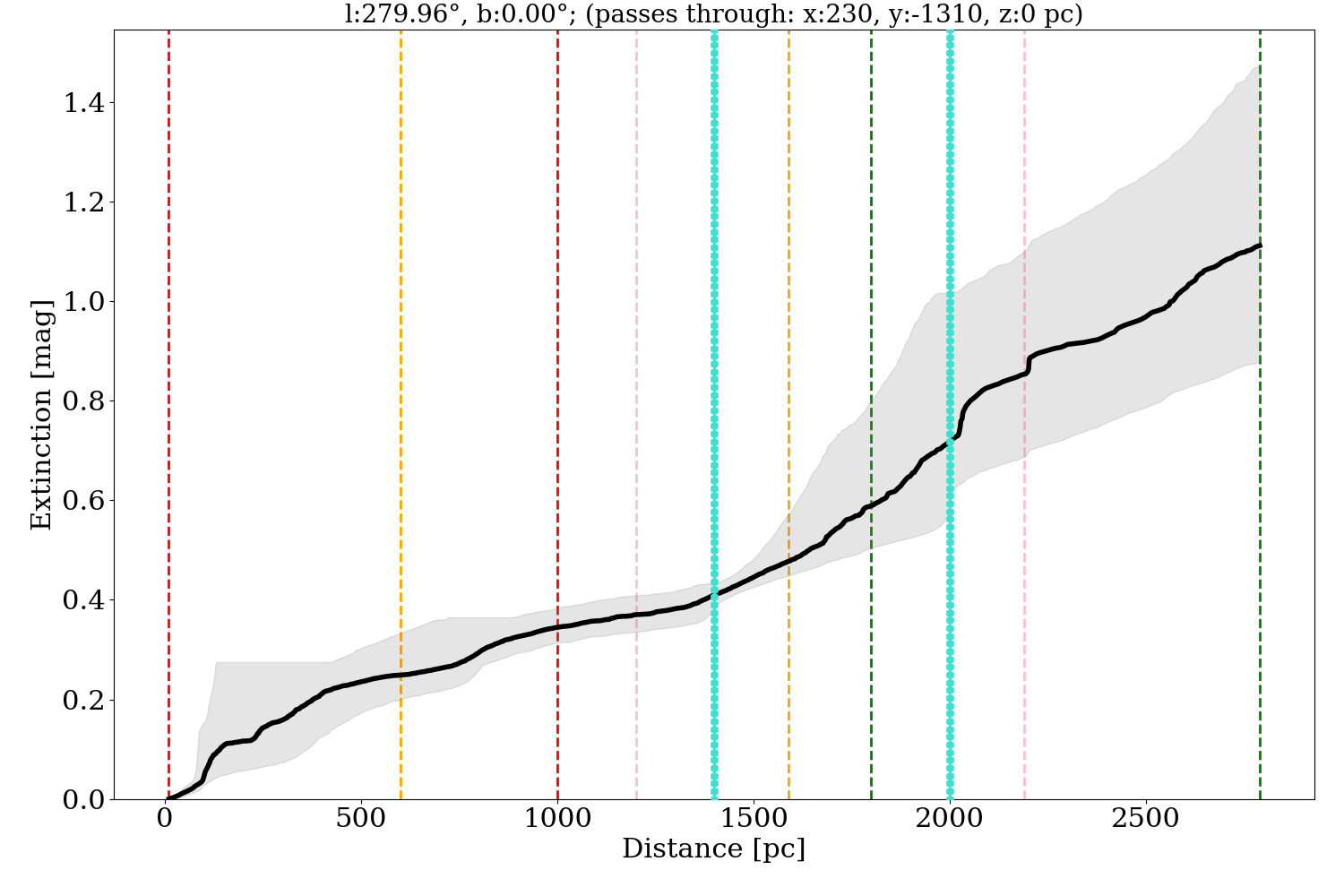}
    
    \includegraphics[width=0.27\textwidth]{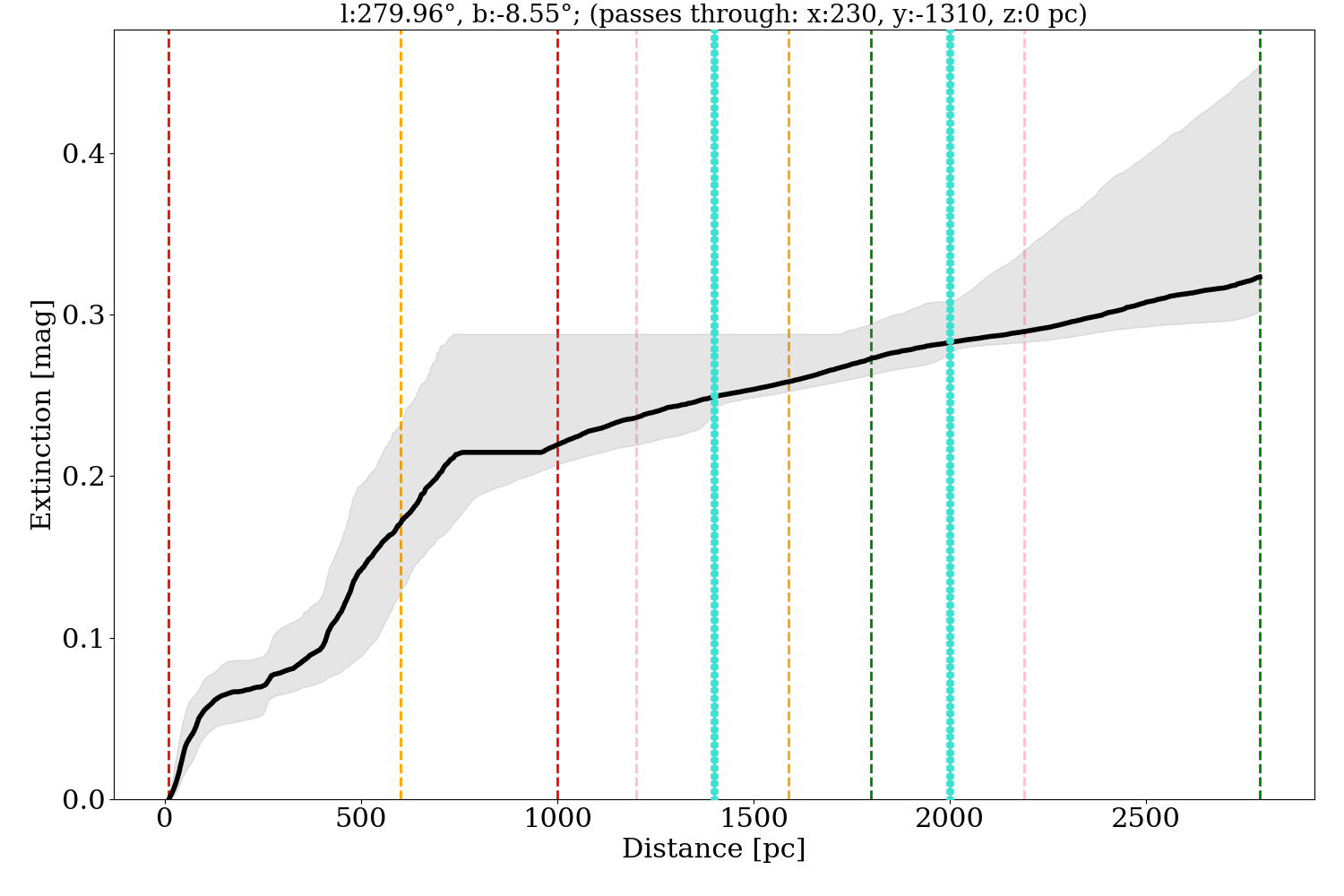}
    \includegraphics[width=0.27\textwidth]{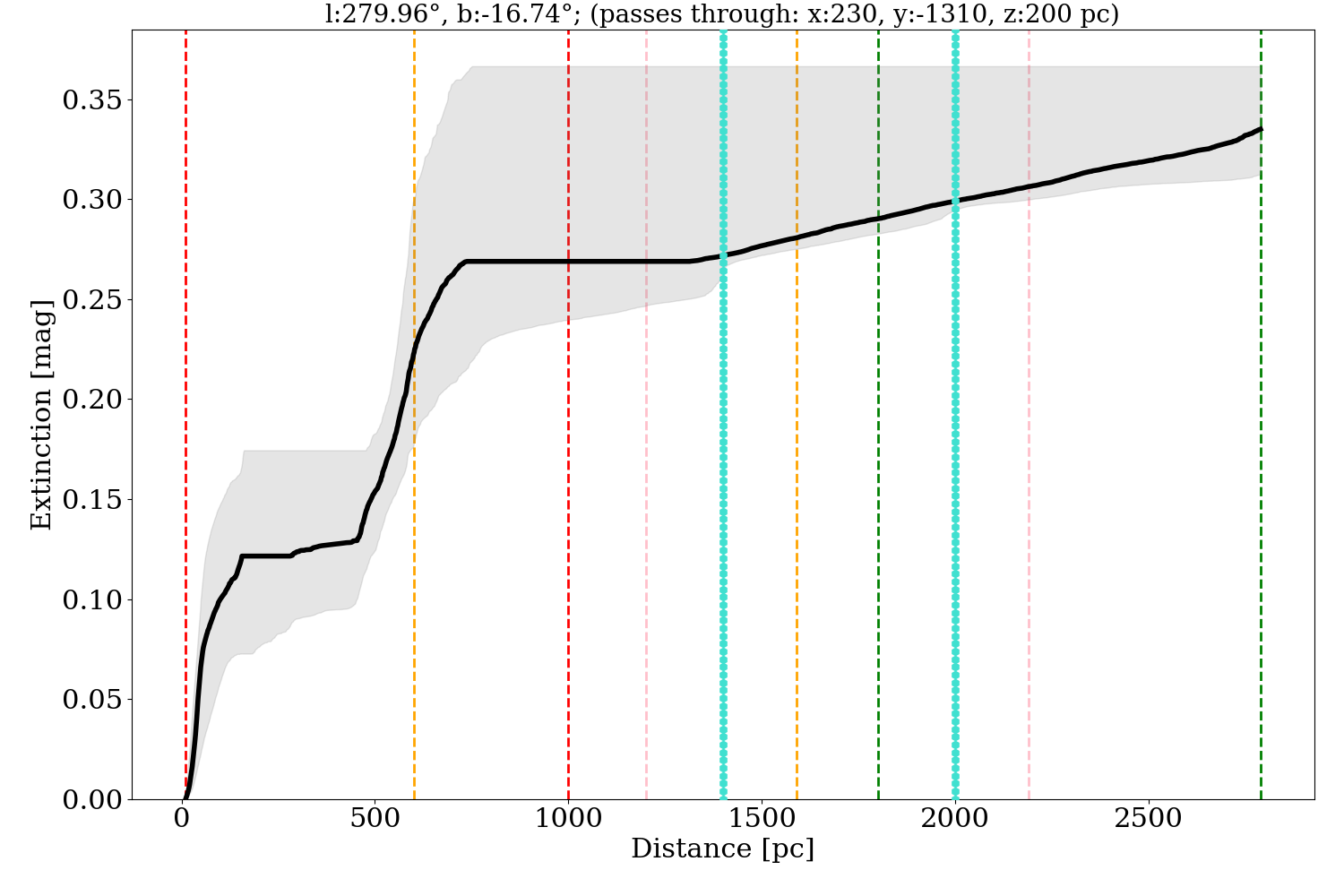}
    \includegraphics[width=0.27\textwidth]{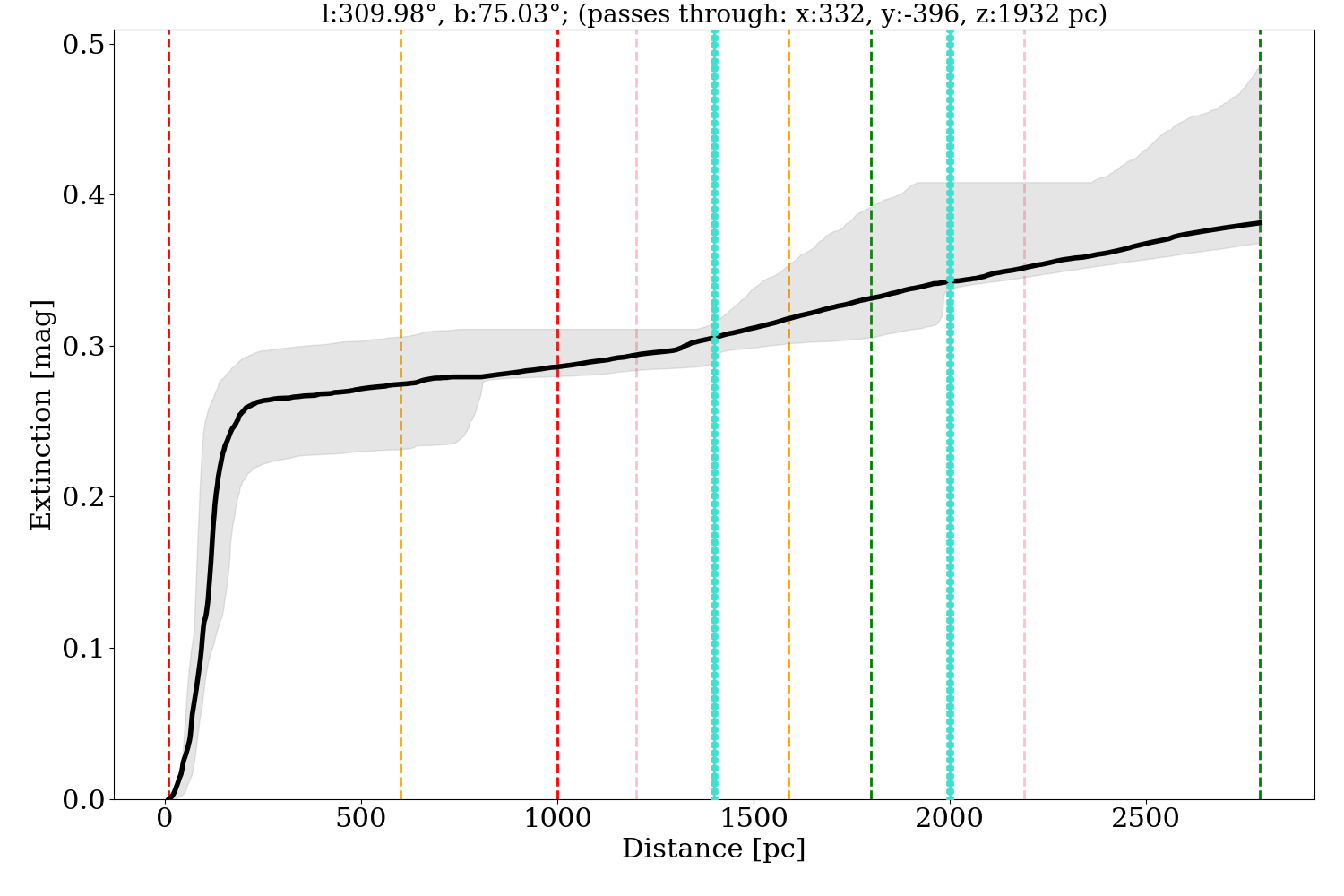}
    
    \caption{Extinction along lines of sight highlighting that any changes in the extinction at the location of the artefact are small compared to the uncertainty on the extinction. The cyan hexagonal lines drawn at $d=1400$ and 2000 pc highlights the possible location of artefacts as shown in Fig~\ref{fig:sat_ext}. The other coloured dashed lines represent the chunk boundaries as shown in Fig.~\ref{fig:MergeBoundaries}.}
    \label{fig:Ext_LOS}
\end{figure*}

% \begin{figure*}
%     \centering
%     \includegraphics[width=0.8\textwidth]%trim: left bottom right top
% {Figures/ExtinctionLoS.png}
%     \caption{The line-of-sight dust extinctions across the multiple lines-of-sight spanning the full 3D range of our map. The dashed vertical lines show the chunk's upper and lower boundaries ($d_{\rm min}$, $d_{\rm max}$).}
%     \label{fig:ExtLos}
% \end{figure*}

To test our merging technique for merging artefacts or any other issues we first choose a single $l,b,d$ chunk and predict new samples from the same trained GP in multiple overlapping $l,b,d$ \emph{mini chunks} to simulate the method carried out for the full Galaxy. This provides samples from the same ground truth to test the merging technique. 

We chose the chunk $l = 72^{\circ} - 90^{\circ}$,  $b = -7^{\circ} - 7^{\circ}$ and $d = 10 - 1000$\,pc to be the $main chunk$ which we split. The first set of $mini chunks$ range from $d= 10 - 500$ and the second set from $d= 300 - 800$; in the $l,b$ direction it is split into $9^{\circ} \times 2^{\circ}$ chunks. We now have a situation where we know exactly what density and extinction results that are expected in the $merged chunk$ because it must be identical to the $main chunk$. We then merge the smaller overlapping $l,b,d$ chunks using our merging method with the merging boundaries matching the ratios used when we merged the full Milky Way. In Fig.\ref{fig:MergeTest}, we see that our merging technique smoothly combines the data in all directions and reproduces the chosen single chunk well with no artefacts. 

Nevertheless, we wish to explore the combined map to see if there is any evidence of discontinuities resulting from the merging process. Figure~\ref{fig:MergeBoundaries} shows the density in the Galactic Plane with the boundaries of the chunks overlaid with dashed lines. At the locations of the boundaries, the map appears smooth on this colour scale, and hence we produce maps on compressed colour scales in figs.~\ref{fig:sat_dens} and \ref{fig:sat_ext}, which correspond to the maps shown in fig.~\ref{fig:xy_dustdense} top and fig.~\ref{fig:xy_dustInteg} in the main paper.  In fig.~\ref{fig:sat_dens} we see a possible weak discontinuity at a distance of 1300~pc where the density is below 0.0003~mag/pc, i.e., 100 times lower than the peak densities in our map, which is marked with purple hexagons. Figure~\ref{fig:sat_ext} shows a possible discontinuity in the integrated density at 1400~pc in the same directions as the weak discontinuity in fig.~\ref{fig:sat_dens}, at a somewhat higher level compared to the peak integrated density, possibly because of integration over dust above and below the plane. However, this difference in distance between the density and integrated density is the opposite of what would be expected if this were an artifact at fixed distance resulting from the merging - since the projected location in x,y of points above the Galactic plane at fixed distance would be inside the ring of points in the Galactic plane, and therefore any discontinuity would appear closer on plots of the integrated density. A second, less significant discontinuity is visible in the integrated density at a distance of roughly 2000~pc with no counterpart in density.

For the avoidance of doubt, we also plot the density and extinction along selected lines-of-sight in figs.~\ref{fig:Dens_LOS} and \ref{fig:Ext_LOS} along with their uncertainties. We see no evidence for discontinuities in these lines of sight, and in addition the uncertainties on both density and extinction are such that the potential discontinuities in figs.~\ref{fig:sat_dens} and \ref{fig:sat_ext} are typically smaller than the uncertainties at those points. Nevertheless, users of the data should remain cautious of over-interpreting features immediately at the boundary of a chunk. Therefore while the possible discontinuities are easily picked up by the human eye they are numerically small and will have little bearing on scientific analysis as demonstrated above.

\section{16$^{\rm th}$ and 84$^{\rm th}$ percentile maps of the \DustT
 produced 3D dust density}
\label{sec:app:Percentiles}

\begin{figure*}
    \centering
    \begin{subfigure}{0.45\textwidth}
    \includegraphics[width=\textwidth, trim=4.5cm 2cm 2.5cm 2cm, clip]%trim: left bottom right top
{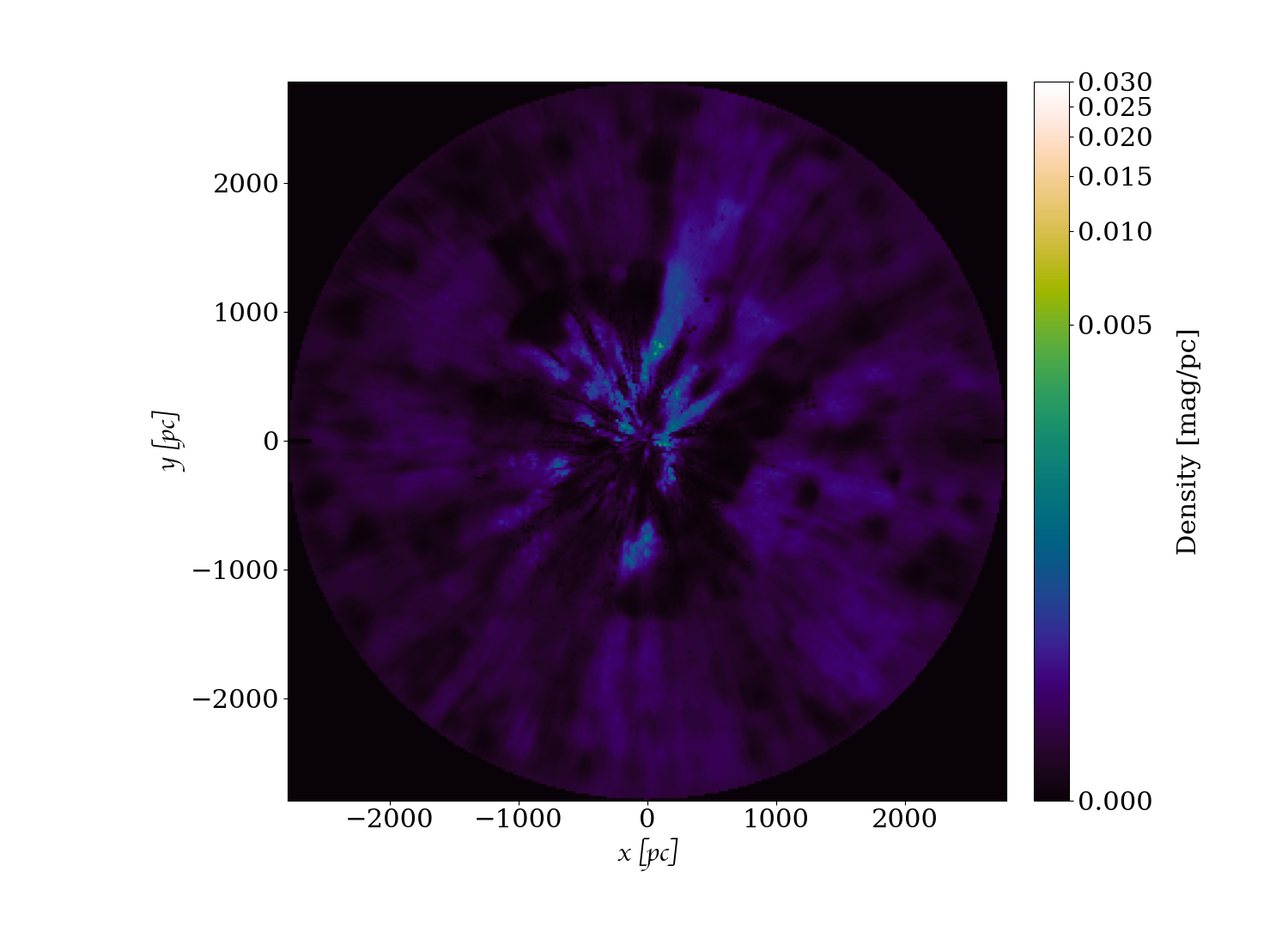}
    \end{subfigure}
    \begin{subfigure}{0.45\textwidth}
    \includegraphics[width=\textwidth, trim=4.5cm 2cm 2.5cm 2cm, clip]%trim: left bottom right top
{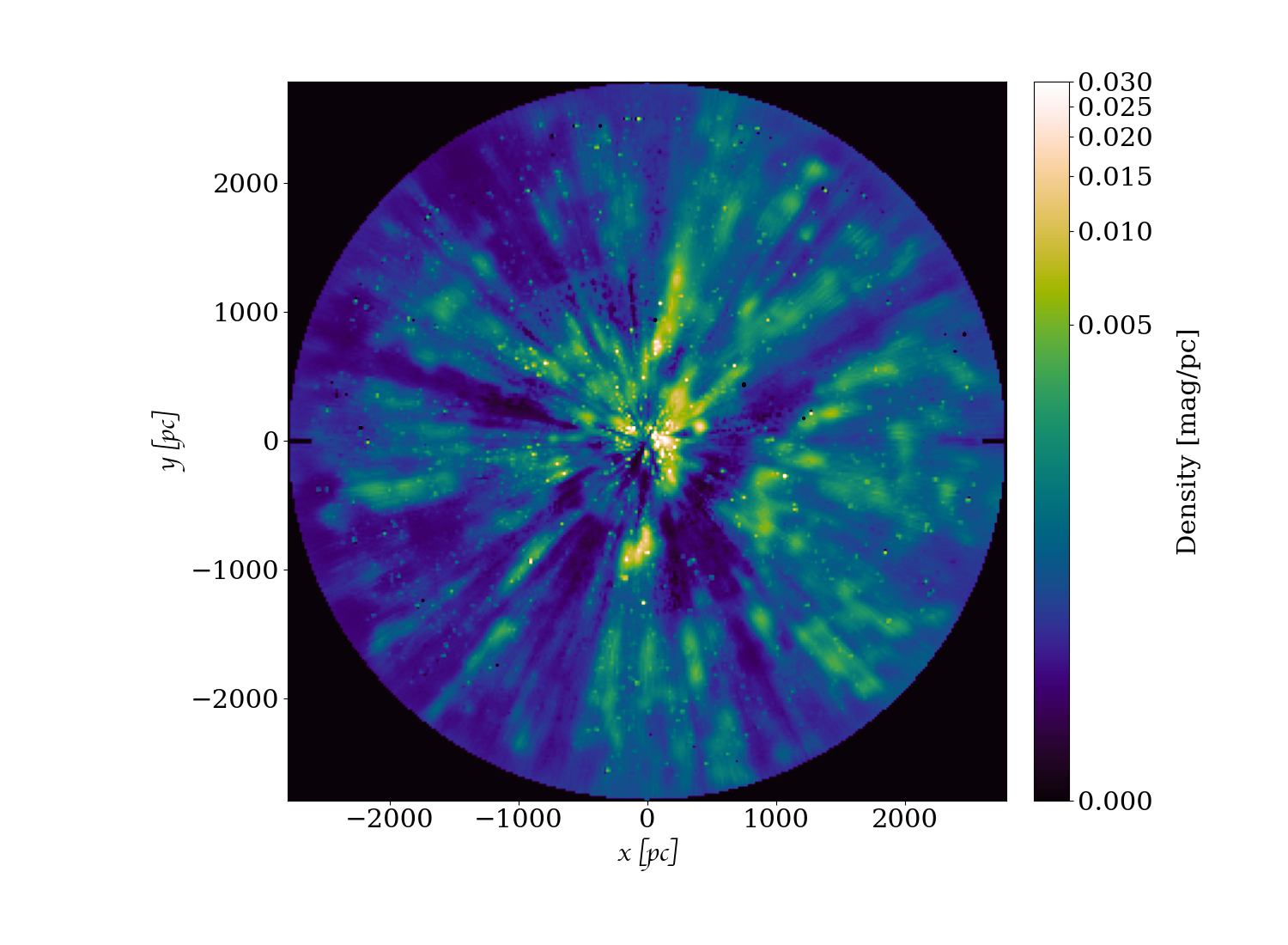}
    \end{subfigure}
    \caption{16$^{\rm th}$ (left) and 84$^{\rm th}$ (right) percentiles maps of the 3D dust density produced by \DustT, on the same colour scale as the median density shown in Fig.~\ref{fig:xy_dustdense}.}
    \label{fig:xy_Percentiles}
\end{figure*}

In the Fig.~\ref{fig:xy_Percentiles}, we show the 16$^{\rm th}$ and 84$^{\rm th}$ percentiles of our density map. This can be interpreted as a point-wise credible interval for the density posterior. Since we compute our GP in the logarithm of density, the credible interval is potentially very asymmetrical in density space.

\section{Further comparison to literature maps}\label{sec:app:litmaps}
\begin{figure*}
  \centering
   \includegraphics[width=\textwidth]{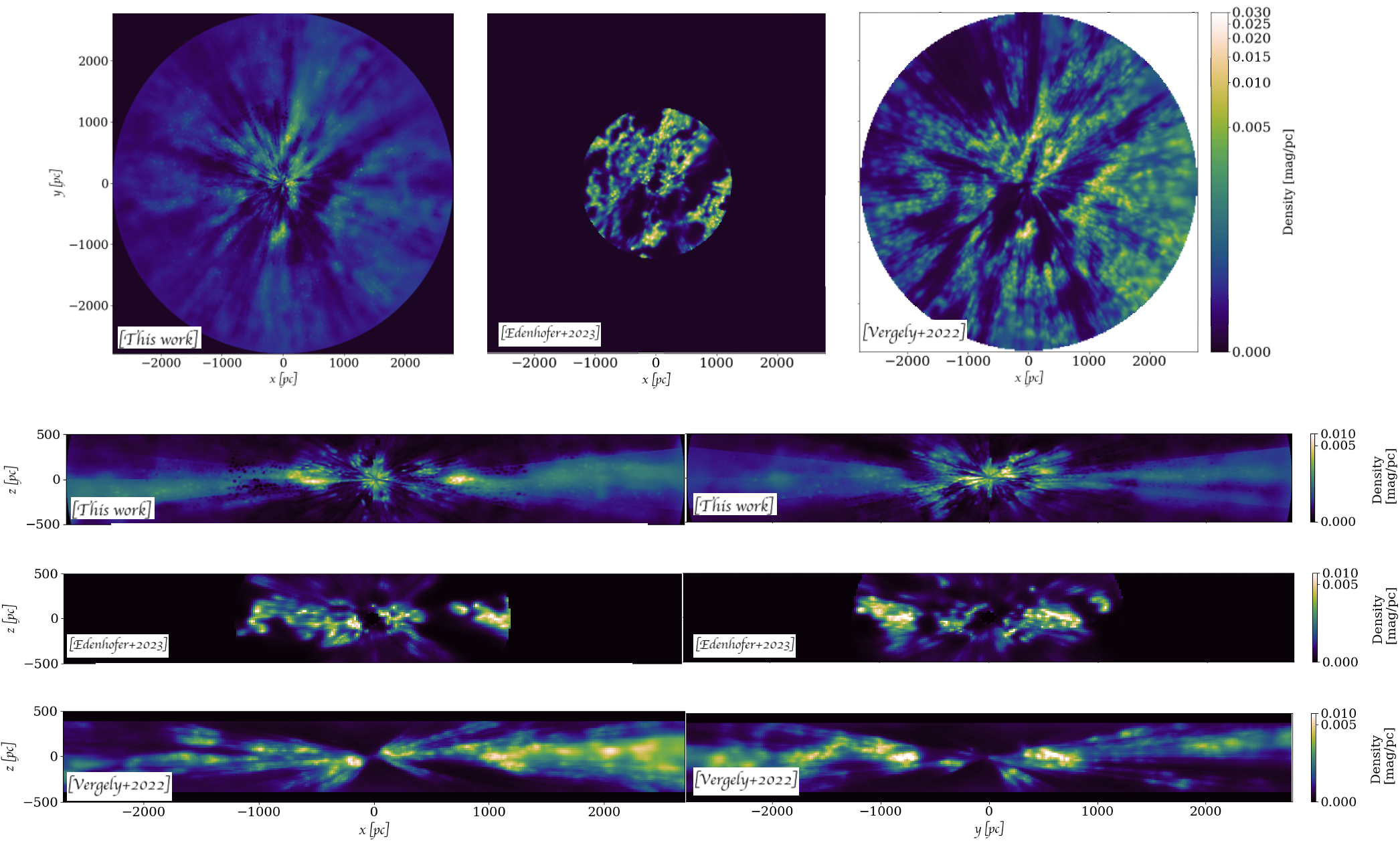}
 \caption{Comparison the measured dust density of our work to \citet{Edenhofer2023} and \citet{Vergely2022} at native sampling of their publicly available datasets prior to interpolating on to the same grid as our work for the numerical comparison in Sec.\ref{sec:LitCompMaps}. Top row: dust density at $z=0$ showing the $x,y$ plane for our work, \citet{Edenhofer2023} and \citet{Vergely2022}. Remaining rows left: dust density at $y=0$ showing the $x,z$ plane; Remaining rows right: dust density at $x=0$ showing the $y,z$ plane.}
    \label{fig:nativres_dens}
\end{figure*}

\begin{figure*}
    \centering
    \includegraphics[width=\textwidth]{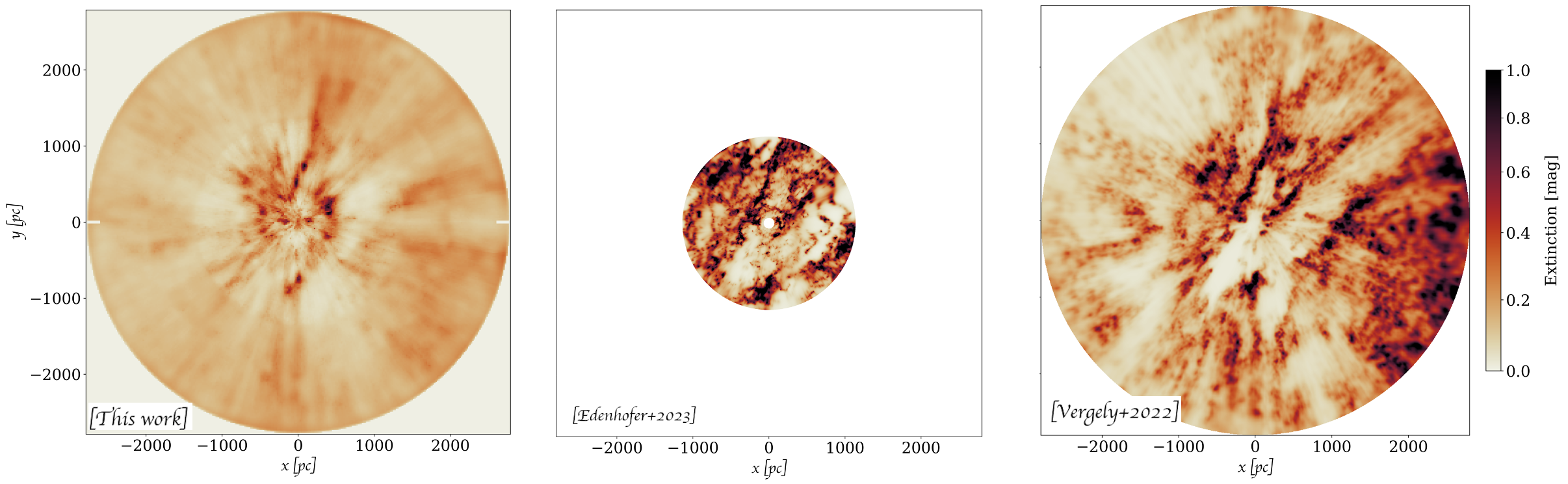}
    \caption{Comparison of the dust density of our work to \citet{Edenhofer2023} and \citet{Vergely2022}, integrated over the z-axis from $z=-500$~pc to $z=500$~pc at their native sampling.}
    \label{fig:nativres_ext}
\end{figure*}

For completeness, we show further comparisons of our map to those of \citet{Edenhofer2023} and \citet{Vergely2022}, where each map is shown with native sampling. In fig.~\ref{fig:nativres_dens} we show slices of the densities, while fig.~\ref{fig:nativres_ext} shows the integral along the z-axis.

%%%%%%%%% END OF PAPER %%%%%%%%%%%%%%%%%%

% Don't change these lines
\bsp	% typesetting comment
\label{lastpage}
\end{document}